\documentclass[twoside,twocolumn,9pt]{article}
\usepackage{extsizes}
\usepackage[super,sort&compress,comma]{natbib} 
\usepackage[version=3]{mhchem}
\usepackage[left=1.5cm, right=1.5cm, top=1.785cm, bottom=2.0cm]{geometry}
\usepackage{balance}
\usepackage{mathptmx}
\usepackage{sectsty}
\usepackage{graphicx} 
\usepackage{lastpage}
\usepackage[format=plain,justification=justified,singlelinecheck=false,font={stretch=1.125,small,sf},labelfont=bf,labelsep=space]{caption}
\usepackage{float}
\usepackage{amsbsy}
\usepackage{cuted}
\usepackage{fancyhdr}
\usepackage{fnpos}
\usepackage[english]{babel}
\addto{\captionsenglish}{%
  
}
\usepackage{array}
\usepackage{droidsans}
\usepackage{charter}
\usepackage[T1]{fontenc}
\usepackage[usenames,dvipsnames]{xcolor}
\usepackage{setspace}
\usepackage[compact]{titlesec}
\usepackage{hyperref}

\usepackage{epstopdf}

\definecolor{cream}{RGB}{222,217,201}

\begin{document}

\pagestyle{fancy}
\thispagestyle{plain}
\fancypagestyle{plain}{
\renewcommand{\headrulewidth}{0pt}
}
\def\etal{\textit{et al.}}
\makeFNbottom
\makeatletter
\renewcommand\LARGE{\@setfontsize\LARGE{15pt}{17}}
\renewcommand\Large{\@setfontsize\Large{12pt}{14}}
\renewcommand\large{\@setfontsize\large{10pt}{12}}
\renewcommand\footnotesize{\@setfontsize\footnotesize{7pt}{10}}
\makeatother

\renewcommand{\thefootnote}{\fnsymbol{footnote}}
\renewcommand\footnoterule{\vspace*{1pt}%
\color{cream}\hrule width 3.5in height 0.4pt \color{black}\vspace*{5pt}} 
\setcounter{secnumdepth}{5}

\makeatletter 
\renewcommand\@biblabel[1]{#1}            
\renewcommand\@makefntext[1]%
{\noindent\makebox[0pt][r]{\@thefnmark\,}#1}
\makeatother 
\renewcommand{\figurename}{\small{Fig.}~}
\sectionfont{\sffamily\Large}
\subsectionfont{\normalsize}
\subsubsectionfont{\bf}
\setstretch{1.125} 
\setlength{\skip\footins}{0.8cm}
\setlength{\footnotesep}{0.25cm}
\setlength{\jot}{10pt}
\titlespacing*{\section}{0pt}{4pt}{4pt}
\titlespacing*{\subsection}{0pt}{15pt}{1pt}

\fancyfoot{}
\fancyfoot[LO,RE]{\vspace{-7.1pt}\includegraphics[height=9pt]{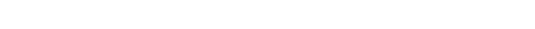}}
\fancyfoot[CO]{\vspace{-7.1pt}\hspace{11.9cm}\includegraphics{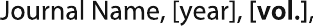}}
\fancyfoot[CE]{\vspace{-7.2pt}\hspace{-13.2cm}\includegraphics{head_foot/RF}}
\fancyfoot[RO]{\footnotesize{\sffamily{1--\pageref{LastPage} ~\textbar  \hspace{2pt}\thepage}}}
\fancyfoot[LE]{\footnotesize{\sffamily{\thepage~\textbar\hspace{4.65cm} 1--\pageref{LastPage}}}}
\fancyhead{}
\renewcommand{\headrulewidth}{0pt} 
\renewcommand{\footrulewidth}{0pt}
\setlength{\arrayrulewidth}{1pt}
\setlength{\columnsep}{6.5mm}
\setlength\bibsep{1pt}

\makeatletter 
\newlength{\figrulesep} 
\setlength{\figrulesep}{0.5\textfloatsep} 

\newcommand{\topfigrule}{\vspace*{-1pt}%
\noindent{\color{cream}\rule[-\figrulesep]{\columnwidth}{1.5pt}} }

\newcommand{\botfigrule}{\vspace*{-2pt}%
\noindent{\color{cream}\rule[\figrulesep]{\columnwidth}{1.5pt}} }

\newcommand{\dblfigrule}{\vspace*{-1pt}%
\noindent{\color{cream}\rule[-\figrulesep]{\textwidth}{1.5pt}} }

\makeatother

\twocolumn[
  \begin{@twocolumnfalse}

\vspace{1em}
\sffamily
\begin{tabular}{m{4.5cm} p{13.5cm} }

\includegraphics{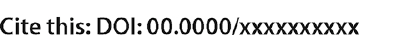} & \noindent\LARGE{\textbf{Temperature dependence of the static permittivity and integral formula for the Kirkwood correlation factor of simple polar fluids$^\dag$}} \\
\vspace{0.3cm} & \vspace{0.3cm} \\

 & \noindent\large{Pierre-Michel D\'{e}jardin\textit{$^{a}$}, Florian Pabst\textit{$^{b}$}, Yann Cornaton\textit{$^{c}$}, Cyril Caliot\textit{$^{d}$}, Robert Brouzet\textit{$^{a}$}, Andreas Helbling\textit{$^{b}$} and Thomas Blochowicz\textit{$^{b}$}}\\
 

\includegraphics{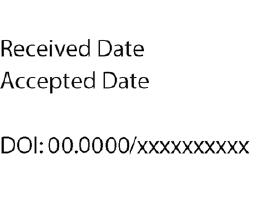} & \noindent\normalsize{An exact integral formula for the Kirkwood correlation factor of isotropic polar fluids $g_{\mathrm{K}}$ is derived from the equilibrium averaged rototranslational Dean equation, which as compared to previous approaches easily lends itself to further analytical approximations. The static linear permittivity of polar fluids $\varepsilon$ is calculated as a function of temperature, density and molecular dipole moment in vacuo for arbitrary pair interaction potentials. Then, using the Kirkwood superposition approximation for the three-body orientational distribution function, we suggest a simple way to construct model potentials of mean torques considering permanent and induced dipole moments. We successfully compare the theory with the experimental temperature dependence of the static linear permittivity of various polar fluids such as a series of linear monohydroxy alcohols, water, tributyl phosphate, acetonitrile, acetone, nitrobenzene and dimethyl sulfoxide, by fitting only one single parameter, which describes the induction to dipole-dipole energy strength ratio. We demonstrate that comparing the value of $g_{\mathrm{K}}$ with unity in order to deduce the alignment state of permanent dipole pairs, as is currently done is in many situations, is a misleading oversimplification, while the correct alignement state is revealed when considering the proper interaction potential. Moreover we show, that picturing  H-bonding polar fluids as polar molecules with permanent and induced dipole moments  without invoking any specific H-bonding mechanism is in many cases sufficient to explain experimental data of the static dielectric constant. In this light, the failure of the theory to describe the experimental temperature dependence of the static dielectric constant of glycerol, a non-rigid polyalcohol, is not due to the lack of specific H-bonding mechanisms, but rather to an oversimplified model potential for that particular molecule.} \\

\end{tabular}

 \end{@twocolumnfalse} \vspace{0.6cm}

  ]

\renewcommand*\rmdefault{bch}\normalfont\upshape
\rmfamily
\section*{}
\vspace{-1cm}


\footnotetext{\textit{$^{a}$~Laboratoire de Math\'{e}matiques et Physique, Universit\'{e} de Perpignan, F-66860 Perpignan, France}}
\footnotetext{\textit{$^{b}$~Institute of Condensed Matter Physics, Technische Universit\"{a}t Darmstadt, D-64289 Darmstadt, Germany }}
\footnotetext{\textit{$^{c}$~Laboratoire de Chimie et Syst\'{e}mique Organo-M\'{e}talliques, Institut de Chimie de Strasbourg, Universit\'{e} de Strasbourg, F-67000 Strasbourg, France}}
\footnotetext{\textit{$^{d}$~Laboratoire de Math\'{e}matiques et de leurs Applications, Universit\'{e} de Pau et des Pays de l'Adour, F-64000 Anglet, France}}



\tableofcontents
\twocolumn

\section{Historical background and motivation}\label{Intro}
The theory of the linear dielectric constant of isotropic polar fluids $\varepsilon$ has a long history which started in the early 20$^{th}$ century and has been developed until nowadays. We recall in this Introduction all the developments and improvements of the theory, starting with the pioneering work of Debye and Lorentz. We further motivate the present work by focusing on a quantity that is vital in the theory of simple isotropic polar fluids, the Kirkwood correlation factor, which essentially contains information on pair dipole ordering and explain why a new theoretical approach is necessary for its estimate.
\subsection{The origins : Debye's and Lorentz's theories}
The theory of the linear static permittivity of isotropic polar fluids $\varepsilon$ was initiated by Debye,\cite{Debye1929Book} who demonstrated that $\varepsilon$ can be linked to the molecular properties of the fluid under study. As is well-known, his theory can be applied to very dilute polar substances but fails at liquid densities because it completely ignores long range intermolecular interactions. 

A first step to include such interactions was accomplished by Lorentz.\cite{Bottcher1973Book} To this purpose, he introduced the concept of internal field at a typical molecule $\mathbf{E}_{i}$, which is made of the field due to all molecules save the one under focus plus external field. In order to calculate this field, he used the following procedure. He selected one molecule (the target or tagged molecule) and drew a macroscopic sphere of radius $R$ centered at the molecule (we alternatively term this sphere the Lorentz cavity or the inner Lorentz sphere throughout), assimilated to a point dipole (this sphere is still smaller than the size of the dielectric itself). All molecules inside the so-formed (Lorentz) cavity are treated on a discrete basis, while the molecules outside the sphere are treated on a continuous one. Then, the local field is calculated as the vector sum of two contributions : the field inside the sphere $\mathbf{E}_{in}$ and that outside the sphere $\mathbf{E}_{out}$. Under quite general conditions, we have from macroscopic electrostatics
\begin{eqnarray}
\mathbf{E}_{out}=\mathbf{E}+\frac{\mathbf{P}}{3\varepsilon_0}
\end{eqnarray}
where $\mathbf{E}$ is the Maxwell field (i.e., the macroscopic electric field inside matter when the latter is treated as continuous), $\mathbf{P}$ is the macroscopic polarization vector and $\varepsilon_0$ the absolute permittivity of vacuum. The computation of $\mathbf{E}_{in}$ is more intricate. Lorentz showed, however, that if the molecules inside the inner sphere are located at the sites of a simple cubic lattice, then whatever its size one has :
\begin{eqnarray}
\mathbf{E}_{in}=\mathbf{0}
\end{eqnarray}
so that in this specific situation, the Lorentz internal field is
\begin{eqnarray}
\mathbf{E}_{i}=\mathbf{E}_{out}=\mathbf{E}+\frac{\mathbf{P}}{3\varepsilon_0}
\label{LorentzField}
\end{eqnarray}
For a spherically shaped dielectric, the Maxwell field is given by 
\begin{eqnarray}
\mathbf{E}=\mathbf{E}_{0}-\frac{\mathbf{P}}{3\varepsilon_0}
\end{eqnarray}
where $\mathbf{E}_{0}$ is the field created by charges external to the dielectric (which we term the externally applied field). Then, the Lorentz field Eq.\eqref{LorentzField} becomes
\begin{eqnarray}
\mathbf{E}_{i}=\mathbf{E}_{0}
\label{LorentzFieldBis}
\end{eqnarray}
that is, no distinction can be made between the internal field and the externally applied field. Assuming in this paragraph polar nonpolarizable molecules for simplicity, the equation of state for linear dielectrics is derived in two simple ways in two subparagraphs, by equating the macroscopic (linear) polarization in the direction of the applied field $P$ to the one calculated by means of statistical mechanics $\Pi$.
\subsubsection{Derivation of the Debye-Lorentz equation : first way}
In this way, the macroscopic polarization of a spherical dielectric specimen is related to the externally applied field $E_{0}$ by the equation \cite{Bottcher1973Book}
\begin{eqnarray}
P=3\varepsilon_{0}\frac{\varepsilon-1}{\varepsilon+2}E_{0}
\label{MacropolE0}
\end{eqnarray}
In order to calculate the microscopic polarization (i.e. the polarization computed by microscopic means), one uses the equation
\begin{eqnarray}
\Pi=\rho_{0}\mu_{g}\langle\mathbf{u}\cdot\mathbf{e}_{i}\rangle
\label{Micropol}
\end{eqnarray}
where $\mu_{g}$ is the molecular dipole moment of a molecule in the ideal gas phase, $\rho_{0}$ the number of molecules per unit volume,$\mathbf{e}_{i}$ a unit vector along the Lorentz internal field that interacts with the dipole moment vector of the tagged molecule (so that in the Lorentz theory, the internal field acts as the directing field \cite{Bottcher1973Book}), and the angular brackets $\langle\rangle$ denote a statistical average over all orientations of the tagged dipole. Indeed, the computation of the average is limited to the linear response of the dipole to $E_{i}$, yielding \cite{Debye1929Book}
\begin{eqnarray}
\langle\mathbf{u}\cdot\mathbf{e}_{i}\rangle\approx\frac{\beta\mu^{2}}{3}E_{i}
\label{DebyeAv}
\end{eqnarray}
where $\beta=(kT)^{-1}$, $k$ is Boltzmann's constant and $T$ the absolute temperature. By combining Eqs. \eqref{LorentzFieldBis}-\eqref{DebyeAv}, we have the Debye-Lorentz (Clausius-Mossotti) equation of state for linear dielectrics, viz.
\begin{eqnarray}
\frac{\varepsilon-1}{\varepsilon+2}=\frac{\lambda}{3}
\label{DebyeLorentzEq}
\end{eqnarray}
where $\lambda=\beta\rho_{0}\mu^{2}/(3\varepsilon_{0})$ is the linear susceptibility of an assembly of polar molecules in the ideal gas phase. The intensivity of $\varepsilon$ is supposed here as it might appear that in deriving Eq. \eqref{DebyeLorentzEq} this result depends on the shape of the sample, in spite of the fact that \textit{there is no mistake in this algebra}.
\subsubsection{Derivation of the Debye-Lorentz equation : second way}
In fact, the shape dependence of Eq. \eqref{DebyeLorentzEq} is only artificial, because one may derive this equation in terms of the Lorentz field as given by Eq. \eqref{LorentzField}. The macroscopic polarization is written in terms of the Maxwell field $\mathbf{E}$ so that $P$ is given by
\begin{eqnarray}
P=\varepsilon_{0}(\varepsilon-1)E
\label{MacropolE}
\end{eqnarray}
so that the Lorentz internal field \eqref{LorentzField} becomes, in terms of the Maxwell field
\begin{eqnarray}
\mathbf{E}_{i}=\frac{\varepsilon+2}{3}\mathbf{E}
\end{eqnarray}
and is collinear with the Maxwell field. The microscopic polarization in terms of the Maxwell field is then given by (linear response to $E_{i}$ is also assumed here)
\begin{eqnarray}
\Pi=\frac{\beta\rho_{0}\mu^{2}}{3\varepsilon_{0}}\frac{\varepsilon+2}{3}E
\label{MicropolE}
\end{eqnarray}
so that combining the two above equations again leads to the Debye-Lorentz equation of state, Eq.\eqref{DebyeLorentzEq}, with the difference however that the sample shape dependence contained in the Maxwell field has been eliminated. For this reason, it is the relation between the internal and Maxwell field which must be specified in order to derive linear dielectric equations of state.
Several objections to the Debye-Lorentz equation of state were made, the most prominent being that it has a ferroelectric Curie point at some temperature. Nevertheless, it is not legitimate that in polar fluids, molecules are located at the sites of a simple cubic lattice, meaning that in general, there is not reason to believe that $\mathbf{E}_{in}=\mathbf{0}$. However, the Lorentz field becomes
\begin{eqnarray}
\mathbf{E}_{i}=\mathbf{E}_{out}=\mathbf{E}+\frac{\mathbf{P}}{3\varepsilon_0}+\mathbf{E}_{in}
\end{eqnarray}
where the relation between $\mathbf{E}_{in}$ and the Maxwell field $\mathbf{E}$ is unknown, and a \textit{simple} equation of state for linear dielectrics can no longer be derived. Moreover, the Lorentz cavity is a virtual one only, by which it is meant that the polarization of the Lorentz cavity does not adapt itself to the polarization of the surrounding continuum. 
\subsection{Onsager's theory} 
We use CGS units throughout this paragraph and the following others for the derivations set in this Introduction, because they are more convenient in reality for theoretical calculations, and restore SI units in the final formula.
In 1936, Onsager \cite{Onsager1936JACS} suggested that Lorentz's approach to the calculation of the internal field was probably not the best one, because the effect of long-range dipole-dipole interactions is not accounted for properly in the Lorentz version. He therefore altered the method to include the effect of the dipole moment vector $\vec{\mu}$ on the local field at that molecule. He used as a model that of a point rigid dipole situated at the centre of an empty spherical cavity of radius $a$ (the radius of the volume available to each molecule) in a dielectric continuum with static permittivity equal to the bulk dielectric constant $\varepsilon$. The radius $a$ of the cavity is given by the close-packing condition
\begin{eqnarray}
\frac{4\pi\rho_{0}a^3}{3}=1
\end{eqnarray}
Now Onsager considers that the dipole $\vec{\mu}$ polarizes the surroundings. The polarization of the surroundings in turn induces a uniform field in the cavity, the reaction field $\mathbf{R}$. For a spherical cavity, $\mathbf{R}$ and $\vec{\mu}$ are collinear so that Onsager writes
\begin{eqnarray}
\mathbf{R}=\zeta\vec{\mu}
\label{ReactionField}
\end{eqnarray}
where $\zeta$ is the reaction field factor. It is given by \cite{Onsager1936JACS,Bottcher1973Book}
\begin{eqnarray}
\zeta=\frac{2(\varepsilon-1)}{(2\varepsilon+1)a^3}
\label{ReactionFieldFactor}
\end{eqnarray}
Furthermore, if $\mathbf{E}$ is the Maxwell field influencing the dipole orientations in the cavity, standard macroscopic electrostatics shows that the field in the \textit{empty} cavity (i.e. with no dipole in it) is not equal to $\mathbf{E}$. This field is termed the cavity field $\mathbf{E}_c$ and is related to the Maxwell field $\mathbf{E}$ via
\begin{eqnarray}
\mathbf{E}_{c}=\eta\mathbf{E}
\end{eqnarray}
where $\eta$ is the cavity field factor given by \cite{Onsager1936JACS,Bottcher1973Book}
\begin{eqnarray}
\eta=\frac{3\varepsilon}{2\varepsilon+1}
\label{CavityFieldFactor}
\end{eqnarray}
By assuming polar and isotropically polarizable molecules, Onsager's internal field can be written \cite{Onsager1936JACS,Bottcher1973Book} 
\begin{eqnarray}
\mathbf{E}_{i}=\frac{\eta\mathbf{E}}{1-\zeta\alpha}+\frac{\zeta\vec{\mu}}{1-\zeta\alpha}
\label{OnsagerInternalField}
\end{eqnarray}
Clearly, the second term is unable to orient the permanent dipole $\vec{\mu}$. Hence, only a part of the internal field is able to orient the permanent dipole in the cavity. This field is referred to by B\"{o}ttcher as the directing field $\mathbf{E}_d$. \cite{Bottcher1973Book}
Because of thermal agitation, the dipole orientations are distributed, so that the internal field at a dipole \eqref{OnsagerInternalField} is a random field. We can write without any approximation \cite{Bottcher1973Book}
\begin{eqnarray}
\mathbf{E}_{i}=\mathbf{E}_{d}+\mathbf{R}_{eff}
\label{OnsagerInternalField2}
\end{eqnarray}
where 
\begin{eqnarray}
\mathbf{E}_{d}=\frac{\eta\mathbf{E}}{1-\zeta\alpha}\label{OnsagerEd}\\
\mathbf{R}_{eff}=\frac{\zeta\vec{\mu}}{1-\zeta\alpha}\label{EffReactField}
\end{eqnarray}
Hence, the reaction field fluctuates because the dipole orientations fluctuate. It follows that the local field fluctuates, because it is made of a deterministic part (the directing field) and a fluctuating part (the reaction field) that is unable to orient the dipole. It follows that the \textit{torque} exerted by the local field is due to the directing field only. 
The macroscopic polarization in the direction of the Maxwell field is given by
\begin{eqnarray}
4\pi P=(\varepsilon-1)E
\label{MacroPolOns}
\end{eqnarray}
while the microscopic polarization is made of two additive contributions : one due to induced moments (generated by the internal field) and one due to permanent moments (which are oriented by the directing field). Because the directing field and the Maxwell field are collinear, we have \cite{Bottcher1973Book}
\begin{eqnarray}
(\varepsilon -1)E=4\pi\rho_{0}(\alpha\langle\mathbf{E}_{i}\cdot\mathbf{e}\rangle+\langle\vec{\mu}\cdot\mathbf{e}\rangle)
\label{EqStateOns1}
\end{eqnarray}
where $\mathbf{e}$ is a unit vector along the Maxwell field $\mathbf{E}$, therefore along the directing field  $\mathbf{E}_{d}$.  Combining the above equation with Eqs. \eqref{OnsagerInternalField2}, \eqref{OnsagerEd} and \eqref{EqStateOns1}, we have
\begin{eqnarray}
(\varepsilon - 1)E=4\pi\rho_{0}\left(\alpha E_{d}+\frac{\langle\vec{\mu}\cdot\mathbf{e}\rangle}{1-\zeta\alpha}\right)
\label{EqStateOns2}
\end{eqnarray}
Indeed, the statistical average in the above equation is evaluated in the linear response limit to $E_{d}$.\cite{Bottcher1973Book} We have
\begin{eqnarray}
\nonumber
\langle\vec{\mu}\cdot\mathbf{e}\rangle&=&\beta\langle(\vec{\mu}\cdot\mathbf{e})^{2}\rangle_{0}E_{d}=\frac{\beta\mu^{2}}{3}E_{d}
\label{LRTDipole}
\end{eqnarray}
where the angular brackets $\langle\rangle_{0}$ denote an average in the absence of (directing) electric field, and the last equality is obtained because all field directions are equivalent, so that Eq.\eqref{EqStateOns2} becomes
\begin{eqnarray}
(\varepsilon - 1)=4\pi\rho_{0}\left(\alpha+\frac{\beta\mu^{2}}{3(1-\zeta\alpha)}\right)\frac{\eta}{1-\zeta\alpha}
\label{EqStateOns3}
\end{eqnarray}
where we have used Eq.\eqref{OnsagerEd}. One could leave Eq.\eqref{EqStateOns3} as it is, since the goal of relating the dielectric constant to molecular parameters has, in principle, been achieved. However, a further modification of it is possible by eliminating $\alpha$ and the size of the cavity $a$ from this equation, with the purpose of using it for extracting permanent dipole moments from  experimental data (in the spirit of the Debye theory). To this aim, Onsager introduced the dielectric constant $\varepsilon_{\infty}$ at a frequency where the permanent dipoles can no longer follow the change of the field, but however for which the atomic and electronic polarizabilities are still the same as in static fields. He used the equation \cite{Onsager1936JACS,Bottcher1973Book}
\begin{eqnarray}
\frac{\varepsilon_{\infty}-1}{\varepsilon_{\infty}+2}=\frac{4\pi\rho_{0}\alpha}{3}=\frac{\alpha}{a^{3}}
\label{LLOns}
\end{eqnarray} 
Furthermore, in practice, the atomic polarizability is negligible so that $\varepsilon_{\infty}=n^2$, where $n$ is the Snell-Descartes optical refractive index of the substance. Combining all the above equations now leads to Onsager's equation
\begin{eqnarray}
\frac{(\varepsilon-\varepsilon_{\infty})(2\varepsilon+\varepsilon_{\infty})}{3\varepsilon}=\lambda\frac{(\varepsilon_{\infty}+2)^{2}}{9}
\label{OnsagerEq}
\end{eqnarray}
where $\lambda$ has the same meaning as in the preceding section (and therefore where MKS units have been restored). The main important feature of Eq.\eqref{OnsagerEq} is that it removes the ferroelectric Curie point predicted by the Debye-Lorentz theory. Yet, it gives $\varepsilon\approx 30$ for water at room temperature, far below the experimental value. The Onsager cavity is a physical one (per opposition to the Lorentz one) because the polarization inside the cavity can adapt to the surroundings thanks to including the reaction field in the expression for the internal field. Strictly, the terminology "internal field" is reserved to the average of the local field at a molecule. Here, we merge the two notions for convenience (hence, throughout, "molecular field", "internal field", "local field" refer to the same concept and when its average is involved, we will mention it explicitly). The Lorentz (with $\mathbf{E}_{in}=\mathbf{0}$) and Onsager theories belong to continuum theories of the dielectric constant of polar fluids.\cite{Bottcher1973Book} These theories do not account for the discrete character of matter at the microscopic level. In order to account for this character, only a statistical-mechanical treatment which includes intermolecular interactions can help.\cite{Bottcher1973Book}

\subsection{The theory of Kirkwood and Fr\"{o}hlich : a step forward to include intermolecular interactions in the statistical mechanical approach to the calculation of $\varepsilon$}

The statistical-mechanical treatment differs in many respects from the one used in continuum theories. At liquid densities, Kirkwood in 1939 \cite{Kirkwood1939JCP} and Fr\"{o}hlich in 1949 devised a method which can make the calculations tractable. \cite{Bottcher1973Book,Frohlich1958Book} The method consists in first considering a very large dielectric of volume $\rm{V}$, of permittivity $\varepsilon$ made of $N_{tot}$ molecules. The number of molecules is then split into two subgroups. One subgroup has $N_c$ molecules which are treated by continuum classical electrostatics, while the other subgroup having $N$ molecules are treated by the methods of classical statistical mechanics. These $N$ molecules occupy a volume $\upsilon<<\rm{V}$ \cite{Bottcher1973Book,Evans1982Book,Dejardin2018Book} in such a way that $N_{tot}/\mathrm{V}=N/\upsilon=\rho_{0}$.
\subsubsection{Derivation of the Kirkwood-Fr\"{o}hlich equation : first way}
In this version, the derivation actually requires little modification with respect to that already given for Onsager's equation. Interestingly, it makes the parallel with continuum theories in a transparent manner. The only modification is to replace $\vec{\mu}$ by the vector sum of the permanent moments contained in $\upsilon$, which we denote by $\mathbf{m}$. From our above definitions the directing field $\mathbf{E}_{d}$ is still given by Eq.\eqref{OnsagerEd}, save that it acts on a cavity that includes $N$ dipoles (so that $\langle\vec{\mu}\cdot\mathbf{e}\rangle$ is replaced by  $\langle\mathbf{m}\cdot\mathbf{e}\rangle$ in Eq. \eqref{EqStateOns2}). Clearly, in linear response, we then have
\begin{eqnarray}
\varepsilon-1=\frac{(\varepsilon_{\infty}-1)\varepsilon}{2\varepsilon+\varepsilon_{\infty}}+\frac{4\pi\beta}{3\upsilon}\frac{\varepsilon(2\varepsilon+1)(\varepsilon_{\infty}+2)^{2}}{3(2\varepsilon+\varepsilon_{\infty})^2}\langle m^{2}\rangle_{0}
\end{eqnarray} 
which results in the Fr\"{o}hlich equation
\begin{eqnarray}
\frac{(\varepsilon-\varepsilon_{\infty})(2\varepsilon+\varepsilon_{\infty})}{3\varepsilon}=\frac{(\varepsilon_{\infty}+2)^{2}}{9}\frac{4\pi\beta}{3\upsilon}\langle m^{2}\rangle_{0}
\label{FEquation}
\end{eqnarray}
One may then define the Kirkwood correlation factor, i.e.
\begin{eqnarray}
g_{\rm K}=\frac{\langle m^{2}\rangle_{0}}{N\mu_{g}^{2}}
\end{eqnarray}
where $\mu_g$ is the permanent dipole modulus in the ideal gas phase, so that
\begin{eqnarray}
{{g}_{\rm K}}=1+\frac{1}{N}\sum\limits_{i=1}^{N}{\sum\limits_{j\ne i}{{{\left\langle {{\mathbf{u}}_{i}}\cdot {{\mathbf{u}}_{j}} \right\rangle }_{0}}}},
\label{KirkGk}
\end{eqnarray}
where $\mathbf{u}_{i}$ is a unit vector in the direction of molecular permanent dipole number $i$. Hence Eq.\eqref{FEquation} becomes the Kirkwood-Fr\"{o}hlich equation, namely
\begin{eqnarray}
\frac{(\varepsilon-\varepsilon_{\infty})(2\varepsilon+\varepsilon_{\infty})}{3\varepsilon}=\frac{(\varepsilon_{\infty}+2)^{2}}{9}\lambda g_{\rm K}
\label{KFEquation}
\end{eqnarray}
Fr\"{o}hlich showed that the summation over $i$ in Eq. \eqref{KirkGk} could be made immaterial by tagging a dipole in the inner sphere, yielding
\begin{eqnarray}
{{g}_{\rm K}}=1+\sum\limits_{j\ne 1}{{{\left\langle {{\mathbf{u}}_{1}}\cdot {{\mathbf{u}}_{j}} \right\rangle }_{0}}},
\label{FrohlichGk}
\end{eqnarray}
but this formulation (although exact) seems cumbersome for numerical simulations for various reasons (it is for example assumed that all correlations yield the same contribution to the $i$ sum, which has to be checked in a numerical simulation because the number of molecules involved is limited). 
Kirkwood further restricted the $j$ summation to nearest neighbors of the tagged molecule, yielding \cite{Kirkwood1939JCP}
\begin{eqnarray}
{{g}_{K}}=1+\left( N-1 \right){{\left\langle {{\mathbf{u}}_{1}}\cdot {{\mathbf{u}}_{2}} \right\rangle }_{0}}
\label{KirkwoodGk}
\end{eqnarray}
where \textit{in this equation only} $N-1$ is the number of nearest neighbors of the tagged molecule labelled $1$. This equation allowed Kirkwood \cite{Kirkwood1939JCP} to conjecture that the nearest neighbors of a water molecules are located at tetrahedral sites. For water and other substances, it is from this equation that dipole pair alignment is deduced from the measurement of the dielectric constant. Namely, if $g_{\rm K}>1$, then dipole pairs prefer parallel alignment, if  $g_{\rm K}<1$, then dipole pairs align antiparallel and when  $g_{\rm K}=1$, no orientational order is preferred and Onsager's equation results. It is needless to say that Eq.\eqref{KirkwoodGk} cannot be correct as the terms in the double sum in Eq.\eqref{KirkGk} undoubtly alternates signs. 
Back to our derivation, it has the merit to demonstrate that the inner sphere of volume $\upsilon$ constitutes a physical cavity. However, we give rapidly below another derivation giving another  insight which will become fruitful later.
\subsubsection{Derivation of the Kirkwood-Fr\"{o}hlich equation : second way}
In the previous subsection, the polarization of the dielectric was split in two mechanisms, the one due to permanent moments and that due to induced moments. Nevertheless, this decomposition is not unique and can be done differently (Felderhof has discussed such decompositions in Reference \citenum{Felderhof1979JPhysC}). In fact, one may decompose the polarization into two mechanisms : the orientational one and the distortional one. The distortional mechanism is governed by induced moments only and macroscopically described by $\varepsilon_{\infty}$, while the orientational one is influenced by both. In order to see this, let us rewrite Eq. \eqref{OnsagerInternalField} more explicitly and adapt it to many dipoles in the cavity. We have
\begin{eqnarray}
\mathbf{E}_{i}=\frac{(\varepsilon_{\infty}+2)\varepsilon}{2\varepsilon+\varepsilon_{\infty}}\mathbf{E}+\frac{2(\varepsilon-1)(\varepsilon_{\infty}-1)}{3\alpha(2\varepsilon+\varepsilon_{\infty})}\mathbf{m}
\label{OnsagerFrohlichLocalField}
\end{eqnarray}
where again $\mathbf{m}$ is the vector sum of all molecular permanent dipole moments in the inner sphere of volume $\upsilon$. We can introduce the effective dipole moment $\mathbf{m}_{eff}$ given by
\begin{eqnarray}
\mathbf{m}_{eff}=\frac{\varepsilon_{\infty}+2}{3}\mathbf{m}=\frac{\varepsilon_{\infty}+2}{3}\mu_{g}\sum_{i=1}^{N}\mathbf{u}_{i}
\label{EffectiveCavDipole}
\end{eqnarray}
and express the local field given by Eq.\eqref{OnsagerFrohlichLocalField} in terms of this effective moment. We have
\begin{eqnarray}
\mathbf{E}_{i}=\frac{(\varepsilon_{\infty}+2)\varepsilon}{2\varepsilon+\varepsilon_{\infty}}\mathbf{E}+\frac{2(\varepsilon-1)}{a^{3}(2\varepsilon+\varepsilon_{\infty})}\mathbf{m}_{eff}
\label{OnsagerFrohlichLocalField2}
\end{eqnarray}
where the relation between the polarizability and $a$ is still given by Eq.\eqref{LLOns}, where \textit{no cavity concept} is required in order to derive it. We can calculate the interaction of one molecular permanent dipole with the above field \eqref{OnsagerFrohlichLocalField2}. In doing this, we have
\begin{eqnarray}
-\vec{\mu}_{i}\cdot\mathbf{E}_{i}=-\vec{\mu}_{i,eff}\cdot\mathbf{E}_{i,F}
\label{EnergEquivalence}
\end{eqnarray}
where of course $\vec{\mu}_{i,eff}=\vec{\mu}_{i}(\varepsilon_{\infty}+2)/3$, and the Fr\"{o}hlich internal field has been introduced,\cite{Evans1982Book} viz.
\begin{eqnarray}
\mathbf{E}_{i,F}=\frac{3\varepsilon\mathbf{E}}{2\varepsilon+\varepsilon_{\infty}}+\frac{2(\varepsilon-1)}{a'^{3}(2\varepsilon+\varepsilon_{\infty})}\mathbf{m}_{eff}
\label{FrohlichInternalField}
\end{eqnarray}
The first term in the right hand side of the above equation is known as the "Fr\"{o}hlich field" $\mathbf{E}_{F}$ \cite{Bottcher1973Book}, and is the field orienting dipoles $\vec{\mu}_{i,eff}$ inside the inner sphere, while we refer to the second term as the Fr\"{o}hlich reaction field \textit{which cannot orient the dipole} $\mathbf{m}_{eff}$ (therefore, it cannot orient the dipole $\mathbf{m}$). The quantity $a'$  has to be interpreted as the radius of the spherical volume available to each molecule with dipole moment modulus $\mu_{eff}=(\varepsilon_{\infty}+2)\mu_{g}/3$ \textit{inside the inner sphere of volume} $\upsilon$, and is linked to $a$ via the relation
\begin{eqnarray}
a'^{3}=\frac{(\varepsilon_{\infty}+2)}{3}a^{3}
\end{eqnarray}
so that the polarizability of the dipoles $\vec{\mu}_{i,eff}$ is $\alpha'$ such that
\begin{eqnarray}
\alpha'=\frac{(\varepsilon_{\infty}+2)}{3}\alpha
\end{eqnarray}
Hence $\alpha'/a'^{3}=\alpha/a^{3}$, ensuring that $\varepsilon_{\infty}$ is not affected by the renormalization of dipole and polarizability (and therefore, the Maxwell field is not affected at all). In fact, Eq.\eqref{EnergEquivalence} is the mathematical expression of B\"{o}ttcher's description of Fr\"{o}hlich's picture of a dielectric. Quoting B\"{o}ttcher (the words between square brackets [] are ours),  "Fr\"{o}hlich introduced a continuum of dielectric constant $\varepsilon_{\infty}$ [immersed in a much larger dielectric of dielectric constant $\varepsilon$] in which point dipoles are embedded. In this model each molecule is replaced by a point dipole having the same non-electrostatic interactions with the other point dipoles as the molecules had [and renormalized electrostatic interactions between them of course] while the polarizability of the molecules can be imagined to be smeared out to form a continuum with  dielectric constant $\varepsilon_{\infty}$". 
This being mentioned, we can proceed to the derivation. The polarization due to the polarizability of the effective dipoles is the distortional polarization $\mathbf{P}_{dst}$ and is given by \cite{Bottcher1973Book}
\begin{eqnarray}
4\pi\mathbf{P}_{dst}=(\varepsilon_{\infty}-1)\mathbf{E}
\end{eqnarray} 
The total polarization $\mathbf{P}$ is the vector sum of the polarization due to the orientational mechanism $\mathbf{P}_{or}$ and that due to the distortional one $\mathbf{P}_{dst}$. Therefore,
\begin{eqnarray}
(\varepsilon-\varepsilon_{\infty})E=4\pi\mathbf{P}_{or}\cdot\mathbf{e}
\label{MacroEq}
\end{eqnarray}
The right-hand side of this equation must now be evaluated by means of statistical mechanics. Since now the distortional polarization mechanism has been eliminated, the potential energy of the assembly consists of that of the effective permanent dipole moments only, made of pair interactions and their interaction with the Fr\"{o}hlich field.\cite{Bottcher1973Book} We have
\begin{eqnarray}
(\varepsilon-\varepsilon_{\infty})E=\frac{4\pi}{\upsilon}\langle\mathbf{m}_{eff}\cdot\mathbf{e}\rangle
\end{eqnarray}
where the statistical average in the right hand side is evaluated in the linear response to the Fr\"{o}hlich field. Clearly, we have
\begin{eqnarray}
(\varepsilon-\varepsilon_{\infty})=\frac{4\pi\beta}{\upsilon}\frac{3\varepsilon}{2\varepsilon+\varepsilon_{\infty}}\langle(\mathbf{m}_{eff}\cdot\mathbf{e})^{2}\rangle_{0}
\end{eqnarray}
or, since $\langle(\mathbf{m}_{eff}\cdot\mathbf{e})^{2}\rangle_{0}=\langle m_{eff}^{2}\rangle_{0}/3$,
\begin{eqnarray}
\frac{(\varepsilon-\varepsilon_{\infty})(2\varepsilon+\varepsilon_{\infty})}{3\varepsilon}=\frac{4\pi\beta}{3\upsilon}\langle m_{eff}^{2}\rangle_{0}
\label{FEquation2}
\end{eqnarray}
which by Eq.\eqref{EffectiveCavDipole} is the same as Eq.\eqref{FEquation}. The Kirkwood correlation factor is
\begin{eqnarray}
g_{\rm K}=\frac{\langle m_{eff}^{2}\rangle_{0}}{N\mu_{eff}^{2}}=\frac{\langle m^{2}\rangle_{0}}{N\mu_{g}^{2}}
\end{eqnarray}
so that the Kirkwood-Fr\"{o}hlich equation \eqref{KFEquation} results. Notice from all what preceeds that in order to define such a factor, \textit{it is essential to separate the distortional polarization from the orientational one}. Moreover, it is clear from this second derivation that one must take $\varepsilon_{\infty}=n^{2}$ as at visible optical frequencies, it is guaranteed the orientational mechanism plays no role, as is well-known. At last, $n$ is always measured with a better experimental uncertainty than $\varepsilon_{\infty}$, which in practice does not differ that much from $n^{2}$ anyway. \cite{Bottcher1973Book} 

\subsection{The experimental use of the Kirkwood-Fr\"{o}hlich equation} 

The use of the Kirkwood-Fr\"{o}hlich equation \eqref{KFEquation} for comparing its outcomes regarding the dielectric constant is not easy in the absence of any theoretical estimate of $g_{\rm K}$. After Kirkwood's seminal work in 1939, this factor was always \textit{bona fide} estimated for various compounds, and all these efforts are summarized in chapter 6 of B\"{o}ttcher's book. \cite{Bottcher1973Book}

Nevertheless, most of the time these theoretical estimates are not used, because they are related to special cases and are semi-empirical. Instead, one deduces $g_{\rm K}$ from measurements of $\varepsilon$ by writing
\begin{eqnarray}
g_{\rm K}=\frac{9\varepsilon_{0}kT}{\rho_{0}\mu_{g}^{2}}\frac{(\varepsilon-\varepsilon_{\infty})(2\varepsilon+\varepsilon_{\infty})}{\varepsilon(\varepsilon_{\infty}+2)^{2}}
\label{gKExperiment}
\end{eqnarray}
and decides which dipolar order arises by comparing the deduced value of $g_{\rm K}$ with $1$. The experimental relative uncertainty on $g_{\rm K}$ noted $(\Delta g_{\rm K}/g_{\rm K})_{exp}$ is, in the least favorable situation,
\begin{eqnarray}
\nonumber
\left(\frac{\Delta g_{\rm K}}{g_{\rm K}}\right)_{exp}&=&\left(\frac{\Delta T}{T}\right)_{exp}+\left(\frac{\Delta\rho_{0}}{\rho_{0}}\right)_{exp}+2\left(\frac{\Delta\mu_{g}}{\mu_{g}}\right)_{exp}\\
\nonumber
&+&\left(\frac{\Delta(\varepsilon-\varepsilon_{\infty})}{\vert\varepsilon-\varepsilon_{\infty}\vert}\right)_{exp}
+\left(\frac{\Delta(2\varepsilon+\varepsilon_{\infty})}{2\varepsilon+\varepsilon_{\infty}}\right)_{exp}+\left(\frac{\Delta\varepsilon}{\varepsilon}\right)_{exp}\\
&+&2\left(\frac{\Delta(\varepsilon_{\infty}+2)}{\varepsilon_{\infty}+2}\right)_{exp}
\label{ExpUncertaintyGeneral}
\end{eqnarray}
Generally, the experimental uncertainty Eq.\eqref{ExpUncertaintyGeneral} is minimized by using $\varepsilon_{\infty}=n^2$, because \textit{per se} the definition of  $\varepsilon_{\infty}$ is too vague. We note that Hill \cite{Hill1970JPC} suggested to use $\varepsilon_{\infty}=4.5$ and $g_{\rm K}=1$ for liquid water at room temperature, because it is found that the Onsager dipole $\mu_{g}\sqrt{g_{\rm K}}$ is practically independent of temperature and the value of $\varepsilon_{\infty}$ then used is well compatible with dielectric relaxation data. \cite{Bottcher1973Book} However, already for monohydroxyalcohols this procedure of fixing  $\varepsilon_{\infty}$ cannot be used. \cite{Bottcher1973Book} We therefore must conclude that this procedure cannot be used in general. At last, B\"{o}ttcher suggests to use $\varepsilon_{\infty}=1.05n^2$. However, this multiplicative factor of $1.05$ is justified on empirical grounds only, and anyway does not correspond to the picture suggested by Onsager \cite{Onsager1936JACS} and Fr\"{o}hlich \cite{Frohlich1958Book}, therefore altering the true value of $g_{\rm K}$ predicted by Eq.\eqref{gKExperiment} for which one must use $\varepsilon_{\infty}=n^{2}$.
This being mentioned, each parameter in the right hand side of Eq.\eqref{ExpUncertaintyGeneral} is deduced from an experimental measurement or directly measured (nowadays, $\mu_{g}$ is sometimes known from quantum chemistry ab initio calculations, but not always). Taking 1\% for each relative uncertainty in the right hand side of Eq.\eqref{ExpUncertaintyGeneral} (a pessimistic view, so that many experimental uncertainties are overestimated) leads to an upper bound to the relative experimental uncertainty on $g_{\rm K}$ , viz.
\begin{eqnarray}
\nonumber
\left(\frac{\Delta g_{\rm K}}{g_{\rm K}}\right)_{exp}\leq 10\%
\label{ExpUncertaintyUpperBound}
\end{eqnarray}
independently of the way by which $g_{\rm K}$ is worked out provided that $\varepsilon$ is measured properly.

The theoretical aspect of the subject would not noticeably evolve before 1971 and Wertheim's approach to the calculation of the dielectric constant.

\subsection{Wertheim's approach to the calculation of the dielectric constant of polar fluids}

In Wertheim's approach \cite{Wertheim1971JCP}, the method of attack  is completely different from the previous ones. In a first approach, he considers polar non-polarizable molecules only. His method may be presented and related to previous approaches as follows. 
\subsubsection{Formula for the statistical linear polarization in the context of Wertheim's 1971 approach}
In the first place, a formula for the polarization in the direction of the field is derived from the equation 
\begin{eqnarray}
P=\frac{\mu}{\upsilon}\sum_{i=1}^{N}\langle\mathbf{u}_{i}\cdot\mathbf{e}\rangle
\label{PolWertheim}
\end{eqnarray}
where in this last equation the statistical average is over the equilibrium solution of the generalized Liouville equation (therefore including \textit{all} interactions like for the derivation of Fr\"{o}hlich's equation) in the presence of an "external field" which will be given later in the text and which we will denote by $\mathbf{E}_{W}$. Of course, in linear response to this field, we have
\begin{eqnarray}
P=\frac{\beta\mu}{\upsilon}\sum_{i=1}^{N}\langle(\mathbf{u}_{i}\cdot\mathbf{e})(\mathbf{u}_{j}\cdot\mathbf{E}_{W})\rangle_{0}
\label{PolWertheimLin}
\end{eqnarray}
One may assume that $\mathbf{E}_{W}$ is along the Maxwell field and that it is uniform (because it does not noticeably change over molecular distances), so that
\begin{eqnarray}
P=\frac{\beta\mu^{2} E_{W}}{\upsilon}\sum_{i=1}^{N}\sum_{j=1}^{N}\langle(\mathbf{u}_{i}\cdot\mathbf{e})(\mathbf{u}_{j}\cdot\mathbf{e})\rangle_{0}
\label{PolWertheimLin2}
\end{eqnarray}
It is then customary to introduce the $m$-body partial densities $n^{(m)}(\mathbf{r}_{1}\cdots\mathbf{r}_{m},\mathbf{u}_{1},\cdots\mathbf{u}_{m})$ from the full solution of the Liouville equation $F(\mathbf{r}_{1}\cdots\mathbf{r}_{N},\mathbf{u}_{1},\cdots\mathbf{u}_{N}),\: N>m$ via the equation (the conjugate momenta can be ignored here)
\begin{eqnarray}
\nonumber
n^{(m)}(\mathbf{r}_{1}\cdots\mathbf{r}_{m},\mathbf{u}_{1},\cdots\mathbf{u}_{m})=\qquad\qquad\qquad\\
\frac{N!}{(N-n)!}\int F(\mathbf{r}_{1}\cdots\mathbf{r}_{N},\mathbf{u}_{1},\cdots\mathbf{u}_{N})d\mathbf{r}_{m+1}\cdots d\mathbf{r}_{N}d\mathbf{u}_{m+1}\cdots d\mathbf{u}_{N}
\end{eqnarray}
so that Eq.\eqref{PolWertheimLin2} may be split in two terms, viz.
\begin{eqnarray}
P=\frac{\beta\mu^{2} E_{W}}{\upsilon}\left(\sum_{i=1}^{N}\langle(\mathbf{u}_{i}\cdot\mathbf{e})^{2}\rangle_{0}+\sum_{i=1}^{N}\sum_{j \neq i}\langle(\mathbf{u}_{i}\cdot\mathbf{e})(\mathbf{u}_{j}\cdot\mathbf{e})\rangle_{0}\right)
\label{PolWertheimLin3}
\end{eqnarray}
The first term contributes $N/3$ as seen before. The second term is written (again because all external field directions are equivalent)
\begin{eqnarray}
\nonumber
\sum_{i=1}^{N}\sum_{j \neq i}\langle(\mathbf{u}_{i}\cdot\mathbf{e})(\mathbf{u}_{j}\cdot\mathbf{e})\rangle_{0})=\frac{N(N-1)}{3}\langle\mathbf{u}\cdot\mathbf{u}'\rangle_{0}
\end{eqnarray}
so that the linear polarization \eqref{PolWertheimLin3} is written as follows :
\begin{eqnarray}
P=\frac{\beta\rho_{0}\mu^{2} E_{W}}{3}\left(1+\frac{1}{N}\int(\mathbf{u}\cdot\mathbf{u}')n^{(2)}(\mathbf{r},\mathbf{r}',\mathbf{u},\mathbf{u}')d\mathbf{r}d\mathbf{r}'d\mathbf{u}d\mathbf{u}'\right)
\label{Pinterm}
\end{eqnarray}
Next, one (legitimately) assumes that in a polar fluid, one has
\begin{eqnarray}
n^{(2)}(\mathbf{r},\mathbf{r}',\mathbf{u},\mathbf{u}')=n^{(2)}(\mathbf{r}-\mathbf{r}',\mathbf{u},\mathbf{u}')
\end{eqnarray}
Introducing the relative position vector of a pair $\vec{\rho}=\mathbf{r}-\mathbf{r}'$ and using $\upsilon=\int d\mathbf{r}$ the polarization \eqref{Pinterm} becomes
\begin{eqnarray}
P=\frac{\beta\rho_{0}\mu^{2} E_{W}}{3}\left(1+\frac{1}{\rho_{0}}\int(\mathbf{u}\cdot\mathbf{u}')n^{(2)}(\vec{\rho},\mathbf{u},\mathbf{u}')d\vec{\rho}d\mathbf{u}d\mathbf{u}'\right)
\label{PwithN2}
\end{eqnarray}
At last, one can introduce the pair distribution function $G_{2}$  and the density pair correlation function $h=G_{2}-1$ via the equations
\begin{eqnarray}
n^{(2)}(\vec{\rho},\mathbf{u},\mathbf{u}')&=&n^{(1)}(\mathbf{u})n^{(1)}(\mathbf{u}')G_{2}(\vec{\rho},\mathbf{u},\mathbf{u}')\label{G2DefW}\\
&=&n^{(1)}(\mathbf{u})n^{(1)}(\mathbf{u}')(1+h(\vec{\rho},\mathbf{u},\mathbf{u}'))\label{hDefW}
\end{eqnarray}
Wertheim then considers that a polar fluid is a simple liquid, which means \cite{Hansen2006Book}
\begin{eqnarray}
n^{(1)}(\mathbf{u})=\frac{\rho_{0}}{4\pi}
\label{SimpleLiquid}
\end{eqnarray}
hence the two alternative final forms for the polarization, namely
\begin{eqnarray}
P=\frac{\beta\rho_{0}\mu^{2} E_{W}}{3}\left(1+\frac{\rho_{0}}{(4\pi)^{2}}\int(\mathbf{u}\cdot\mathbf{u}')h(\vec{\rho},\mathbf{u},\mathbf{u}')d\vec{\rho}d\mathbf{u}d\mathbf{u}'\right)
\label{PwithH}
\end{eqnarray}
or
\begin{eqnarray}
P=\frac{\beta\rho_{0}\mu^{2} E_{W}}{3}\left(1+\frac{\rho_{0}}{(4\pi)^{2}}\int(\mathbf{u}\cdot\mathbf{u}')G_{2}(\vec{\rho},\mathbf{u},\mathbf{u}')d\vec{\rho}d\mathbf{u}d\mathbf{u}'\right)
\label{PwithG2}
\end{eqnarray}
Here, we notice that the partial densities are the solutions of the so-called Yvon-Born-Green (YBG) hierarchy, \cite{Hansen2006Book} which is an (infinite) set of \textit{nonlinear} differential-integral equations. Here we can wonder about two points :
\begin{itemize}
\item it is not clear whether Eq.\eqref{SimpleLiquid} applies for a polar fluid,
\item it is also not clear whether Eqs.\eqref{PolWertheimLin3} and \eqref{PwithN2} are equivalent in linear response theory. This is because Eqs.\eqref{PolWertheimLin3} is derived from a linear equation (and therefore linear response theory applies), while $n^{(1)}$ and $n^{(2)}$ obey coupled nonlinear differential-integral equations and here, \textit{linear response theory does not apply} (of course, the linear response can be calculated by first order perturbation theory, but definitely not by Kubo's method). 
\end{itemize}
We will criticize these hypotheses later when we come to the motivation of our work in a later section.

Now, Wertheim \cite{Wertheim1971JCP} used Eq.\eqref{PwithH}, while Madden and Kivelson (for example) used Eq.\eqref{PwithG2} \textit{but do not numerically compute} $G_{2}$.\cite{Madden1984Book} We are therefore left with finding a method for calculating $h$, which is the originality of Wertheim's approach.

\subsubsection{Calculation of P by using the Ornstein-Zernike (OZ) equation for a simple liquid}
For a simple liquid with molecules having rotational and translational degrees of freedom, the OZ equation is
\begin{eqnarray}
\nonumber
h(\vec{\rho},\mathbf{u},\mathbf{u}')=c^{(2)}(\vec{\rho},\mathbf{u},\mathbf{u}')+\qquad\qquad\\
\frac{\rho_{0}}{4\pi}\int c^{(2)}(\vec{\rho}-\vec{\rho}'',\mathbf{u},\mathbf{u}'')h(\vec{\rho}'',\mathbf{u}'',\mathbf{u}')d\vec{\rho}''d\mathbf{u}''
\label{OZSimpleLiquid}
\end{eqnarray}
where $c^{(2)}$ is the direct pair correlation function.\cite{Hansen2006Book} In order to calculate $h$ from this equation, one must supply an extra relation linking $h$ (or $G_{2}$), the pair interaction potential and $c^{(2)}$. This extra equation is called the closure of the OZ equation, and when provided, allows to calculate $h$ from Eq.\eqref{OZSimpleLiquid}. Several closures exist and give the name of the approximation. Very few solutions are available in closed form, hence generally, one must achieve the calculations numerically. 
In his 1971 paper, Wertheim uses the Mean Spherical Approximation (MSA) closure, viz.
\begin{eqnarray}
c^{(2)}(\vec{\rho},\mathbf{u},\mathbf{u}')=-\beta U_{int}(\vec{\rho},\mathbf{u},\mathbf{u}'), \vert\vec{\rho}\vert \geq R_{H}
\label{MSA}
\end{eqnarray}
$R_H$ being a hard sphere radius, and in this precise model, $U_{int}$ is the usual dipole-dipole interaction. Below the hard sphere radius, $h=-1$ and the interactions are purely repulsive. Because of this specific form of the closure, one may write \cite{Wertheim1971JCP} ($\rho=\vert\vec{\rho}\vert$)
\begin{eqnarray}
h(\vec{\rho},\mathbf{u},\mathbf{u}')&=&h_{s}(\rho)+h_{\Delta}(\rho)\Delta(\mathbf{u},\mathbf{u}')+h_{D}(\rho)D(\hat{\rho},\mathbf{u},\mathbf{u}')\qquad\qquad\label{hExpn}\\
c^{(2)}(\vec{\rho},\mathbf{u},\mathbf{u}')&=&c_{s}^{(2)}(\rho)+c_{\Delta}^{(2)}(\rho)\Delta(\mathbf{u},\mathbf{u}')+c^{(2)}_{D}(\rho)D(\hat{\rho},\mathbf{u},\mathbf{u}')\qquad\label{C2Expn}
\end{eqnarray}
where $\hat{\rho}=\vec{\rho}/\rho$, and where \cite{Wertheim1971JCP}
\begin{eqnarray}
\Delta(\mathbf{u},\mathbf{u}')&=&\mathbf{u}\cdot\mathbf{u}'\\
D(\hat{\rho},\mathbf{u},\mathbf{u}')&=&3(\mathbf{u}\cdot\hat{\rho})(\mathbf{u}'\cdot\hat{\rho})-\mathbf{u}\cdot\mathbf{u}'
\end{eqnarray}
By combining Eqs. \eqref{PwithH} and \eqref{hExpn}, one arrives at
\begin{eqnarray}
P=\frac{\beta\rho_{0}\mu^{2} E_{W}}{3}\left(1+\frac{\rho_{0}}{3}\tilde{h}_{\Delta}(0)\right)
\label{PolWertheimFinal}
\end{eqnarray}
where the tilde denotes the three-dimensional space Fourier transform, viz.
\begin{eqnarray}
\tilde{h}_{\Delta}(\mathbf{q})=\int h_{\Delta}(r)\exp(i\mathbf{q}\cdot\mathbf{r})d\mathbf{r}
\end{eqnarray}
Although with the expansions \eqref{hExpn} and \eqref{C2Expn} it would be straightforward to calculate $h$ using the Fredholm theory of integral equations for separable kernels, this procedure however leads to a $\delta$ function singularity for $h$, resulting from the singularity of the dipole-dipole interaction at $\rho=0$.\cite{Wertheim1971JCP} To remove this singularity, he introduced linear transformations of the functions $h_D$, $h_{\Delta}$,  $c_{D}^{(2)}$ and $c_{\Delta}^{(2)}$ and sought integral equations for these new functions (we do not give these transformations here, the reader is referred to Wertheim's original work). Wertheim \textit{naturally} found that the functions $h_{s}$ and $c_{s}^{(2)}$ obey an OZ equation with Percus-Yevick (PY) closure for hard spheres (i.e., the PY closure is \textit{not imposed} there). For the linear transformations of the aformentioned functions, he also was lead \textit{naturally} to solve OZ integral equations with PY closure for hard spheres (again, the PY closure for hard spheres is not \textit{imposed} but rather arises from the development of the calculations). Then, the computation of the polarization \eqref{PolWertheimFinal} becomes easy. The outcomes of Wertheim's original calculations are compared with ours in Appendix F. 
We just mention that if the Lorentz field \eqref{LorentzField} is used for $E_{W}$, we have, by combining Eqs. \eqref{PolWertheimFinal} and \eqref{MacroPolOns},
\begin{eqnarray}
\frac{\varepsilon-1}{\varepsilon+2}=\frac{\lambda}{3}\left(1+\frac{\rho_{0}}{3}\tilde{h}_{\Delta}(0)\right)
\label{ClausiusMossottiWertheim}
\end{eqnarray}
This result is \textit{independent of the sample shape}. This equation is, as we have seen, flawed if the right hand side is much larger than unity, since the left hand side \textit{cannot exceed unity}. Hence, this equation is correct for weak densities only. Nevertheless, if $E_{W}$ is replaced by Onsager's cavity field (which is the directing field for purely polar molecules \cite{Bottcher1973Book}), one has instead the Kirkwood-Wertheim equation, viz.
\begin{eqnarray}
\frac{(\varepsilon-1)(2\varepsilon+1)}{3\varepsilon}=\lambda\left(1+\frac{\rho_{0}}{3}\tilde{h}_{\Delta}(0)\right)
\label{KirkwoodWertheim}
\end{eqnarray}
in which a correlation factor can be introduced, viz.
\begin{eqnarray}
g=1+\frac{\rho_{0}}{3}\tilde{h}_{\Delta}(0)
\label{WertheimgKdef}
\end{eqnarray}
As we have seen, this expression for $g$ assumes that a polar fluid can be assimilated to a simple liquid. However, it is not the Kirkwood correlation factor $g_{\rm K}$, \textit{save if the molecular polarizability is neglected}.  If nevertheless the hypothesis that a polar fluid behaves as a simple liquid is maintained, the ways to change the $g$ values with respect to that obtained from Wertheim's 1971 work are :
\begin{itemize}
\item to change the interaction potential $U_{int}$,
\item to maintain $U_{int}$ but use a different closure than MSA (which is an art by itself),
\item to change both,
\item to change the OZ equation \eqref{OZSimpleLiquid} into another OZ equation which relaxes the assumption that a polar fluid is not a simple liquid, but a more complex system (this is never done currently because the mathematics become quite involved).
\end{itemize} 
Yet, the four above points do not yet reply to the question so as to include induced dipole moments in Eq.\eqref{KirkwoodWertheim}, which necessarily contribute much more to the dielectric constant of polar fluids than multipoles higher than the dipole (For example, Onsager does not use multipoles). The situation in 1971 was unclear, but however, Wertheim's effort was a breakthrough at the time, which generated an enormous hope for solving the problem of how to compute the dielectric constant of isotropic polar fluids from first principles.  

\subsection{Wertheim's fluctuation law}
From 1973, and being aware that including the effect of induced dipoles in the calculation of the dielectric constant of polar fluids is crucial, Wertheim started to develop his theory of polar fluids from the nucleus he had published in 1971.\cite{Wertheim1973aMolPhys,Wertheim1973bMolPhys,Wertheim1977aMolPhys,Wertheim1977bMolPhys,Wertheim1978MolPhys} In these papers, he dropped the pedagogical language of his 1971 work to adopt the diagrammatic one commonly used in quantum electrodynamics to handle perturbation series. Besides trying to introduce the polarizability of the molecules in an as exact way as possible in the calculation of the dielectric constant, he derived a fluctuation law that in actual fact slightly differs from the Fr\"{o}hlich Eq. \eqref{FEquation}. This equation is \cite{Wertheim1978MolPhys}
\begin{eqnarray}
\frac{(\varepsilon-\varepsilon_{\infty})(2\varepsilon+\varepsilon_{\infty}^{-1})}{3\varepsilon}=\frac{4\pi\beta}{3\upsilon}\langle m_{W}^{2}\rangle_{0}
\label{WertheimFluctuationLaw}
\end{eqnarray}
where the introduced subscript "W" anticipates that the Wertheim and Fr\"{o}hlich dipole moments are not the same. This was demonstrated by Felderhof on a macroscopic basis \cite{Felderhof1979JPhysC} and by Madden and Kivelson \cite{Madden1984Book} on a microscopic one. 
\subsection{Development of Wertheim's theory : Stell and co-workers}
The theory of Wertheim has subsequenly been developed by a number of authors. The culminating point of the usable form of the theory has been summarized in the review paper by Stell et al. \cite{Stell1981Book} Particularly, in these works, the equation of state for linear dielectrics is written as follows :
\begin{eqnarray}
\frac{(\varepsilon-1)(2\varepsilon+1)}{3\varepsilon}=\frac{4\pi\beta\rho_{0}\mu_{W}^{2}}{3}\left(1+\frac{\rho_{0}}{3}\tilde{h}_{\Delta}(0)\right)
\label{StellEqState}
\end{eqnarray}
where $\mu_{W}$ is a renormalized dipole moment modulus which in particular depends on the hard sphere radius and on an effective polarizability which also depends on the hard sphere radius. Most of the focus of the review by Stell et al.\cite{Stell1981Book} is on using closures of hypernetted chain (HNC) type together with model potentials consisting in a spherically symmetric part, the dipole-dipole interaction, the dipole-quadrupole interaction and the quadrupole-quadrupole interaction (of course, other approximations are discussed). The numerical solutions of the OZ equation so obtained are asserted by Monte-Carlo simulations of $h$ and/or its projections on rotational invariants. Then, the dielectric constant is computed in terms of these effective parameters. At present, it is argued that disagreement between simulations and experimental data on any polar fluid is only apparent and that such disagreement may be diminished by properly chosing the pair interaction and the closure, nevertheless maintaining the hypothesis that a simple polar fluid is a simple liquid. In particular, Stell et al.\cite{Stell1981Book} show that increasing the quadrupolar contribution  allows the obtaining of Onsager's result without polarizability and destroys orientational correlations, in the context of the linear HNC (LHNC) closure of the OZ equation. At the time of writing, nevertheless, the theory was not able to reproduce the temperature dependence of the dielectric constant of water, for which Kirkwood and Fr\"{o}hlich basically developed their theory, nor other simple polar fluid such as acetone or acetonitrile, as is apparent from Figure 22 of the review of Stell et al.\cite{Stell1981Book}

\subsection{Comparison of the OZ approach with experiment}
One year later, Carnie and Patey \cite{Carnie1982MolPhys} performed numerical simulations of "waterlike" molecules, which carry axial quadrupoles. They wrote an adequate model for pair interactions and considered 3 closures, namely MSA, LHNC and QHNC (Q meaning quadratic) closures. The outcomes of Eq. \eqref{StellEqState} were successfully compared with the temperature dependence of the dielectric constant of water across a wide temperature range ($0-300^{\circ}$C) using Eq.\eqref{StellEqState}. However, they do moderate their success because the dielectric constant is calculated within 10\% theoretical uncertainty there. Within the variation of 7\% for the hard sphere radius $R_H$, $\mu_{eff}$ is also computed within 10\% theoretical uncertainty, which means also that $g$ as defined by
\begin{eqnarray}
g=1+\frac{\rho_{0}}{3}\tilde{h}_{\Delta}(0)
\label{StellPateyg}
\end{eqnarray}
is also computed within 10\% theoretical uncertainty. Left apart the fact that $g$ differs from $g_{\rm K}$ defined by Eq.\eqref{gKExperiment} (therefore, again, $g$ is \textit{not} the Kirkwood correlation factor), such theoretical uncertainties \textit{superimpose onto experimental ones}. Since for liquid water, $\varepsilon>>1$, the overall experimental uncertainty on $g$ is around 60\%, in spite of a numerically correct calculation. We do not deny the progress made at the time of writing. However, this experimental uncertainty means that it is meaningless to \textit{quantitatively} compare the outcomes of such theory (at this stage of development) with experimental data. Moreover, Carnie and Patey refrain to compare the structure of water with experimental data, because they quote their pair potential (the so-called microscopic model) as a too rough one. We do not deny the theoretical effect of quadrupoles as a destructive effect on the correlations in $g$ (which is not $g_{\rm K}$) \textit{in their simulations}. 
\noindent
Since the OZ approach to the problem has little evolved since, we conclude that such approaches, although certainly indicate some trends that can guide the experimentalist, are not amenable to comparison with experimental data in reality. Therefore, a theory of the Kirkwood correlation factor that is amenable to comparison with experiment is still missing.

\subsection{Motivation of this work}

After having exposed the long history of the subject, we can summarize the drawbacks of previous work on the subject as follows. 
\begin{itemize}
\item The Kirkwood correlation factor as defined by Eq.\eqref{KirkGk} is not easy to calculate from first principles, while deducing dipolar order from Eq.\eqref{KirkwoodGk} is only a \textit{bona fide} criterion.
\item In the OZ approach a polar fluid is assimilated to a simple liquid. This, in our opinion, is a working hypothesis that is not really justified. In effect, the electrostatic interactions strongly depend on dipole orientations. It follows that $n^{(1)}$, which is the solution of the first member of the Yvon-Born-Green hierarchy, obeys the equation ($V_{ext}$ is a potential arising from externally applied fields)
\begin{eqnarray}
\nonumber
\nabla_{\mathbf{u}}(\ln n^{(1)}(\mathbf{u})+\beta V_{ext}(\mathbf{u}))=\qquad\qquad\\
-\beta\int \nabla_{\mathbf{u}}U_{int}(\vec{\rho},\mathbf{u},\mathbf{u}')n^{(1)}(\mathbf{u}')G_{2}(\vec{\rho},\mathbf{u},\mathbf{u}')d\vec{\rho}d\mathbf{u}'
\label{YBG1}
\end{eqnarray}
and the OZ equation must write more generally as \cite{Hansen2006Book}
\begin{eqnarray}
\nonumber
h(\vec{\rho},\mathbf{u},\mathbf{u}')=c^{(2)}(\vec{\rho},\mathbf{u},\mathbf{u}')+\qquad\qquad\\
\int c^{(2)}(\vec{\rho}-\vec{\rho}'',\mathbf{u},\mathbf{u}'')n^{(1)}(\mathbf{u}'')h(\vec{\rho}'',\mathbf{u}'',\mathbf{u}')d\vec{\rho}''d\mathbf{u}''
\label{OZPolarFluid}
\end{eqnarray}
which is the \textit{true} OZ equation for a dense polar fluid. Since $\beta U_{int}$ can be large, especially for water, there is no justification to believe that $n^{(1)}$ is a constant even in the simplest polar fluids. However, the mathematical problem becomes quite intricate compared to the use of such a method for a simple liquid.
\item The closures of the OZ equation \eqref{OZSimpleLiquid} are always approximate ones, the range of validity of which is unknown. Besides this, Hu et al.\cite{Hu1992Physica} have demonstrated using the YBG hierarchy that such closures may be derived from an other well-known approximation, which is the Kirkwood superposition approximation (KSA). Moreover, relatively recently Singer \cite{Singer2004JCP} demonstrated that KSA (and its generalizations to the higher members of the YBG hierarchy) describes a state of \textit{maximal entropy} and of minimal Helmholtz free energy in a very simple fashion. By the corresponding H theorem, it follows that KSA describes a state of statistical-mechanical equilibrium \textit{exactly}. Since all the standard approximations of simple liquids can be derived from KSA, it follows that the accuracy by which MSA, PY and HNC describe statistical equilibrium is unclear, which indeed add to the theoretical uncertainty of models which are used in the current OZ formulation of the problem.
\item The equivalence between Eqs. \eqref{PolWertheimLin3} and \eqref{PwithN2} in linear response theory  is also questionable, because $n^{(1)}$ obeys Eq.\eqref{YBG1} which is \textit{nonlinear}. If Eq.\eqref{PwithN2} is the correct representation of the polarization from YBG (to which $n^{(2)}$ obeys), then it must be possible to derive it from the YBG hierarchy directly. But in fact, we will see that this is not generally possible, unless the assumption that a polar fluid is a simple liquid is made (see Appendix C). 
\item When the molecules are polarizable, $g$ defined by Eq.\eqref{StellPateyg} is \textit{not} the Kirkwood correlation factor as determined by experiment.
\item Sometimes, for the need of fulfilling the criterion of comparing $g_{\rm K}$ with $1$ in order to deduce dipole alignment, $\varepsilon_{\infty}$ is taken much larger than $n^{2}$, the square of the Snell-Descartes refractive index (a recent example is that of Saini et al. in Reference\citenum{Saini2017JCP} for Tributyl Phosphate data, but this is not the only one). In fact, because Eq.\eqref{KirkwoodGk} cannot describe the true value of $g_{\rm K}$ given by Eq.\eqref{KirkGk}, we believe that this criterion is a \textit{bona fide} one, but has no real theoretical justification.  Actually, we shall see that this criterion which is still in current use has nothing to do with dipole alignment.
\end{itemize}

Very recently, based on the work of Kawasaki \cite{Kawasaki1994PA} and Dean \cite{Dean1996JPA}, Cugliandolo et al.\cite{Cugliandolo2015PRE} derived a stochastic nonlinear integro-differential equation governing the dynamics of the microscopic density of collective modes for Brownian dipoles. In doing so, they ignored inertial effects, but included translational as well as rotational degrees of freedom of the molecules. Furthermore, their equation when averaged over the probability density of realizations of the local noises, reduces at equilibrium to the first member of a generalized Yvon-Born-Green hierarchy. Moreover, when ignoring specific intermolecular interactions the Debye theory of dielectric relaxation \cite{Debye1929Book} is recovered in a transparent manner. Then, by considering that the dipoles have fixed positions in space but can rotate under the action of externally applied fields, D\'{e}jardin et al.\cite{Dejardin2018JCP} derived an analytical expression for $g_{\rm K}>1$ and one for $g_{\rm K}<1$ as a function of the molecular dipole moment in vacuo $\mu_g$, the molecular density $\rho_0$ and temperature $T$. They further qualitatively compared the outcomes of their theory with the experimental temperature dependence and numerical simulations of $\varepsilon$ of water and methanol and found that agreement between their theoretical findings and  experimental data was relatively satisfactory. In order to derive their analytical formula, they used both the Ornstein-Zernike route\cite{Dejardin2018JCP} and the Kirkwood superposition approximation applied to the orientational pair distribution function\cite{Dejardin2019PRB} together with the averaged rotational Dean equation in order to derive the relevant Kirkwood potential of mean torques. The moment method used in References \citenum{Dejardin2018JCP,Dejardin2019PRB} is a general method of attack when the interaction potential is specified. However, it makes the detailed comparison of the theory  with experiment rather cumbersome, due to its restriction to one specific interaction potential.

In order to improve on this first approach, it is the purpose of the present work to derive a formula for the Kirkwood correlation factor that does not depend on any approximation made in solving the first member of the  Yvon-Born-Green hierarchy, but itself represents a good starting point for further approximations. An integral formula will then be obtained in the context of Kirkwood's superposition approximation (which, as analytically shown by Singer,\cite{Singer2004JCP} describes statistical equilibrium exactly), allowing $g_{\rm K}$ to be calculated for arbitrary pair interaction potentials of forces and torques. Then, our theoretical results will be compared with experimental data concerning a series of primary linear alcohols, water, glycerol, tributyl phosphate (TBP), acetonitrile, acetone, nitrobenzene and dimethyl sulfoxide (DMSO). The general theory is developed in Appendices, while we keep  only useful derivations in the main body text.

\section{Kirkwood correlation factor from the equilibrium averaged rotational Dean equation}
We consider an assembly of interacting polar molecules that are subjected both to thermal agitation and to uniform externally applied DC electric fields. The averaged rotational Dean equation at statistical equilibrium (time-independent regime) is : \cite{Cugliandolo2015PRE,Dejardin2018JCP}
\begin{eqnarray}
\nonumber
{{\nabla }_{\mathbf{u}}}\cdot \left[ {{\nabla }_{\mathbf{u}}}W\left( \mathbf{u} \right)+\beta W\left( \mathbf{u} \right){{\nabla }_{\mathbf{u}}}{{V}_{1}}\left( \mathbf{u} \right) \right]+\\
\beta {{\nabla }_{\mathbf{u}}}\cdot \int{{{\nabla }_{\mathbf{u}}}{{U}_{m}}\left( \mathbf{u},\mathbf{{u}'} \right){{W}_{2}}\left( \mathbf{u},\mathbf{{u}'} \right)d\mathbf{{u}'}}=0
\label{eqDK}
\end{eqnarray}
where $\mathbf{u}$ is a unit vector along a molecular dipole moment of constant magnitude $\mu$, $W(\mathbf{u})$ is the one-body orientational probability density, $V_{1}(\mathbf{u})=-\mu\mathbf{u}\cdot\mathbf{E}$ is a one-body potential containing the effect of the directing uniform electric field $\mathbf{E}$, $U_{m}(\mathbf{u},\mathbf{u}')$ is a space averaged orientational pair interaction potential, $W_{2}(\mathbf{u},\mathbf{u}')$ is the orientational pair probability density. The integral in Eq.\,\eqref{eqDK} is extended to the unit sphere of representative points of a dipole with constant magnitude $\mu$ and orientation $\mathbf{u}'$. It is demonstrated in Appendix A that Eq.\,\eqref{eqDK} is an \textit{exact} one under the assumption of a translationally invariant system made of many interacting molecules, while relation of Eq. \eqref{eqDK} -or its dynamic version- to well-established results in liquids, nematic liquid crystals and solids is discussed in Appendix B. Then, using first-order perturbation theory it may easily be demonstrated that an integral representation of the Kirkwood correlation factor $g_{\rm K}$ can be derived from Eq.\,\eqref{eqDK} (see Appendix C). Thus, on fairly general grounds, we have:  
\begin{eqnarray}
{{g}_{\rm K}}=1+\frac{\beta }{6}\iint{{{\nabla }_{\mathbf{u}}}{{U}_{m}}(\mathbf{u},\mathbf{{u}'})\cdot\Phi(\mathbf{u},\mathbf{{u}'})d\mathbf{u}d\mathbf{{u}'}}
\label{KirkwoodGkGeneral}
\end{eqnarray}
with
\begin{eqnarray}
\Phi(\mathbf{u},\mathbf{u}')=W_{2}^{(0)}(\mathbf{u},\mathbf{u}')\nabla_{\mathbf{u}}P_{2}(\mathbf{u})-9W_{2}^{(1)}(\mathbf{u},\mathbf{u}')\nabla_{\mathbf{u}}P_{1}(\mathbf{u}),
\label{Phi01}
\end{eqnarray}
where $W_{2}^{(0)}$ is the field-free equilibrium pair probability density and $W_{2}^{(1)}$ its linear response counterpart, while $P_n(\mathbf{u})$ denotes the Legendre Polynomial of order $n$. Equation \eqref{KirkwoodGkGeneral} is the rotational Dean (in fact, the rotational Yvon-Born-Green \cite{Hansen2006Book}) representation of the Kirkwood correlation factor, and is a central result of our paper. We note that our result for $g_{\rm K}$ does not depend on the number of neighbors of a "tagged" molecule and is therefore totally equivalent to Eq.\,\eqref{KirkGk}. It is nevertheless impossible to obtain explicit results if one does not link $W_{2}^{(1)}(\mathbf{u},\mathbf{u}')$ to $W_{2}^{(0)}(\mathbf{u},\mathbf{u}')$. The general task is made complicated by the fact that the equation governing $W_{2}(\mathbf{u},\mathbf{u}')$ involves the three-body orientational probability density $W_{3}(\mathbf{u},\mathbf{u}',\mathbf{u}'')$, the governing equation of which involves the four-body orientational probability density $W_4$ and so on, and for these distributions, the respective linear response to external fields must be calculated. As a result, in principle the Kirkwood correlation factor not only depends on pair correlations, but also on higher many-body correlations. These higher many-body correlations are extremely difficult to compute for an arbitrary substance in general. Hence, one must make a choice in order to obtain explicit results. In fact, it was shown analytically by Singer\cite{Singer2004JCP} that the KSA  describes statistical equilibrium exactly. Moreover, it was shown recently that under this approximation, we have \cite{Dejardin2019PRB}
\begin{eqnarray}
W_{2}^{\left( 1 \right)}\left( \mathbf{u},\mathbf{{u}'} \right)=W_{2}^{\left( 0 \right)}\left( \mathbf{u},\mathbf{{u}'} \right)\left( \mathbf{u}+\mathbf{{u}'} \right)\cdot \mathbf{e}
\label{W21Kirkwood}
\end{eqnarray}
where $\mathbf{e}$ is a unit vector along the directing field, and where 
\begin{eqnarray}
W_{2}^{\left( 0 \right)}\left( \mathbf{u},\mathbf{{u}'} \right)={{Z}^{-1}}{{e}^{-\beta V_{2}^\text{eff}\left( \mathbf{u},\mathbf{{u}'} \right)}},
\label{eq9}
\end{eqnarray}
where $Z$ is the partition function defined by
\begin{eqnarray}
\nonumber
Z=\iint e^{-\beta V_{2}^\text{eff}\left( \mathbf{u},\mathbf{{u}'} \right)}d\mathbf{u}d\mathbf{u}',
\end{eqnarray}
$V_{2}^\text{eff}\left( \mathbf{u},\mathbf{{u}'} \right)$ is an effective (rotational) pair potential given by \cite{Dejardin2019PRB}
\begin{eqnarray}
V_{2}^\text{eff}\left( \mathbf{u},\mathbf{{u}'} \right)=U_{m}\left( \mathbf{u},\mathbf{{u}'} \right)+U_{an}\left(\mathbf{u}\right)+U_{an}\left(\mathbf{u}'\right)
\label{V2effK}
\end{eqnarray}
while $U_{an}\left(\mathbf{u}\right)$ is obtained by solving the differential equations\cite{Dejardin2019PRB}
\begin{eqnarray}
{{\nabla }_{\mathbf{u}}}{{U}_{an}}\left( \mathbf{u} \right)={{\left. {{\nabla }_{\mathbf{u}}}{{U}_{m}}\left( \mathbf{u},\mathbf{{u}''} \right) \right|}_{\mathbf{{u}''}=\mathbf{u}}}.
\label{Uan}
\end{eqnarray}
\noindent
Yet, in spite of its apparent simplicity, Eqs.\,\eqref{V2effK} and \eqref{Uan} must be used with caution because the stationary points of $V_{2}^\text{eff}$  must at least approximately, if not exactly, be located at the same angles and must be of the same nature as those of $U_m$ so that both potentials describe the same physics. This was so far only vaguely described in the original work of D\'{e}jardin et al. \cite{Dejardin2019PRB} Therefore, the necessary decorrelation procedure is described in Appendix E.
We can further use the expressions for the Legendre polynomials $P_{n}(\cos\vartheta)$ in order to obtain a tractable version of Eq.\,\eqref{KirkwoodGkGeneral}. This results in the following expression for $g_{\rm K}$:
\begin{eqnarray}
{{g}_{\rm K}}=1+\beta \iint{G(\mathbf{u},\mathbf{u}')W_{2}^{\left( 0 \right)}\left( \mathbf{u},\mathbf{{u}'} \right)d\mathbf{u}d\mathbf{{u}'}}.
\label{UsableGk}
\end{eqnarray}
where we have used Eq.\,\eqref{W21Kirkwood}, $\cos \vartheta =\mathbf{u}\cdot \mathbf{e}$ in conjunction with Eq.\,\eqref{KirkwoodGkGeneral} and where we have defined $G(\mathbf{u},\mathbf{u}')$ via the equation
\begin{eqnarray}
\nonumber
G(\mathbf{u},\mathbf{u}')=\sin\vartheta\left( \cos \vartheta +\frac{3}{2}\cos {\vartheta }' \right)\frac{\partial {{U}_{m}}}{\partial \vartheta }\left( \mathbf{u},\mathbf{{u}'} \right)
\end{eqnarray}
By steepest descents arguments, if the pair intermolecular interactions is large with respect to $kT$, the value of $g_{\rm K}$ rendered by Eq.\,\eqref{UsableGk} depends on the location of the minima of $V_{2}^\text{eff}$, therefore on the state of alignment of dipole pairs at equilibrium. This strong mathematical argument is clearly different from the empirical criterion which compares $g_{\rm K}$ with 1 in order to deduce dipole pair alignment. However, in order to use this equation, $U_m$ needs to be specified.

\section{Construction of a model potential for electrostatic interactions}

It is well-known that the inclusion of the effect of the polarizability of the molecules is a necessity in order to describe the polarization state at the molecular level. This means in particular that inclusion of the translational fluctuations (i.e. coupling between translational effects and the induced moment), makes it impossible to apply the Kirkwood-Fr\"{o}hlich theory,\cite{Bottcher1973Book} because then the back action of the reaction field is unknown. We therefore suggest, as an intermediate point of view between these two extreme situations, i.e., no polarizability effects and full inclusion of the latter, to average the true intermolecular interaction potential over translational degrees of freedom of the molecules before using the Fr\"{o}hlich internal field, so that the potential effectively becomes a function of the permanent dipole moment orientations only, and that this average still keeps a trace of polarizability effects. In other  words, the task is therefore to encode, at least approximately, the molecular  physical effects in the potential $U_m$. To this purpose, we write the pair interaction potential $U_m$ as follows :
\begin{eqnarray}
\beta U_{m}(\mathbf{u},\mathbf{u}')=\beta\int U_{int}(\vec{\rho},\mathbf{u},\mathbf{u}')G(\rho)d\vec{\rho}
\label{DefUmStatic}
\end{eqnarray}
where $U_{int}(\vec{\rho},\mathbf{u},\mathbf{u}')$ is the true pair intermolecular interaction potential and $G(\vec{\rho})$ is the probability density that a pair of molecules is distant of $\rho=\vert\vec{\rho}\vert$ with orientation $\hat{\rho}$. The precise result of integration indeed depends on the system under study. Formally, however, and without any loss of generality, we can assume that the result of integration can formally be written as :  
\begin{eqnarray}
\beta U_{m}(\mathbf{u},\mathbf{u}')=\sum_{n=1}^{+\infty}a_{n}\varphi_{n}^{*}(\mathbf{u})\varphi_{n}(\mathbf{u}')
\label{DefUmSPoly}
\end{eqnarray}
where $\varphi_{n}(\mathbf{u})$ is a polynomial function of the direction cosines of $\mathbf{u}$ of degree $n$ (in general, these functions are spherical harmonics or Wigner $D$ functions), the starred quantity $f^{*}$ denotes the complex conjugate of $f$ and the expansion coefficients $a_{n}$ are parameters which are chosen to match the physical reality as much as possible. In order to exploit further Eq. \eqref{DefUmSPoly}, we also require that $\varphi_{n}(\mathbf{u})$ have the parity of their degree, i.\,e.
\begin{eqnarray}
\varphi_{n}(-\mathbf{u})=(-1)^{n}\varphi_{n}(\mathbf{u})
\label{Parity}
\end{eqnarray}
Hence, we may remark using Eqs.\eqref{DefUmSPoly} and \eqref{Parity} that $U_{m}(\mathbf{u},\mathbf{u}')$ has the necessary property of global rotational invariance, i.\,e.
\begin{eqnarray}
U_{m}(-\mathbf{u},-\mathbf{u}')=U_{m}(\mathbf{u},\mathbf{u}')
\label{RotInv}
\end{eqnarray}
The simplest choice for $\varphi_{1}(\mathbf{u})$ which encodes the correct dipole physics is
\begin{eqnarray}
\varphi_{1}(\mathbf{u})=\cos\vartheta
\end{eqnarray}
so that the leading term of the series Eq.\,\eqref{DefUmSPoly} is $a_{1}\cos\vartheta\cos\vartheta'$. According to the sign of $a_{1}$, this term has minima for parallel order or antiparallel order of the permanent dipoles, and represents the dipole-dipole interactions $V_{dd}$. Hence, following for example Refs.\citenum{Berne1975JCP,Dejardin2018JCP,Dejardin2019PRB} we have (see Appendix D for a full justification of this):
\begin{eqnarray}
a_{1}=\mp\lambda
\label{DefA1}
\end{eqnarray}
where
\begin{eqnarray}
\lambda=\frac{\beta\rho_{0}\mu^2}{3\varepsilon_{0}}
\end{eqnarray}
is the Debye susceptibility of ideal dipolar gases with individual permanent molecular dipole moment modulus $\mu$. Thus we have
\begin{eqnarray}
\beta {{V}_{dd}}\left( \mathbf{u},\mathbf{{u}'} \right)=\mp \lambda \cos \vartheta \cos {\vartheta }'
\label{DipDip}
\end{eqnarray}
In order to account for the effect of the polarizability of the molecules and its probable coupling with the permanent dipole, we add the term $n=2$ to Eq.\,\eqref{DipDip}, and use $\varphi_{2}(\mathbf{u})=\cos^{2}\vartheta$, which is the simplest choice we can make. This results in a term
\begin{eqnarray}
a_{2}\cos^{2}\vartheta\cos^{2}\vartheta'
\end{eqnarray}
This term loosely represents induction and dispersion terms. Nevertheless, because these interaction energy terms are in general not individually additive,\cite{Stone2013Book} this is very difficult to specify $a_{2}$ in terms of the polarizability exactly. Nevertheless, we may still write that $a_{2}$ is proportional to $a_{1}$, so that we have:
\begin{eqnarray}
a_{2}=\mp\kappa\lambda
\label{DefA2}
\end{eqnarray}  
resulting in the interaction energy term:
\begin{eqnarray}
\beta {{V}_{indisp}}\left( \mathbf{u},\mathbf{{u}'} \right)=\mp\kappa \lambda {{\cos }^{2}}\vartheta {{\cos }^{2}}{\vartheta }'
\label{InducDisp}
\end{eqnarray}
where $\kappa$ is a dimensionless parameter that may depend on the molecular density and temperature. However, in the following, we will consider it as a constant, the value $\lvert\kappa\rvert$ giving the deviation to \textit{pure} dipole-dipole interactions. The parameter $\kappa$ can be taken positive or negative, and $\lvert\kappa\rvert$ may exceed unity, meaning in the latter situation that the dipole-dipole interaction is not the most significant interaction in a given substance, which may happen if a given molecule has a tiny permanent dipole, typically less than 1 Debye. The overall electrostatic interaction $U_m$ between dipole pairs is then written as follows:
\begin{eqnarray}
\beta {{U}_{m}}\left( \mathbf{u},\mathbf{{u}'} \right)=\beta {{V}_{dd}}\left( \mathbf{u},\mathbf{{u}'} \right)+\beta {{V}_{indisp}}\left( \mathbf{u},\mathbf{{u}'} \right)
\label{UmNeedles}
\end{eqnarray}
A generic expression for the Kirkwood potential of mean torques $V_2^\text{eff}$ is not possible to obtain, see Appendix E for the practical determination of $V_2^\text{eff}$ from $U_m$ and $U_{an}$. For $\kappa = 0$, we obtain the analytical results already derived elsewhere \cite{Dejardin2018JCP}. For $\kappa\neq0$, this leads to 4 possible numerical values of $g_{\rm K}$. The notation for these values together with their corresponding interaction potentials are summarized in Table \ref{TableKirk} below. In the next section, we discuss the theoretical $g_{\rm K}$ values rendered by these functions.

\begin{table*}
\caption{Notations for the Kirkwood correlation factors, interaction potential $U_m$ and potential of mean torques $V_{2}^\text{eff}$. The shorthand notations $z=\cos\vartheta$ and $z'=\cos\vartheta'$ have been used.}
\label{TableKirk}
\begin{tabular*}{\textwidth}{@{\extracolsep{\fill}}cccc}
\hline 
 & $\beta U_m$ & $\beta V_{2}^\text{eff}$ \\ 
\hline
$g_{\rm K}^{1(-)}$ & $-\lambda zz'-\kappa\lambda z^2 z'^2$ & $-\frac{\lambda}{2}(z+z')^2+\frac{\kappa\lambda}{2}(z^2-z'^2)^2$ \\ 
\\
\hline
\\ 
$g_{\rm K}^{2(-)}$ & $-\lambda zz'+\kappa\lambda z^2 z'^2$ & $-\frac{\lambda}{2}(z+z')^2+\frac{\kappa\lambda}{2}(z^2+z'^2)^2$ \\ 
\\
\hline 
\\
$g_{\rm K}^{1(+)}$ & $\lambda zz'+\kappa\lambda z^2 z'^2$ & $-\frac{\lambda}{2}(z-z')^2-\frac{\kappa\lambda}{2}(z^2-z'^2)^2$ \\ 
\\
\hline 
\\
$g_{\rm K}^{2(+)}$ & $\lambda zz'-\kappa\lambda z^2 z'^2$ & $-\frac{\lambda}{2}(z-z')^2-\frac{\kappa\lambda}{2}(z^2+z'^2)^2$ \\ 
\\
\hline 
\end{tabular*}
\end{table*}

The choice we have made in Table \ref{TableKirk} is such that when the theory is compared with experiment it generally renders a positive value of $\kappa$, an exception being made in case of water, as will be shown in Paragraph \ref{TheoryExperiment}. Unphysical situations have been eliminated according to the criteria mentioned in Ref.~\citenum{Dejardin2019PRB} and exposed in detail in Appendix E.

\section{Theoretical results}

As already pointed out previously, the integral representation Eq.\,\eqref{KirkwoodGkGeneral} of the Kirkwood correlation factor is equivalent to Eq.\,\eqref{KirkGk}. The two equations differ in mathematical form simply because the starting point for their derivation is different. For example, Eq.\,\eqref{KirkGk} is obtained from the equilibrium linear response solution of the generalized Liouville equation, while our Eq.\,\eqref{KirkwoodGkGeneral} is derived from the first member of the (rotational) Yvon-Born-Green hierarchy, which is a representation of the generalized Liouville equation when interactions are represented by pair interactions only. \cite{Hansen2006Book} Therefore, Eq.\,\eqref{KirkwoodGkGeneral} is an \textit{exact} one, provided that only pair interactions are considered. Although it is as difficult as Eq.\,\eqref{KirkGk} to evaluate exactly, it nevertheless lends itself to approximations in a much easier manner since it does not explicitly depend on the number of molecules in the cavity. As an example of a possible approximation, one may choose the mean field one for which we have $W_{2}(\mathbf{u},\mathbf{u}')=W(\mathbf{u})W(\mathbf{u}')$ and use Eq.\,\eqref{KirkwoodGkGeneral} for $\kappa = 0$, which yields:\cite{Dejardin2011JAP,Dejardin2014JCP}
\begin{eqnarray}
g_{\rm K}=\left(1\mp\frac{\lambda}{3}\right)^{-1}
\label{GkMF}
\end{eqnarray}
where the minus sign holds for parallel alignment and the plus sign holds for antiparallel alignment. Indeed, for parallel alignment, Eq.\,\eqref{GkMF} produces (as is common in usual mean field approaches) a Curie point at $\lambda = 3$ which is undesirable here. Indeed, it has been shown elsewhere that Eq.\,\eqref{GkMF} is valid for $\lambda<<1$, leading, for parallel alignment, to:\cite{Dejardin2018JCP}
\begin{eqnarray}
g_{\rm K}\approx1+\frac{\lambda}{3}
\label{GklowLbda}
\end{eqnarray}
In this context, the dielectric constant is given by:\cite{Dejardin2019PRB}
\begin{eqnarray}
\varepsilon\approx 1+\lambda\left(1+\frac{2\lambda}{3}\right).
\label{epsMF}
\end{eqnarray}
so that the Debye theory is recovered at weak densities, i.e., when $\lambda<<1$. If one uses the Kirkwood superposition approximation one obtains Eq.\,\eqref{UsableGk}, the explicit evaluation of which in terms of the error function of imaginary argument has been given elsewhere.\cite{Dejardin2018JCP}

\begin{figure}[h!]
\centering
\includegraphics[height=7cm]{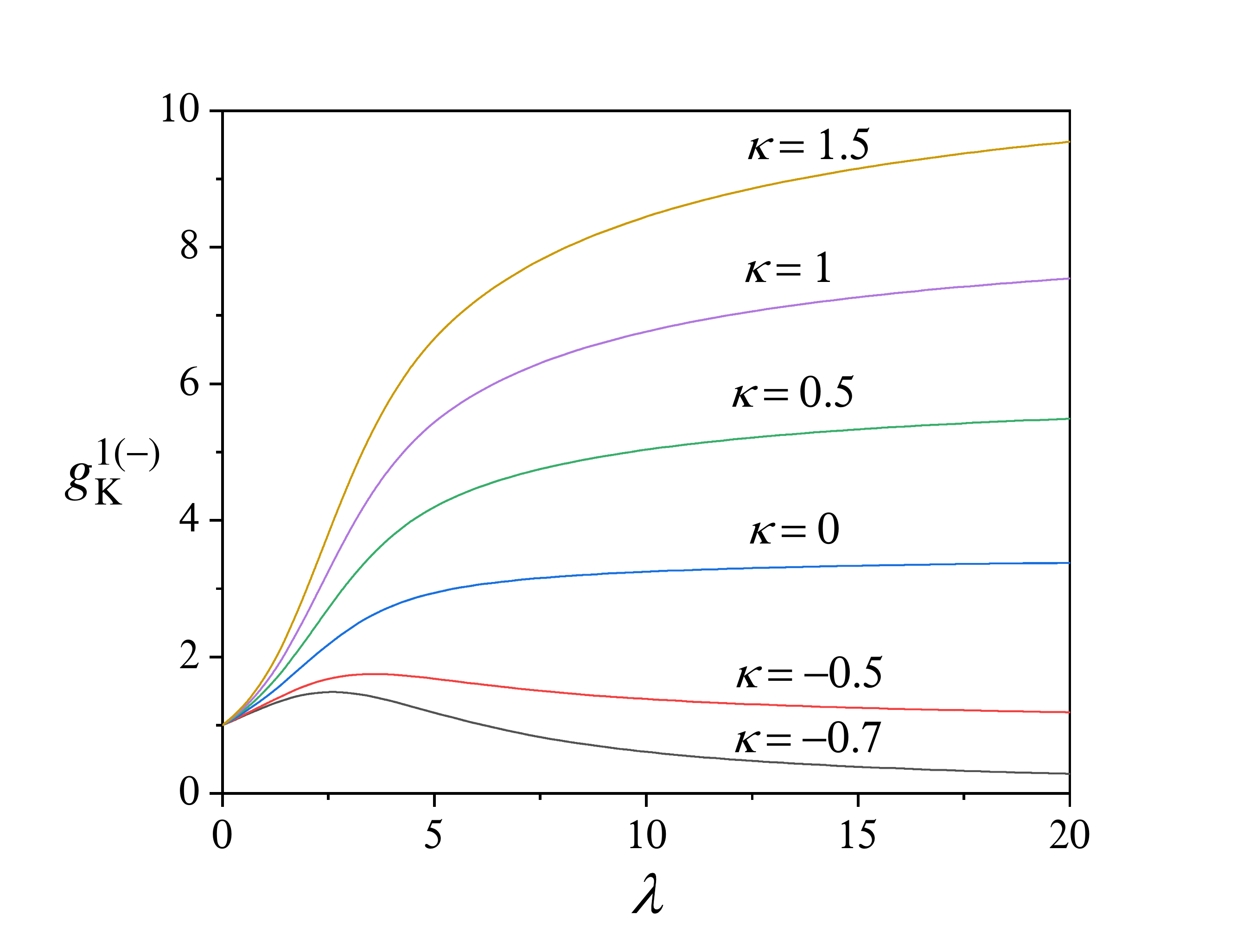}
\caption{Kirkwood correlation factor $g_{\rm K}^{1(-)}$ as a function of $\lambda$ for various values of $\kappa$.} 
\label{plots1}
\end{figure}

\begin{figure}[h!]
\centering
\includegraphics[height=7cm]{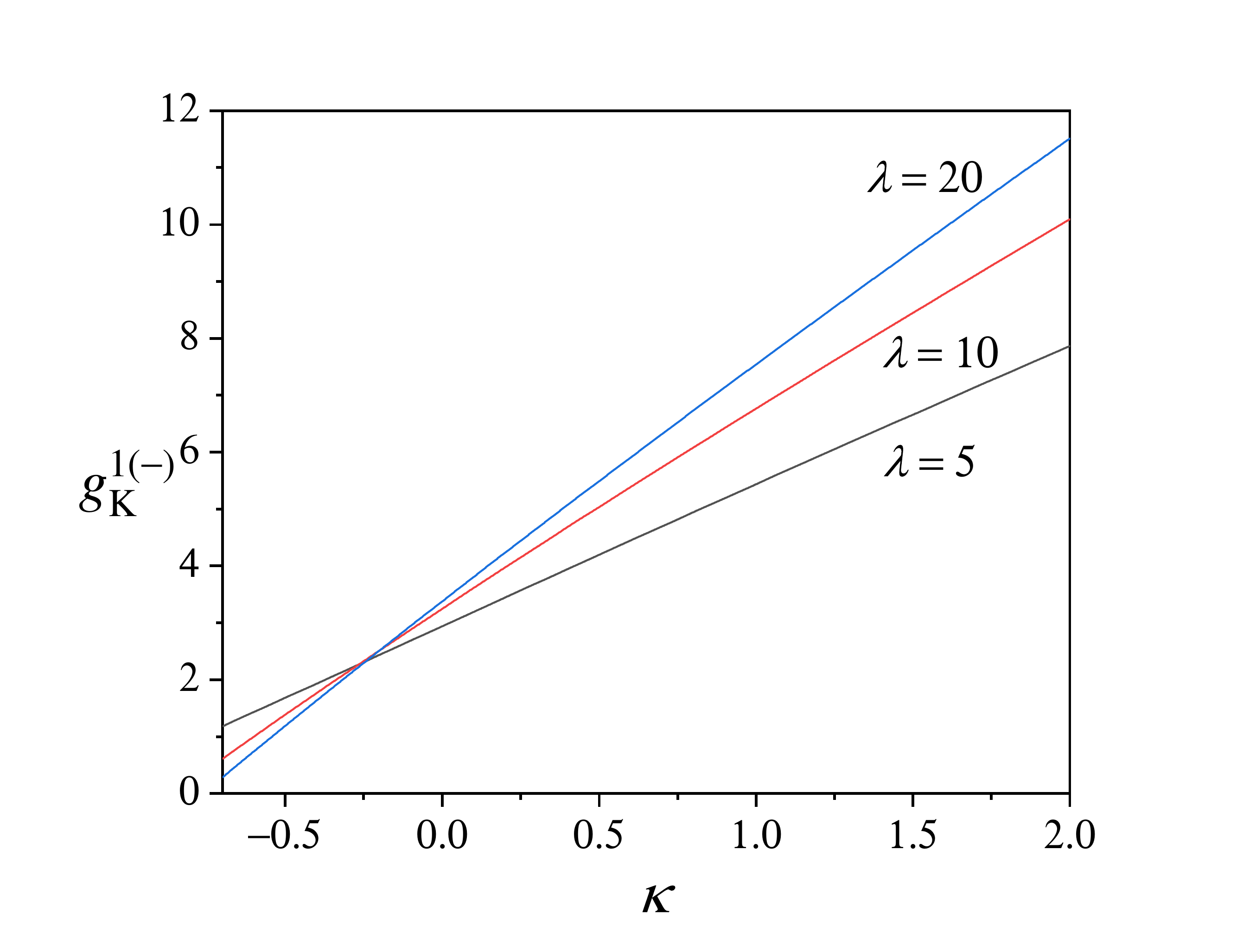}
\caption{Kirkwood correlation factor $g_{\rm K}^{1(-)}$ as a function of $\kappa$ for various values of $\lambda$.} 
\label{plots2}
\end{figure}

Thus, we essentially have interaction energies $U_m$, and four corresponding Kirkwood potentials of mean torques given in Table \ref{TableKirk}. This leads to four values $g_{\rm K}^{1(-)}$, $g_{\rm K}^{2(-)}$, $g_{\rm K}^{1(+)}$ and $g_{\rm K}^{2(+)}$ that reduce to previously derived results for $g_{\rm K}$ when $\kappa = 0$,\cite{Dejardin2018JCP} i.e., when $V_{indisp}$ is neglected. The variation of $g_{\rm K}^{1(-)}$ as a function of $\lambda$ and $\kappa$ is represented in Figures ~\ref{plots1} and ~\ref{plots2}. One notices the substantial increase of the Kirkwood correlation factor as $\kappa$ is increased from $0$. The explanation is that in this situation, the induction term neither affects the location of the minima $(0,0)$ and $(\pi,\pi)$, nor the location of the saddle point $(\frac{\pi}{2},\frac{\pi}{2})$ of both $U_m$ and $V_{2}^\text{eff}$, but \textit{increases} the energy barrier separating the two multidimensional minima in $V_{2}^\text{eff}$, which in turn governs the pair equilibrium statistics. As a result, the parallel states $(0,0)$ and $(\pi,\pi)$ are made even more (respectively less) probable for $\kappa>0$ (respectively $\kappa<0$) than for $\kappa = 0$. This results in an increase (respectively a decrease) in the Kirkwood correlation factor with respect to the situation where $\kappa = 0$. As illustrated in Figure ~\ref{plots2}, the variation of $g_{\rm K}^{1(-)}$ with $\kappa$ for given $\lambda$ is linear. This means that in this situation, the dipolar field has a trend to induce a dipole in the same direction as that of the alignment of the molecular permanent dipole moments. Thus, the bonds are slightly stretched, so the atomic charge distributions are more distant than in the absence of induced dipoles. The result is simply a proportion of $g_{\rm K}$ with $\kappa$. We also note from Figures \ref{plots1} and \ref{plots2} that values of $g_{\rm K}<1$ are possible \textit{in spite of preferred parallel alignment of the permanent dipoles}. Now, if too large negative $\kappa$ values are used here, this causes $g_{\rm K}^{1(-)}$ to take unphysical negative or null values. The higher transcendental nature of the functions representing the integrals makes it difficult to precisely state the limiting $\kappa$ value at which this occurs, nevertheless these integrals can straightforwardly be computed numerically. Therefore, if any negative $\kappa$ value is to be applied when comparing the present theory with experiments, then one must guarantee the positiveness of $g_{\rm K}^{1(-)}$ in the whole temperature range where the species under study is in its liquid phase.

\begin{figure}[h!]
\centering
\includegraphics[height=7cm]{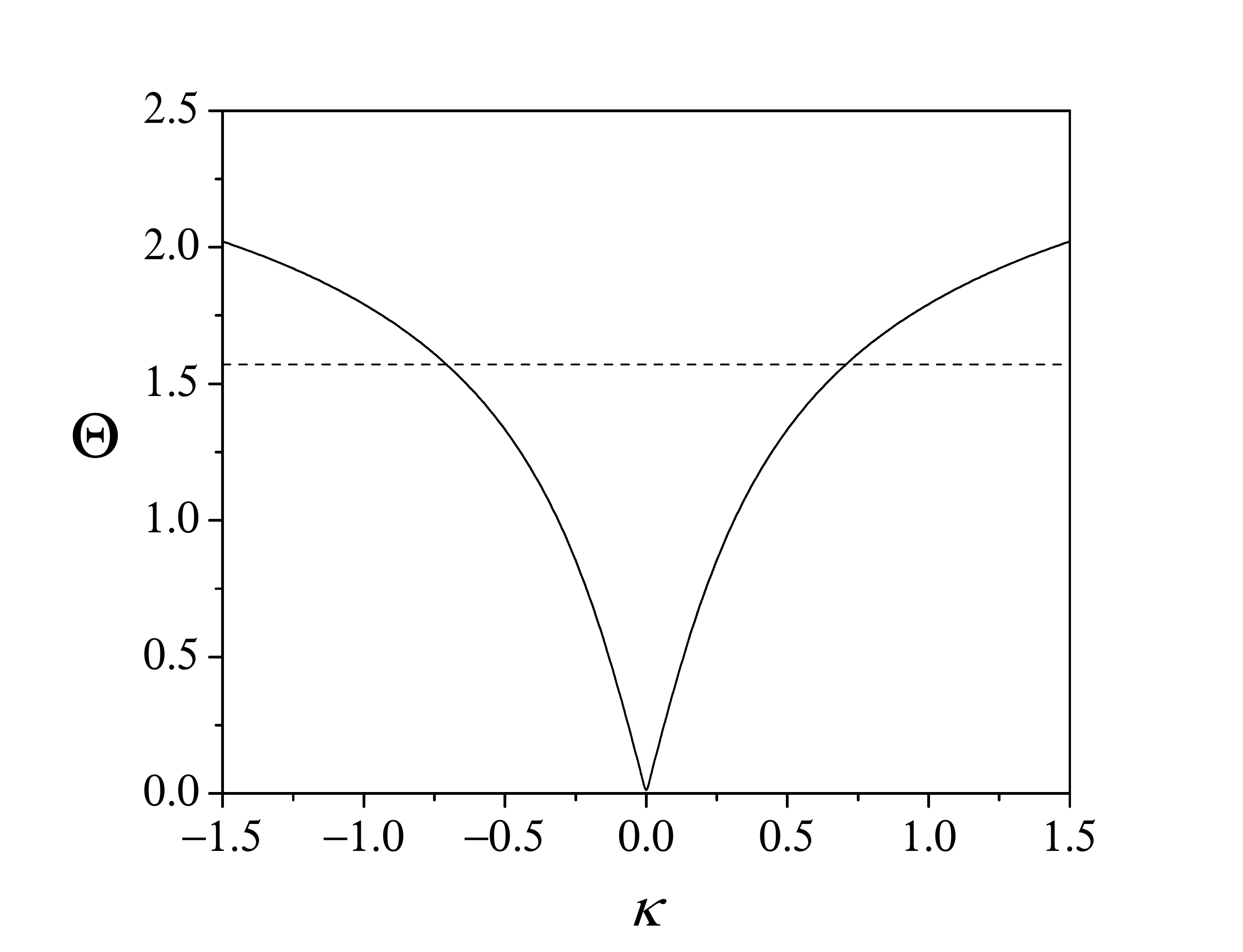}
\caption{The maximal relative orientation of dipole pairs $\Theta$ as a function of $\kappa$. The dashed line is the $\pi/2$ relative orientation.} 
\label{Theta}
\end{figure}

\begin{figure}[h!]
\centering
\includegraphics[height=7cm]{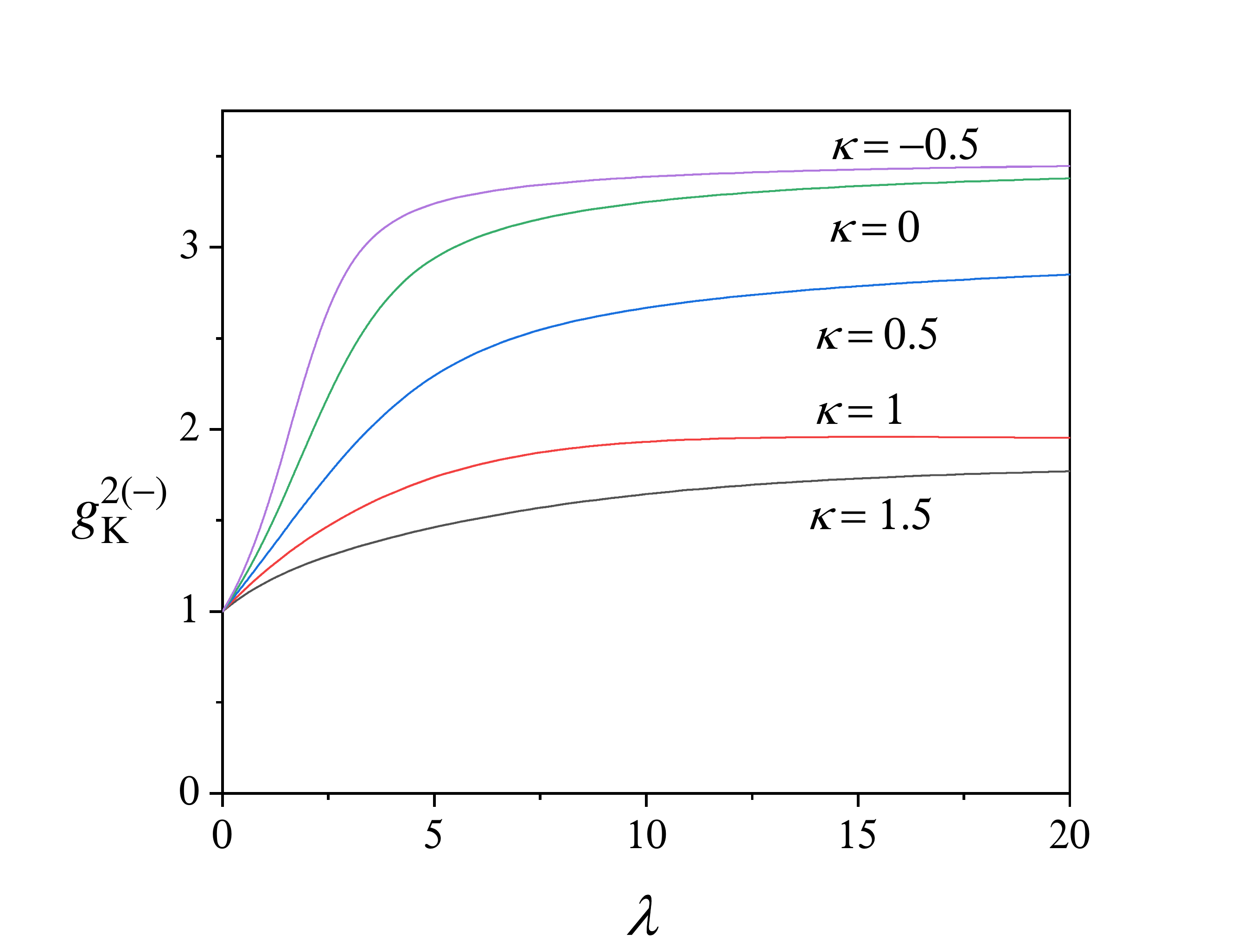}
\caption{Kirkwood correlation factor $g_{\rm K}^{2(-)}$ as a function of $\lambda$ for various values of $\kappa$.} 
\label{plots3}
\end{figure}

\begin{figure}[h!]
\centering
\includegraphics[height=7cm]{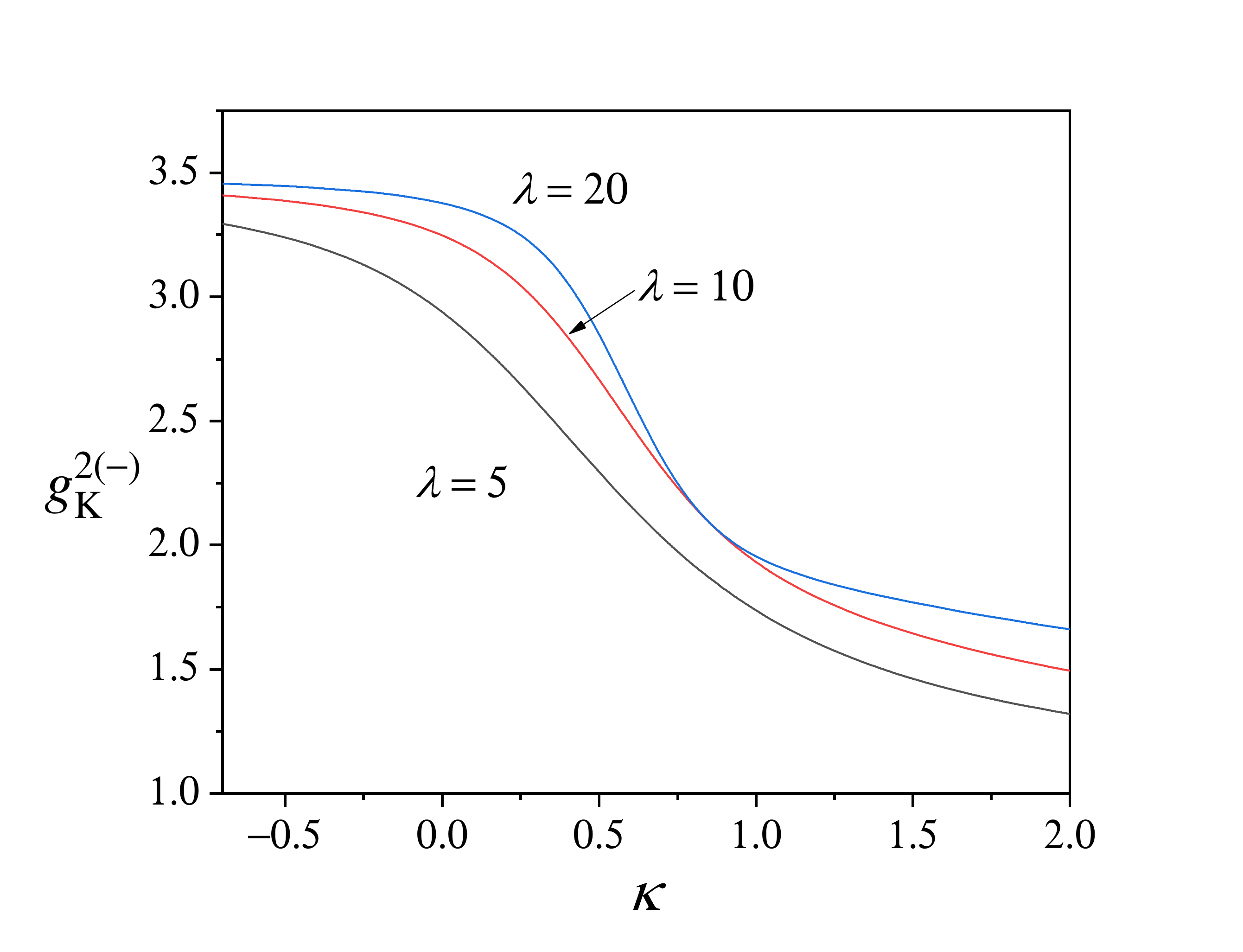}
\caption{Kirkwood correlation factor $g_{\rm K}^{2(-)}$ as a function of $\kappa$ for various values of $\lambda$.} 
\label{plots4}
\end{figure}

Figures~\ref{plots3} and ~\ref{plots4} show the behavior of $g_{\rm K}^{2(-)}$ when $\lambda$ and $\kappa$ are varied. In this situation, the locations of the minima of $V_{2}^\text{eff}$ are affected in raising $\kappa$, while the saddle point remains unchanged. Thus, the strictly parallel equilibrium states are affected, and pairs of dipoles form an angle at equilibrium, so that the pair alignment state is \textit{a canted one}. The energy barrier separating the two minima is furthermore lowered and therefore the equilibrium states are less populated with respect to the situation where $\kappa = 0$. Altogether, this results in a decrease of $g_{\rm K}$. Unlike for $g_{\rm K}^{1(-)}$, the behavior of $g_{\rm K}^{2(-)}$ with $\kappa$ is not linear at all. Here, a tentative explanation may be that the term $V_{indisp}$ fights non-trivially against the aligning effect of the permanent dipole moments due to $V_{dd}$. Altogether, the equilibrium parallel alignment of permanent dipoles is affected. The angle between a pair of dipoles in the wells is not so well-defined in this situation, as our simplified interaction potentials are azimuth-independent, so that in the present model transverse modes are energy costless modes. Nevertheless, according to our model, we may state that the relative orientation of dipole pairs at equilibrium obeys the double inequality:
\begin{eqnarray}
0\leq\vartheta_{(\mathbf{u},\mathbf{u}')}\leq 2\arctan{\frac{\sqrt{\sqrt{1+16\kappa^{2}}-1}}{\sqrt{2}}}
\label{EqAngle}
\end{eqnarray}
where the upper bound in Eq.\,\eqref{EqAngle} is equal to $\Theta=\vartheta_{min}+\vartheta'_{min}=2\vartheta_{min}$ and $(\vartheta_{min},\vartheta'_{min})$ is  the location of one of the deepest symmetric minima of the corresponding Kirkwood potential of mean torques, while the lower bound is given by $\vartheta_{min}-\vartheta'_{min}=0$. Thus, the relative orientation of dipole pairs may be \textit{larger} than $\pi/2$, in spite of the fact that in this situation, $g_{\rm K}>1$. In order to illustrate this, we have plotted the quantity $\Theta$ as a function of $\kappa$ in Figure ~\ref{Theta}, where it becomes clear that $\Theta$ may be larger than $\pi/2$ at some $\kappa$ values.

\begin{figure}[h!]
\centering
\includegraphics[height=7cm]{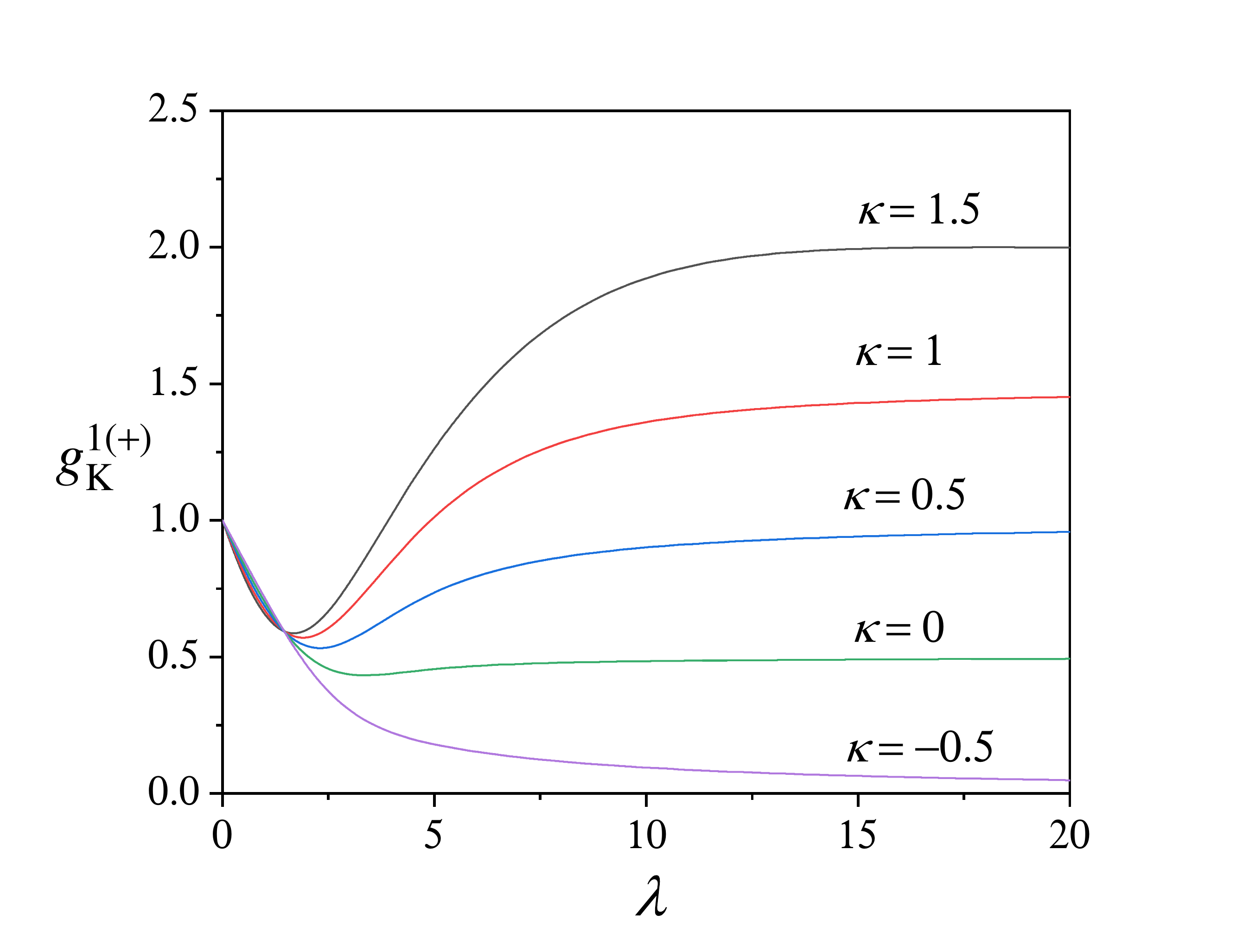}
\caption{Kirkwood correlation factor $g_{\rm K}^{1(+)}$ as a function of $\lambda$ for various values of $\kappa$.} 
\label{plots5}
\end{figure}

\begin{figure}[h!]
\centering
\includegraphics[height=7cm]{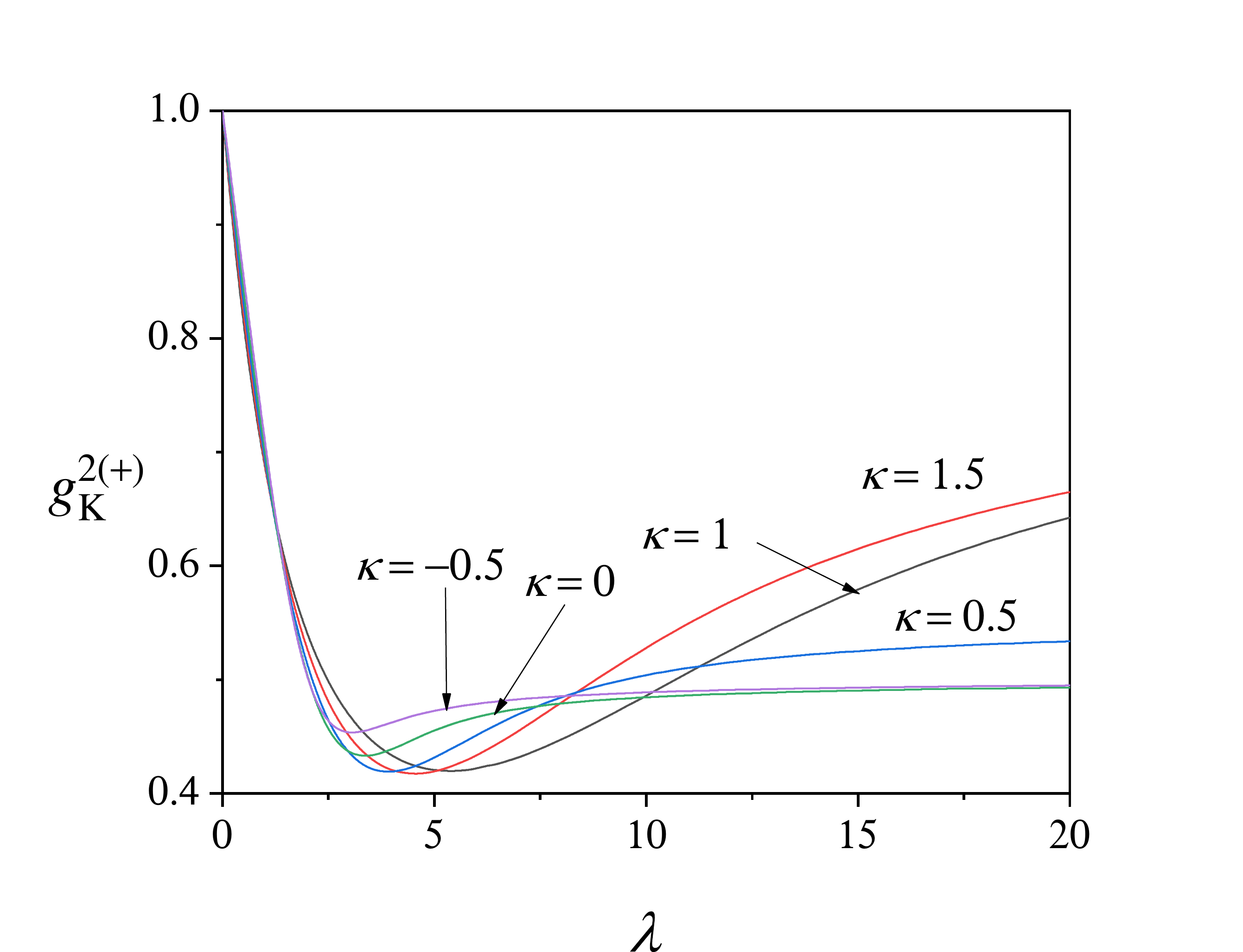}
\caption{Kirkwood correlation factor $g_{\rm K}^{2(+)}$ as a function of $\lambda$ for various values of $\kappa$.} 
\label{plots6}
\end{figure}

\noindent
This unusual result is explained by the very definition of $g_{\rm K}$, showing that Eq.\,\eqref{KirkwoodGk} is an over-idealization of the real value of $g_{\rm K}$ given by Eq.\,\eqref{KirkGk}. Hence, the Kirkwood estimate for $g_{\rm K}$ only applies to very special cases such as liquid water. Thus, in particular, $g_{\rm K}>1$\textit{ does not guarantee the parallel alignment of dipole pairs at equilibrium}. In the next section we give a comparison of our calculations with the experimental temperature dependence of the static linear permittivity of tributyl phosphate in order to illustrate the situation we just described.
The variation of $g_{\rm K}^{1(+)}$ and $g_{\rm K}^{2(+)}$ with $\lambda$ for various values of $\kappa$ are shown in Figures~\ref{plots5} and ~\ref{plots6}. These values of the Kirkwood correlation factor correspond to preferred antiparallel alignment when $\kappa=0$. The most remarkable feature of $g_{\rm K}^{1(+)}$ is that in this situation, the Kirkwood correlation factor is able to exhibit both $g_{\rm K}$ values that are smaller and larger than 1 (this effect is similar with the "quadrupolar effect" dealt with by Stell et al. \cite{Stell1981Book}), and that this happens at moderate values of $\lambda$. Furthermore, for $\kappa>0$, $g_{\rm K}^{1(+)}$ is able to render negative values of $g_{\rm K}$ if $\vert\kappa\vert$ takes too large values, so that the same prescriptions as those given above for $g_{\rm K}^{1(-)}$ apply to $g_{\rm K}^{1(+)}$ when attempting a comparison with experimental data.

\begin{figure}[h!]
\centering
\includegraphics[height=7cm]{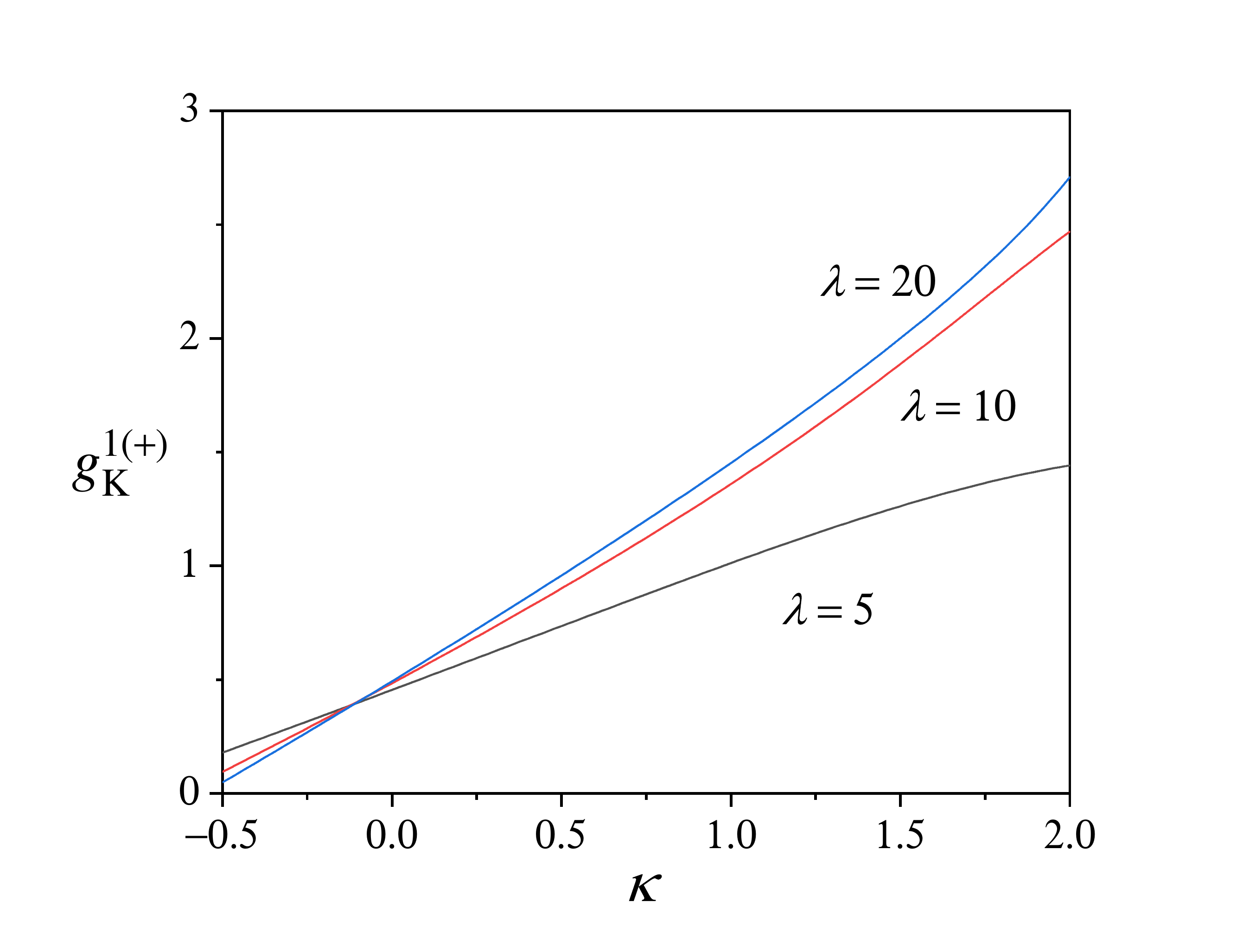}
\caption{Kirkwood correlation factor $g_{\rm K}^{1(+)}$ as a function of $\kappa$ for various values of $\lambda$.} 
\label{plots7}
\end{figure}

\begin{figure}[h!]
\centering
\includegraphics[height=7cm]{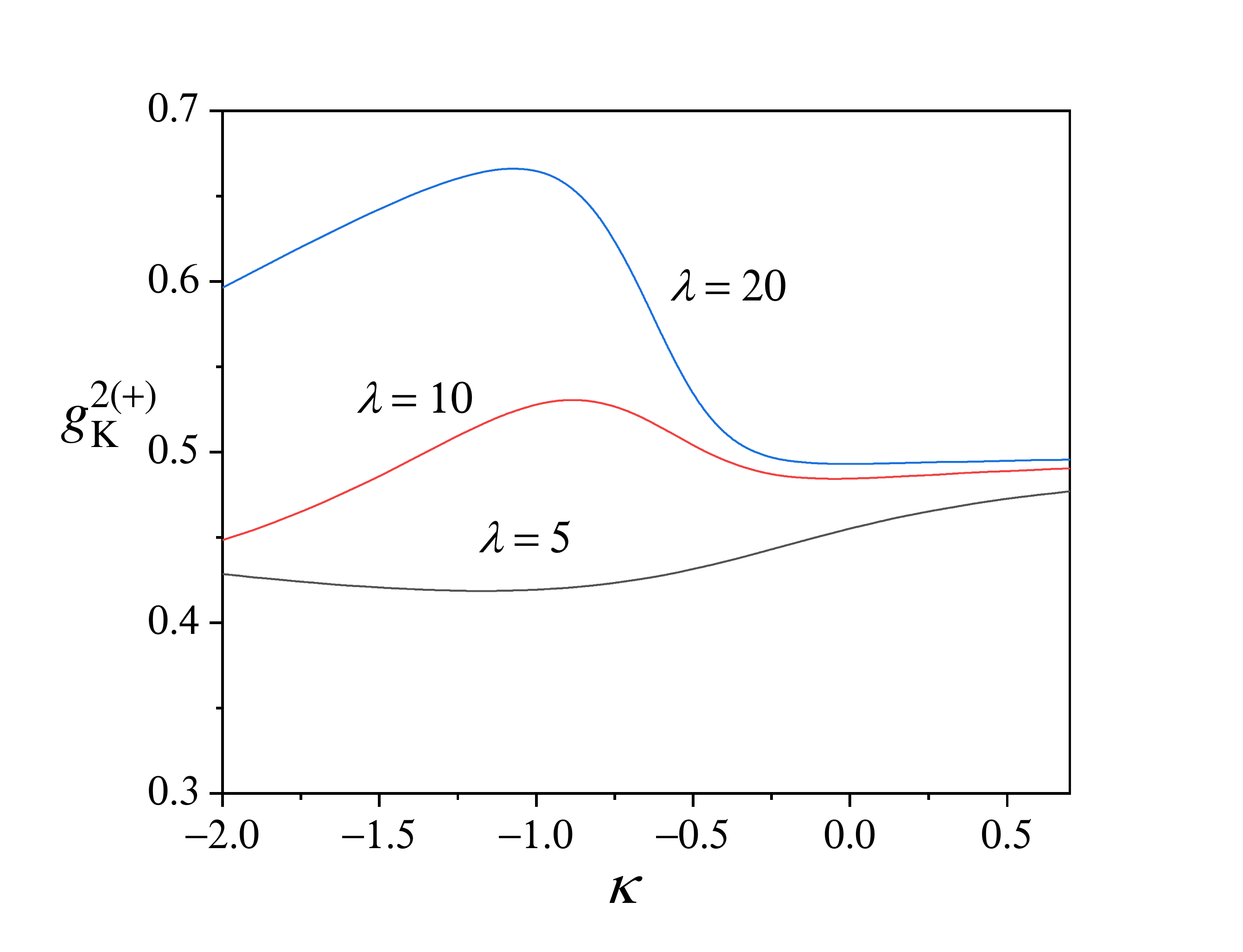}
\caption{Kirkwood correlation factor $g_{\rm K}^{2(+)}$ as a function of $\kappa$ for various values of $\lambda$.} 
\label{plots8}
\end{figure}

The variation of $g_{\rm K}^{1(+)}$ and $g_{\rm K}^{2(+)}$ with $\kappa$ is shown on Figures ~\ref{plots7} and \ref{plots8}. As for $g_{\rm K}^{1(-)}$, the variation of $g_{\rm K}^{1(+)}$ with $\kappa$ is linear, so that the stretching of molecular bonds has the same effect as that for $g_{\rm K}^{1(-)}$. In fact, here, the extra dipole is induced in the direction opposite to the permanent dipole alignment direction, leading to an overall increase of $g_{\rm K}$, therefore to an increase of the dielectric constant with respect to the situation where $\kappa=0$. At last, in this situation, the minima of the  potential $V_{2}^\text{eff}$ are those of antiparallel alignment.

In contrast, the variation of $g_{\rm K}^{2(+)}$ with $\kappa$ is not linear at all. Here, the explanation is different from the $\kappa$ behavior of variation of $g_{\rm K}^{2(-)}$. In effect, for positive $\kappa$, the Kirkwood potential of mean torques $V_{2}^\text{eff}$ exhibits $2$ pairs of unequal minima in a cycle of the motion of dipole pairs, located both at the parallel and antiparallel states. This altogether affects the $g_{\rm K}$ value in a non-trivial way, depending on the $\lambda$ values. For negative $\kappa$, the equilibrium orientations of the permanent moments are spread over the range:

\begin{eqnarray}
\pi-2\arctan{\frac{\sqrt{\sqrt{1+16\kappa^{2}}-1}}{\sqrt{2}}}\leq\vartheta_{(\mathbf{u},\mathbf{u}')}\leq\pi.
\label{EqAngleBis}
\end{eqnarray}
\noindent
This is similar with the behavior of $g_{\rm K}^{2(-)}$ as in this situation, dipoles are induced in such a way that they are parallel. Here, $g_{\rm K}$ is near $0.5$, as if the induction term did not significantly affect orientational correlations. 

\section{Comparison with experimental data}\label{TheoryExperiment}

In this section we compare our theoretical findings with experimental data. In order to do so, we use static dielectric permittivity values either from the literature, i.e., unless stated otherwise, values from Wohlfarth's Landolt-Bornstein Tables,\cite{Wohlfarth2008Book} or from our own measurements and compare them to calculated values employing the theory described in the foregoing sections. 
In the Kirkwood-Fr\"{o}hlich theory, the dielectric constant is given by: 
\begin{eqnarray}
\varepsilon=\frac{1}{4}\left(3\lambda g_{\rm K}+\varepsilon_{\infty}+\sqrt{8\varepsilon_{\infty}^{2}+(3\lambda g_{\rm K}+\varepsilon_{\infty})^{2}}\right)
\label{FrohlichFEps}
\end{eqnarray}
where
\begin{eqnarray}
\lambda(T)=\frac{M_{v}(T)N_{A}\mu_{g}^{2}(n^{2}(T)+2)^{2}}{27M_{\text{mol}}\varepsilon_{0}kT}
\label{ExpLambda}
\end{eqnarray}
Here, $n$ is the mean refractive index of the fluid measured for the Sodium D spectral line and $M_{v}(T)$ is the experimentally measured temperature-dependent mass density of the polar fluid. Both quantities are sometimes extrapolated to the temperature of interest either via the equations given in the respective references or via a linear law fitted to the measured values. Furthermore, in Eq.\,\eqref{ExpLambda}, following Onsager, Kirkwood and Fr\"{o}hlich, \cite{Onsager1936JACS,Kirkwood1939JCP,Frohlich1958Book} we set
\begin{eqnarray}
\varepsilon_{\infty}(T)= n^{2}(T).
\label{epsinfty}
\end{eqnarray}
For some polar fluids we compute it from the Lorenz-Lorentz equation, i.e. :
\begin{eqnarray}
\frac{n^{2}(T)-1}{n^{2}(T)+2}=\frac{M_{v}(T)N_{A}\bar{\alpha}}{3M_{mol}\varepsilon_{0}}
\label{LorenzLorentz}
\end{eqnarray}
where $\bar{\alpha}$ is the mean molecular polarizability, taken from the literature. 

The Kirkwood correlation factor $g_K$ in Eq.~\eqref{FrohlichFEps} is, according to our theory, dependent on $\lambda(T)$ and $\kappa$, and four different functions for $g_K(\lambda,\kappa)$ are possible according to Table~\ref{TableKirk}. By substituting the respective $U_m$ and $V_{2}^\text{eff}$ as well as Eq.~\eqref{eq9} into Eq.~\eqref{UsableGk}, the Kirkwood correlation factor is calculated by numerical integration.

As mentioned above, $\kappa$ can be regarded as a measure of the strength of the induction/dispersion-type interaction and is the only unknown parameter which is needed to calculate the theoretical Kirkwood correlation factor. It is expected that $\kappa$ is somehow related to the molecular polarizability $\bar{\alpha}$, however, in the current state of our theory, it can not be determined explicitly and thus it is left as the only fitting parameter to achieve agreement between theory and experiment. The choice between the four different representations of $g_K(\lambda,\kappa)$ is based upon some possibly existing foreknowledge about the preferred alignment from the literature and/or based upon the comparison of the theoretical and experimental temperature dependences of the static permittivity. Since the four $g_K(\lambda,\kappa)$ have distinct slopes depending on $\lambda(T)$, as can be seen in Figures~\ref{plots1},\ref{plots3},\ref{plots5},\ref{plots6}, \textit{this results in an unambiguous assignment of one}  $g_K(\lambda,\kappa)$ to the respective polar fluid.

In the following subsections we discuss the comparison of theory and experiment for different classes of polar liquids. An overview of all substances under study, including all values needed to calculate the Kirkwood correlation factor is given in Table~\ref{ElectricParameters}.

\begin{table*}[h!]
\small
\caption{Parameters used in the computation of the static permittivity Eq.\eqref{FrohlichFEps}. Mean molecular polarizabilities from Ref.\citenum{Gussoni1998JMS}. Molecular dipole moments from Ref.\citenum{Maryott1967Book} except $^{(a)}$ from Ref. \citenum{Malecki1984JPC} and $^{(b)}$ from Ref.\citenum{petkovic1973JPhysChem}, which is the value of the dipole moment of TBP in decalin, which is a nonpolar solvent that has no influence on the molecular TBP dipole. $^{(c)}$ We performed refractive index measurements between 10$^\circ$C and 50$^\circ$C using an Abbe refractometer.}
\label{ElectricParameters}
\begin{tabular*}{\textwidth}{@{\extracolsep{\fill}}lllllll}
\hline 
 & $\mu_g$ (D) & $\bar{\alpha}(\text{\AA{}}^{3})$ & $\kappa$ & $g_{\rm K}$ & $M_{v}(T)$ & $n(T)$ \\ 
\hline 
\hline
Methanol & 1.68 & - & 0.04 & $g_{\rm K}^{1(-)}$ & Ref.\citenum{DDBST2020} & Ref.\citenum{ortega1982densities} \\ 
\hline 
Ethanol & 1.68 & - & 0.05 & $g_{\rm K}^{1(-)}$  & Ref.\citenum{DDBST2020} & Ref.\citenum{ortega1982densities}\\ 
\hline 
Propan-1-ol & 1.68 & - & 0.22 & $g_{\rm K}^{1(-)}$ & Ref.\citenum{Daubert1997Book} & Ref.\citenum{ortega1982densities} \\ 
\hline 
Butan-1-ol & 1.68 & - & 0.35 & $g_{\rm K}^{1(-)}$ & Ref.\citenum{DDBST2020} & Ref.\citenum{ortega1982densities} \\ 
\hline 
Pentan-1-ol & 1.68 & - & 0.5 & $g_{\rm K}^{1(-)}$ & Ref.\citenum{boned2008liquid} & Ref.\citenum{ortega1982densities} \\ 
\hline 
Hexan-1-ol & 1.68 & - & 0.65 & $g_{\rm K}^{1(-)}$ & Ref.\citenum{alaoui2012liquid} & Ref.\citenum{ortega1982densities} \\ 
\hline 
Heptan-1-ol & 1.68 & - & 1.05 & $g_{\rm K}^{1(-)}$  & Ref.\citenum{alaoui2013liquid} & Ref.\citenum{ortega1982densities} \\ 
\hline 
Octan-1-ol & 1.68 & - & 1.5 & $g_{\rm K}^{1(-)}$ & Ref. \citenum{ye2012} & Ref.\citenum{ortega1982densities} \\ 
\hline 
Water & 1.845 & 1.501 & -0.15 & $g_{\rm K}^{1(-)}$ & Ref.\citenum{DDBST2020} & L.-L. \\ 
\hline 
Acetonitrile & 3.92 & 4.44 & 0.345 & $g_{\rm K}^{1(+)}$ & Ref.\citenum{DDBST2020} & L.-L. \\ 
\hline 
Nitrobenzene & $4.02^{a}$ & 12.26 & 0.67 & $g_{\rm K}^{1(+)}$ & Ref.\citenum{Mazur1931Nature} & L.-L. \\ 
\hline 
Acetone & 2.88 & 6.27 & 0.83 & $g_{\rm K}^{1(+)}$ & Ref.\citenum{DDBST2020} & L.-L. \\ 
\hline 
DMSO & 3.96 & 7.97 & 0.73 & $g_{\rm K}^{1(+)}$ & Ref.\citenum{Schlafer1960AngeChem} & Ref.\citenum{Schlafer1960AngeChem} \\ 
\hline 
TBP & $2.6^{b}$ & - & 0.85 & $g_{\rm K}^{2(-)}$ & Ref.\citenum{tian2007densities} & own$^{(c)}$ \\ 
\hline 
Glycerol & 2.67 & 7.80 & -0.3 & $g_{\rm K}^{1(-)}$ & Ref.\citenum{Blazhnov2004JCP} & L.-L.  \\ 
\hline 
\end{tabular*} 
\end{table*} 

\subsection{Parallel alignment -- Linear primary alcohols}

We start with a series of linear primary alcohols with different alkyl-chain length, for which preferred parallel alignment of the dipole moments, which are located at the O$-$H group at one end of the carbon chain, is well known. 
Different values for this dipole moment of linear primary alcohols are found in the literature, and these values usually range between 1.65 and 1.70 D \cite{Maryott1967Book}. Since the total dipole of a molecule is the sum of the dipole moments of its chemical bonds, and the C$-$H bonds are almost apolar,  the permanent dipole moment of all linear primary alcohols should be the same in a first approximation. An average value of 1.68~D has thus be chosen as the value of $\mu_g$ for all the considered linear primary alcohols. 

\begin{figure}[h!]
\centering
\includegraphics[height=6.5cm]{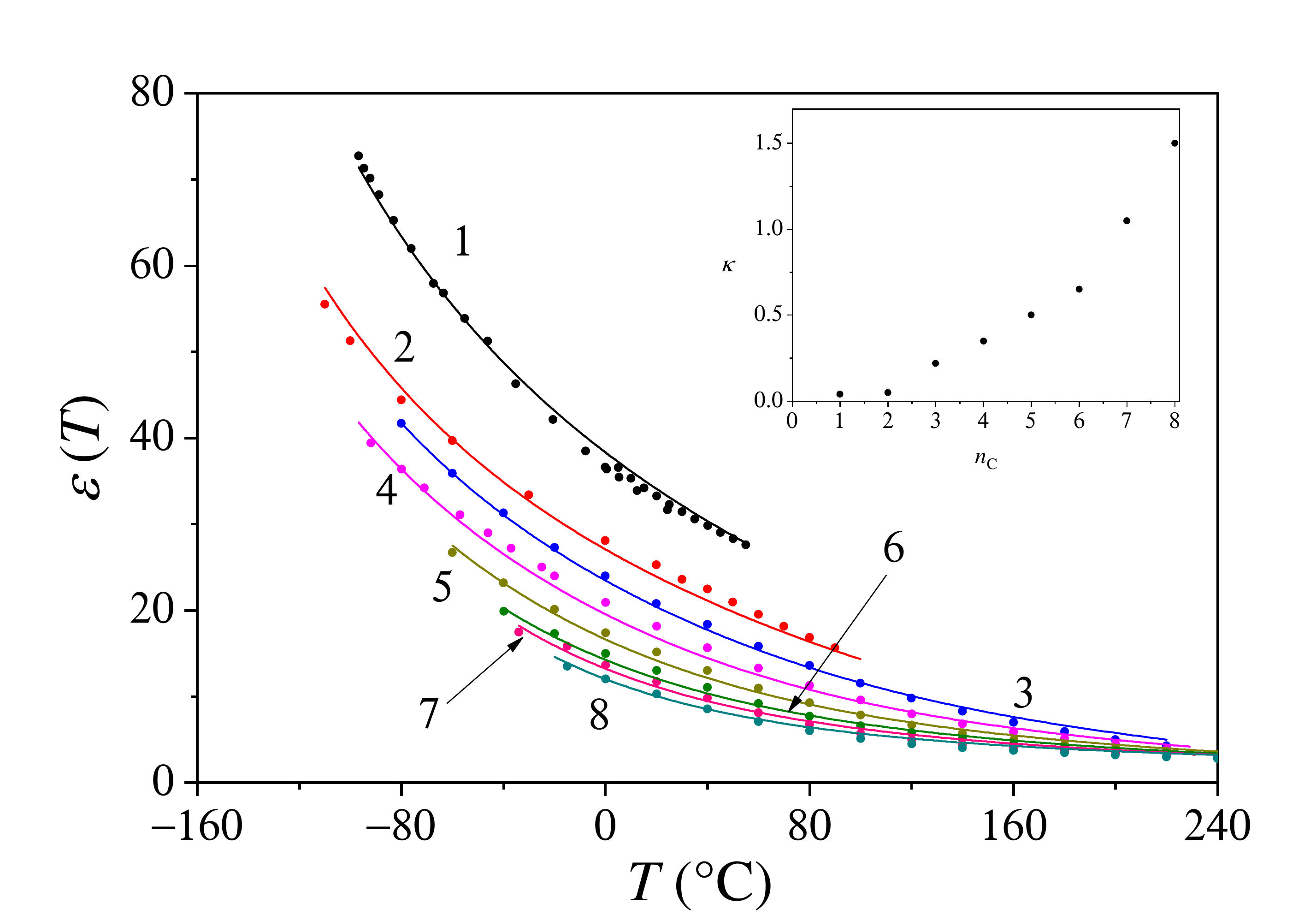}
\caption{Experimental temperature dependence of the linear static permittivity of Methanol (1), Ethanol (2), Propan-1-ol (3), Butan-1-ol (4), Pentan-1-ol (5), Hexan-1-ol (6), Heptan-1-ol (7) and Octan-1-ol (8). Solid line : Theory. Dots : Experimental data from Reference \citenum{Wohlfarth2008Book}. For heptan-1-ol, the experimental data are the same as those published by Vij el al.\cite{Vij1978JPhysD} at normal pressures. Inset : variation of $\kappa$ with the number of carbon atoms $n_{\rm C}$ in the alkyl chain. } 
\label{plots9}
\end{figure}

In Figure~\ref{plots9} the experimental static permittivities for all alkyl-chain lengths from methanol to octan-1-ol are shown as plain circles, together with the theoretical values calculated using $g_K^{1(-)}$ as solid lines.  

As one can see, the agreement of the theoretical values with the experimental ones is excellent for all linear primary alcohols over the whole temperature range where experimental data are available. The values of $\kappa$, which are chosen in order to achieve this agreement, are shown in the inset of Figure~\ref{plots9}. It is obvious that $\kappa$ increases with increasing number of carbon atoms in the alkyl-chain, which indicates the increasing strength of the induction/dispersion-type interaction. Since the polarizability of a molecule increases with its molecular mass while  the permanent dipole moment is the same for all molecules of this series, this finding is perfectly reasonable and underlines the importance of the induction/dispersion-type interaction for larger molecules.  However, it is clear that the $\kappa$ parameter does not depend linearly on the number of carbon atoms in the alkyl-chain, which shows that the latter parameter is not a trivial function of the polarizability, particularly as a result of non-additivity of induction-dispersion energies \cite{Stone2013Book}. Therefore, the determination of $\kappa$ from molecular properties is beyond the scope of this work and thus is left as a fitting parameter. 

\begin{figure}[h!]
\centering
\includegraphics[height=7cm]{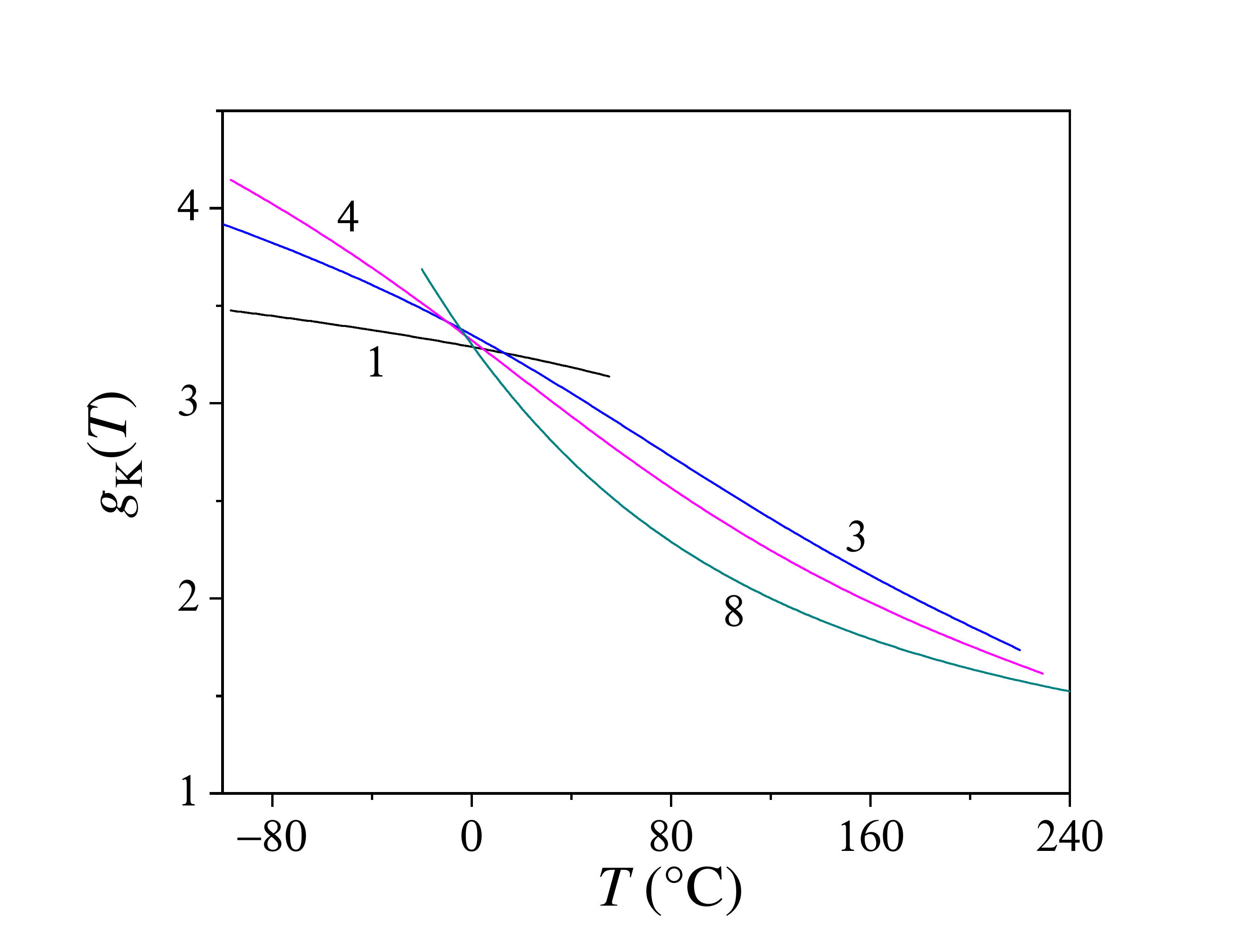}
\caption{Experimental temperature dependence of the Kirkwood correlation factor of Methanol (1),1-Propanol (3), 1-Butanol (4) and 1-Octanol (8)}
\label{plots12}
\end{figure}

As indicated by the use of $g_{\rm K}^{1(-)}$, the preferred dipolar order in these substances is, as is well-known, the parallel one. The temperature dependence of the calculated Kirkwood correlation factor is shown in Figure \ref{plots12}, only for some of these substances for clarity.

It is obvious that the slope of $g_K(T)$ is non-trivial and behaves distinctly different for various linear alcohols and it agrees with those found experimentally in the literature.\cite{Bottcher1973Book} Therefore, by adjusting the strength of the induction/dispersion-type interaction via $\kappa$, our theory is able to calculate the correct Kirkwood correlation factor and thus reproduces the experimental static permittivities. At last, since the graphical representation of $g_{\rm}^{1(-)}(\kappa)$ is a straight line, there is a one for one correspondance between a selected value of $\kappa$ and $\varepsilon$, so that our theoretical uncertainty on all calculated parameters \textit{is zero}. 

\subsection{Antiparallel alignment}
\begin{figure}[h!]
\centering
\includegraphics[height=7cm]{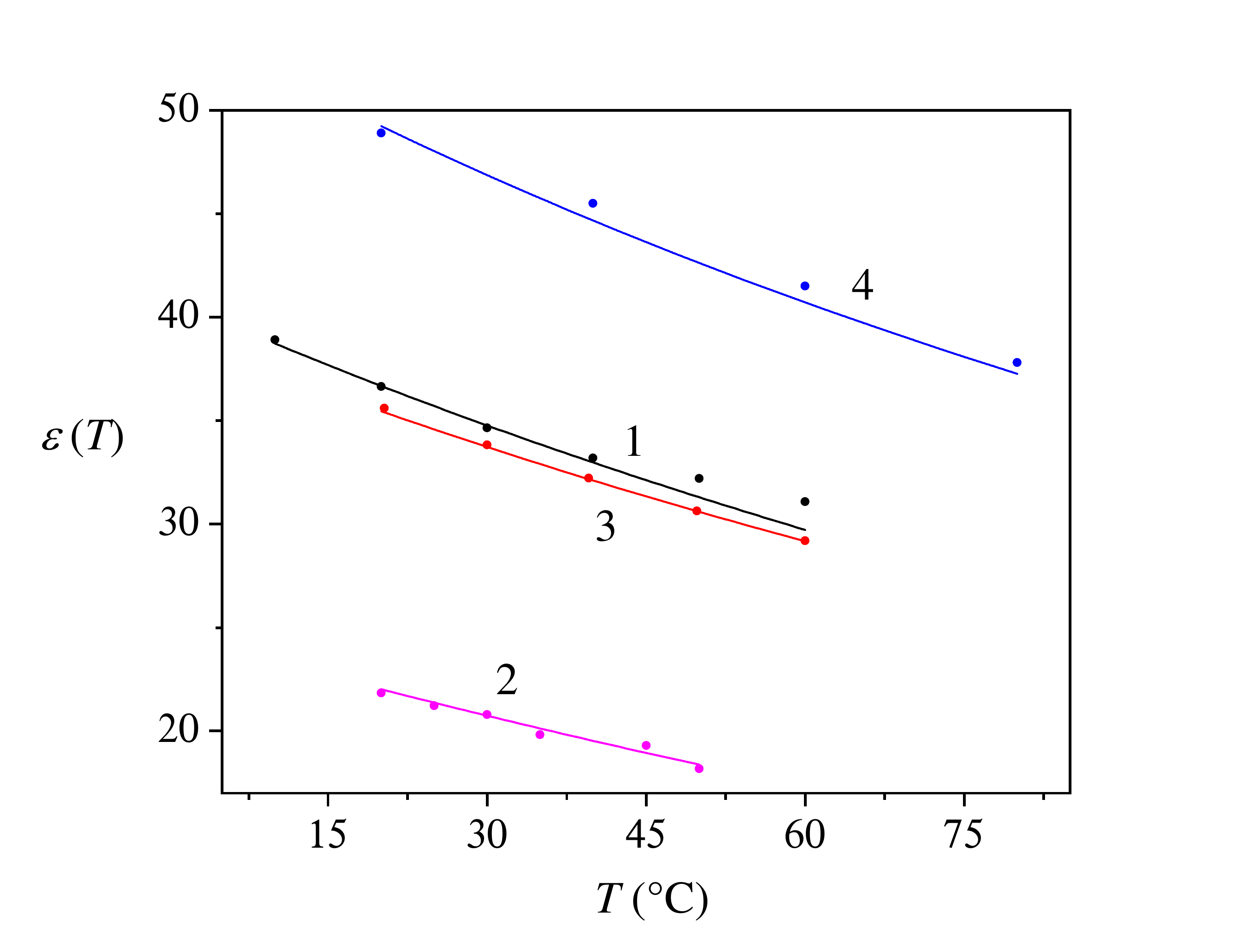}
\caption{Experimental temperature dependence of the linear static permittivity of Acetonitrile(1), Acetone (2), Nitrobenzene (3) and DMSO (4). Solid line : Theory. Dots : Experimental points \cite{Wohlfarth2008Book}. DMSO data, including density and refractive index from Schl\"{a}fer et al. \cite{Schlafer1960AngeChem}} 
\label{plots11}
\end{figure}

\begin{figure}[h!]
\centering
\includegraphics[height=7cm]{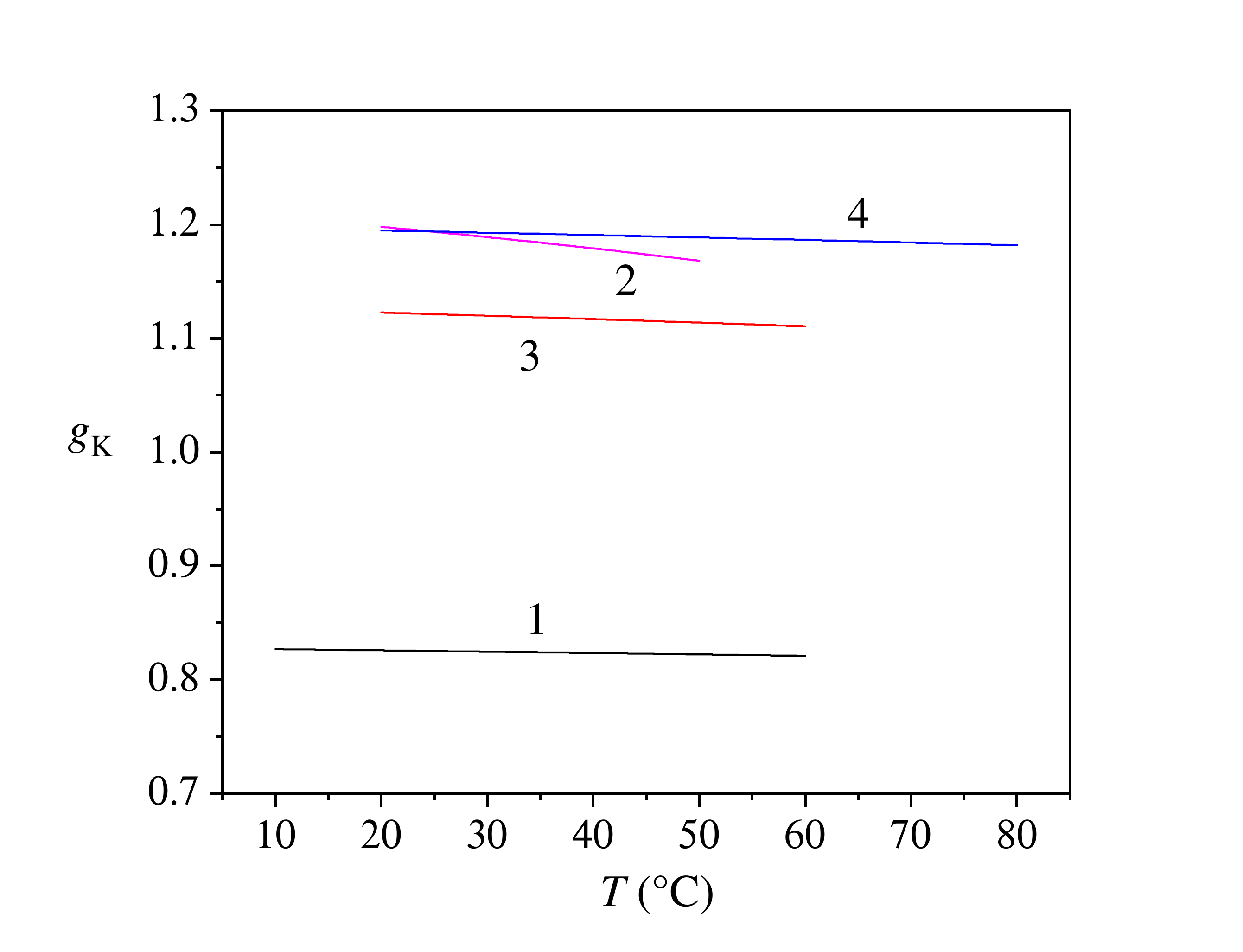}
\caption{Theoretical temperature dependence of Kirkwood correlation factor of acetonitrile(1), acetone(2), nitrobenzene(3), and DMSO(4).}
\label{plots14}
\end{figure}

In this subsection, we compare our theory with experimental static permittivities of substances, for which it is known from techniques other than dielectric spectroscopy, that they exhibit preferred antiparallel dipolar ordering. These substances are acetonitrile,\cite{hu2010acetonitrile} nitrobenzene,\cite{shelton2016orientation} acetone\cite{mclain2006orientational} and dimethyl sulfoxide (DMSO)\cite{mclain2006orientational} and the comparison between experiment and theory is shown in figure~\ref{plots11}.
Experimental data for these substances are only available over a narrow temperature range. However, as can be seen 
in figure~\ref{plots14}, the Kirkwood correlation factors hardly depends on temperature, thus this is not too great a drawback.

\subsubsection{Acetonitrile and Acetone}  
Our theoretical estimates of the static permittivity of Acetonitrile (ACN) apparently deviate from the experimental data of Stoppa et al. \cite{Buchner2015JMolLiq} at high temperatures, of at most $4.7\%$, while yielding good agreement at the lowest ones. Here, this is difficult to believe that the deviation between theory and experiment is due to a poor representation of intermolecular interactions as $\lambda$ takes rather low values at high temperatures. 
Yet, our theoretical findings remains not too far from the experimental data, and agree to some extent with the molecular dynamics data on the Kirkwood correlation factor of Koverga et al. \cite{Idrissi2017JMolLiq} 

For ACN, the Kirkwood correlation factor remains almost temperature-independent between $10^{\circ}$ and $60^{\circ}$ Celsius, yielding $g_{\rm K}\approx 0.82$. Since $g_{\rm K}^{1(+)}$ is used, the dipolar order is strictly antiparallel, as expected. These values agree reasonably well with the experimentally deduced values of Helambe et al. \cite{Helambe1995Pramana} in the pure liquid phase. 

Our theoretical estimates of the static permittivity of acetone are in good agreement with the experimental ones. We also find antiparallel order for acetone, using $g_{\rm K}^{1(+)}$ as a representative of $g_{\rm K}$. This substance exhibits the strongest temperature dependence of $g_{\rm K}$ out of the four substances discussed in this subsection, as illustrated in Figure \ref{plots14}. Our values range between $1.22$ at $20^{\circ}$ Celsius and decreases to $1.19$ at $50^{\circ}$. Our values are slightly above the value at $25^{\circ}$ Celsius of pure acetone by Kumbharkhane \etal\cite{Kumbharkhane1996Pramana} which is $1.02$, while Vij \etal\cite{Vij1991JPhysChem} found the value $1.38$. Our values are framed between both experimentally determined ones, and therefore, our theoretical findings may be considered as satisfactory for this substance in the considered temperature range. We emphasize that due to the relatively large value of $\kappa = 0.83$, the $g_{\rm K}$ values of acetone are above unity, despite preferred anti-parallel alignment.

\subsubsection{Nitrobenzene and DMSO}
The same notion is true for Nitrobenzene and DMSO, where a Kirkwood correlation factor of larger than one (see figure~\ref{plots14}) reproduces the experimental data in figure~\ref{plots11} quite well, employing $g_{\rm K}^{1(+)}$, i.e. antiparallel alignment.

We emphasize here again that the expectation that antiparallel dipolar alignment has to result in a Kirkwood correlation factor of less than unity based on Eq.\,\eqref{KirkwoodGk}, has led for example Shikata et al.,\cite{Shikata2014AIPAdv} like many authors, to use a too high value 
of $\varepsilon_{\infty}=3.5$, in order to obtain $g_K = 0.65 <1$ for nitrobenzene. This procedure is misleading, because  Eq.\,\eqref{KirkwoodGk} is most of the time a poor approximation of Eq. \eqref{KirkGk} and results in some cases in  somewhat arbitrary choices of $\varepsilon_{\infty}$, just to fulfill the expectations about the value of the Kirkwood factor in comparison with unity.

We also note here that great care must be taken regarding the frequency at which the dielectric constant is measured. If measurements are performed at a fixed frequency instead of measuring a spectrum over several orders of magnitude in frequency, one has to be sure that this frequency is sufficiently low to neglect relaxation effects but also sufficiently high so that one also can neglect electrode polarization effects stemming from ionic impurities, which might be present in some occasions.

For example, in the case of DMSO we have compared our theoretical findings with the data of Schl\"{a}fer et al., \cite{Schlafer1960AngeChem} who report measurements of the static permittivity at a measuring frequency of 100 kHz. We were quite surprised that the data of Schl\"{a}fer et al. were \textit{the only ones} (see Reference \citenum{Wohlfarth2008Book}) that we were able to interpret. Yet, they are the sole data of Reference \citenum{Wohlfarth2008Book} which, in our opinion, truly reflect the static permittivity of DMSO, because all data but Schl\"{a}fer's were recorded at least at a ten times higher frequency, indicating that dipolar relaxation might play a role, so that the measured permittivities can no longer be considered as the static ones.

We note in passing that Schl\"{a}fer et al. quote a dipole value of DMSO $\mu_{g}=4.3\pm0.1$ D, using Onsager's equation.\cite{Onsager1936JACS} In effect, we find that the Onsager dipole $\mu_{g}\sqrt{g_{\rm K}}$ varies between $4.28$ and $4.32$ D, in agreement with the experimental one.


Finally, we remark that Onsager's equation \cite{Onsager1936JACS} is generally most successful in polar substances with antiparallel order (one exception being liquid water) because as illustrated in Figure \ref{plots14}, generally $g_{\rm K}$ has almost no temperature dependence. However, as explained by Coffey \cite{Evans1982Book} and later in Ref.~\citenum{Dejardin2018Book}, this equation is difficult to understand from a microscopic point of view. Yet, it is useful because it yields a relatively good estimate of the dipole moment $\mu_{g}$ in many cases, for example, using Malecki's method.\cite{Malecki1984JPC}

\subsection{Special cases -- Water, TBP, Glycerol}
\begin{figure}[h!]
\centering
\includegraphics[height=7cm]{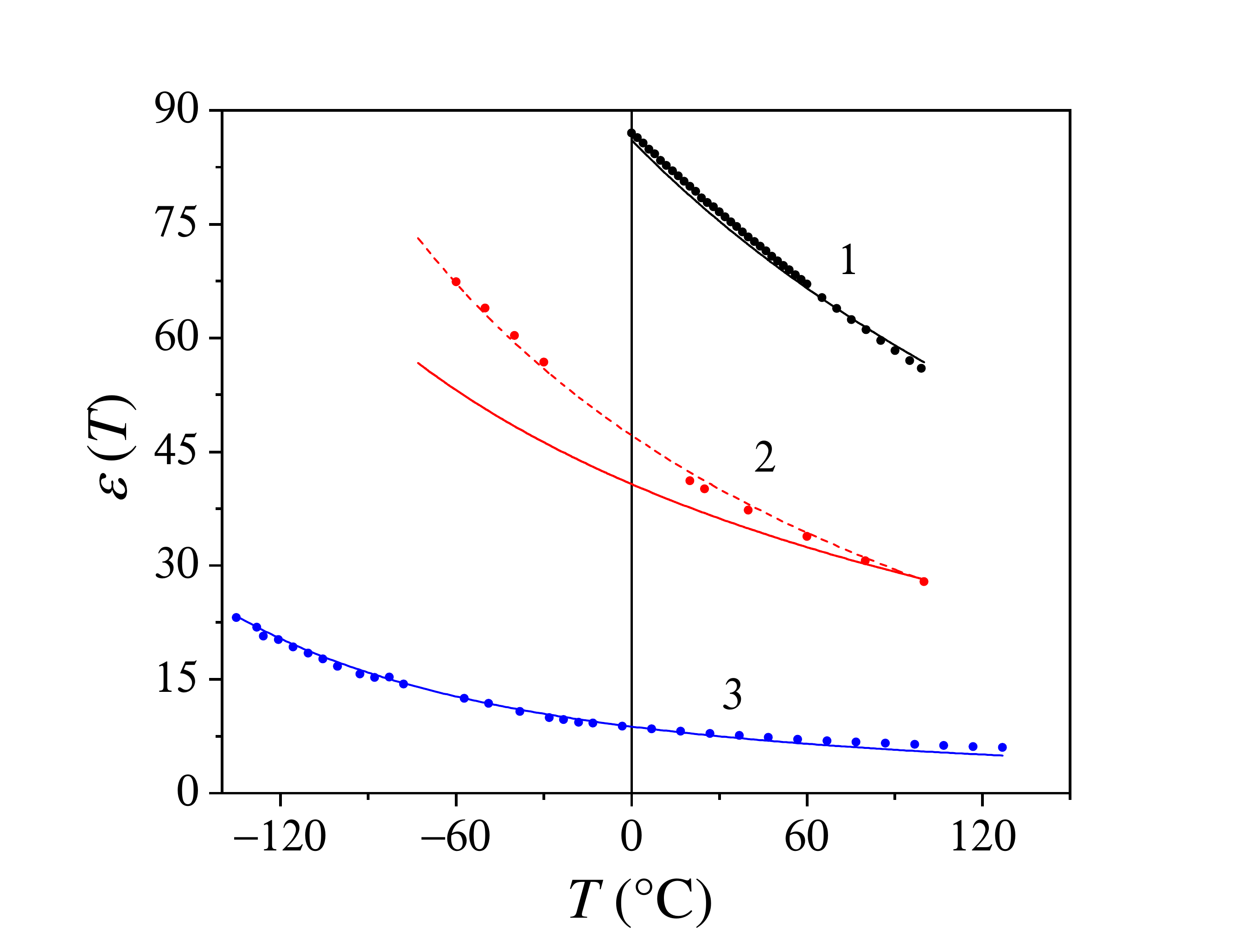}
\caption{Experimental temperature dependence of the linear static permittivity of Water (1), Glycerol (2) and TBP (3). Solid line : Theory. Dots : Experimental points. Dashed line : empirical equation of Matyushov and Richert \cite{Matyushov2016JCP} for glycerol.} 
\label{plots10}
\end{figure}

In this subsection we compare our theory to experimental values of three special liquids, namely water, glycerol and tributyl phosphate (TBP). The specialties of these substances will become clear in the following.
Figure~\ref{plots10} displays the experimental $\epsilon_s$ values as points and the theoretical ones as solid lines for these three liquids. 

\subsubsection{Water}
A comparison of experimental static permittivities of water with an earlier stage of our theory was already given in Reference~\citenum{Dejardin2018JCP}. Therein, the induction/dispersion-type interaction was not yet accounted for, the refractive index was kept temperature independent and $\varepsilon_{\infty}=1.03n^{2}$ was chosen. This leads to a disagreement with the experimental data at temperatures above $80^{\circ}$ Celsius. Here, the induction/dispersion effects together with inclusion of the temperature dependence of $n$ allows our theoretical findings to agree with experimental data across the whole temperature range. The $\kappa$ parameter was adjusted to -0.15 to achieve this agreement, indicating a slight reduction of the total effective dipole moment $(\varepsilon_{\infty}+2)\mu_{g}/3$. Moreover it indicates a specific equilibrium geometry of the water molecules in the liquid phase, which, however, is impossible to specify precisely in the present context. 

\begin{figure}[h!]
\centering
\includegraphics[height=7cm]{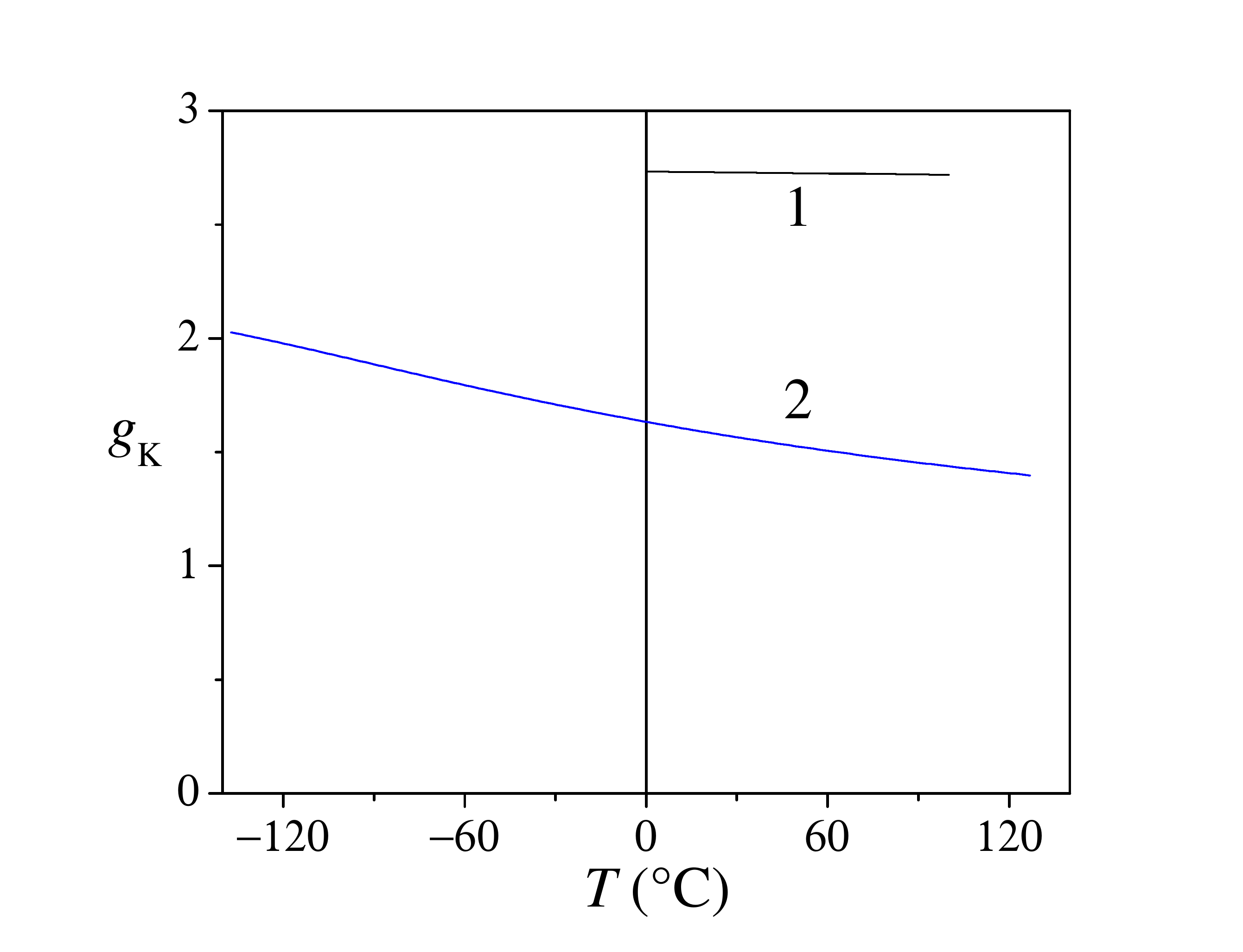}
\caption{Theoretical temperature dependence of Kirkwood correlation factor of Water (1) and TBP (2).} 
\label{plots13}
\end{figure}

The Kirkwood correlation factor of liquid water as a function of temperature is shown on Figure \ref{plots13}.
For water, it is known that the experimental Kirkwood correlation factor is $g_{\rm K}=2.75$ at $0^{\circ}$C \cite{Bottcher1973Book} and decreases to $g_{\rm K}=2.49$ at $T=83^{\circ}$C, under the conditions that $\varepsilon_{\infty}=1.05n^2$ and $\varepsilon_{\infty}$ is temperature-independent.\cite{Bottcher1973Book} In the present work, we find $g_{\rm K}=2.73$ at $T=0^{\circ}$C and $g_{\rm K}=2.72$ at $T=83^{\circ}$C, however, under the condition $\varepsilon_{\infty}(T)=n^2(T)$, with $n^2$ obeying the Lorenz-Lorentz Eq.~\eqref{LorenzLorentz}. Since we use $g_{\rm K}^{1(-)}$ as a representative of $g_{\rm K}$ for this substance, the dipolar order in liquid water is the parallel one, in agreement with Kirkwood's predictions \cite{Kirkwood1939JCP}. We also remark, that, incidentally, the $g_{\rm K}$ is basically independent of temperature, which clearly explains why Onsager's equation works at room temperature for liquid water with values of $\varepsilon_{\infty}$ as large as 4.5 \cite{Bottcher1973Book,Hill1970JPC}. This exaggerated value of $\varepsilon_{\infty}$ has led many authors, including some of us,\cite{Dejardin2018JCP,Dejardin2019JP} to treat $\varepsilon_{\infty}$ as a fitting parameter, in order to obtain values of $g_{\rm K}$ that comply with what is believed about dipolar order in water based on Kirkwood's formula Eq.\,\eqref{KirkwoodGk}. Again, we insist that this procedure is misleading, because  Eq.\,\eqref{KirkwoodGk} is most of the time a poor approximation of Eq.\,\eqref{KirkGk}. 
Finally we note, that our calculations for $g_{\rm K}$ of water are also in agreement with the molecular dynamics (SPC/E) numerical simulations of van der Spoel et al.\cite{vanderSpoel1998JCP}

\subsubsection{TBP}
Tributyl phosphate (TBP) is special in so far as it is the only substance -- out of all we tested so far -- where $g_K^{2(-)}$ has to be employed to achieve agreement between theory and experiment.
The experimental static permittivities, which are shown in Figure~\ref{plots10}, were obtained in our laboratory. Details of the experimental setup are described elsewhere.\cite{Pabst2020PRL} 
As can be seen in this Figure, the theory is able to describe the experimental data over a temperature range of more than 260\,K,
and since the glass transition temperature of TBP is about $T_{g}=-132^{\circ}$C,\cite{Saini2017JCP} we may say that unlike what was stated in reference \citenum{Dejardin2018JCP}, the theory is sometimes able to predict correct values of the static permittivity even below the calorimetric $T_{g}$.

The temperature variation of $g_{\rm K}$ for TBP is shown in Figure \ref{plots13}. Clearly, for this substance, $g_{\rm K}>1$. However, since $g_{\rm K}^{2(-)}$ is used here with $\kappa=0.85$, the permanent dipole pair relative orientations continuously spread between $0$ and $97$ degrees, as obtained from Eq.~\eqref{EqAngle}. This means that both, parallel and antiparallel alignment of dipolar pairs are present in this substance. 

A Kirkwood correlation factor of less than unity was obtained in a different study by Saini et al.\cite{Saini2017JCP} and thus needs a comment:
The value of the molecular dipole moment $\mu_g$ of tributyl phosphate (TBP) used in their study is $3.1$~D, which is the value of TBP dissolved in carbon tetrachloride. Although this solvent is non-polar, it still affects the value of $\mu_g$ as it has a non-negligible effect on the phosphoryl group.\cite{petkovic1973JPhysChem} We used the value of 2.60~D, which is obtained in an octane solution and is almost identical to the value obtained in a decalin solution,\cite{petkovic1973JPhysChem} both unpolar solvents without influence on the TBP molecules.

Moreover, in the work of Saini et al., $\varepsilon_{\infty}\approx 5$ was used, which is far off from $\varepsilon_{\infty} = n^2$. This value was read off the spectrum at frequencies lower than the strong secondary relaxation, which is clearly due to molecular reorientation. Thus, this choice is not justified in our opinion and leads together with the too high dipole moment to a $g_{\rm K}$ value less than unity. 

The value $3.32$~D of undiluted TBP quoted by Petkovic et al.\cite{petkovic1973JPhysChem} is the one compatible with Onsager's equation at room temperature. If we use the Onsager dipole $\mu_{g}\sqrt{g_{\rm K}}$ with our calculated $g_{\rm K}$, we find $3.27$~D at room temperature, which is rather close to Petkovic's result.

\subsubsection{Glycerol}
As can be seen in Figure~\ref{plots10}, the experimental data points of glycerol cannot be described by our theory at all. Here, we show the calculated values for $g_K^{1(-)}$, however, also no other representation of $g_K$ is able to reproduce the experimental values with physically reasonable values of the parameters.

Often, the specificity of H-bonding is invoked in order to explain disagreement between theory and experiment. This is not so here, since  H-bonding specific mechanisms are not needed at all in order to obtain agreement between theory and experiment for linear primary alcohols and water, both prominent examples of H-bonding liquids. Rather, we believe that the disagreement is explained by the oversimplification of the interaction potential Eq.~\eqref{UmNeedles} which, in effect, pertains to molecules having their permanent dipole moment fixed with respect to a given axis of symmetry of the molecule. Thus, due to the floppyness of the glycerol molecules, and due to the fact that comparable contributions to the overall dipole moment are located in different positions in the molecule, the situation for glycerol is quite different. Owing to this reason, we believe that the interaction energy landscape is much too simple to capture the main physics which is necessary for the theoretical description of the temperature dependence of the dielectric constant of this polar fluid. We note, that seemingly good agreement between theory and experiment with the potential \eqref{UmNeedles} can be obtained across the whole temperature range using the unphysical assumption $\varepsilon(T)=0.5n^{2}(T)$ together with $g_{\rm K}=g_{\rm K}^{1(-)}$ and $\kappa=0.45$. The relation $\varepsilon(T)=0.5n^{2}(T)$ used in such a fit actually reveals that the reason of our failure indeed lies in the oversimplification of the intermolecular interaction potential Eq.\,\eqref{UmNeedles} and the resulting Kirkwood potential of mean torques $V_{2}^\text{eff}$ rather than in the specific H-bonding mechanism, which is not accounted for. Therefore, we state that glycerol is a non-simple polar fluid (and even less a simple liquid), where the intermolecular interaction is not appropriately represented in our theory and thus, the substance is out of scope of the present work.

\section{Summary of results and perspectives}

In this work, we have derived an integral formula for the Kirkwood correlation factor of polar fluids, Eq.\,\eqref{KirkwoodGkGeneral} from the equilibrium averaged Dean equation Eq.\,\eqref{eqDK}. This equation has a lot of advantages over Eq.\,\eqref{KirkGk}, the first one being that it is independent on the number of neighbors (and where no molecule is tagged to achieve the calculation), the second one being the fact that it easily lends itself to tractable approximations, and the third one being that our theory is quantitatively amenable to comparison with experiment. For example, in the Kirkwood superposition approximation, one immediately obtains a tractable expression for the correlation factor, namely Eq.\,\eqref{UsableGk}. Then we suggest how to construct a model potential for the electrostatic interaction that not only includes the permanent electric dipoles but also in the next order some induction/dispersion-like effects. Finally, for each case of preferred parallel or antiparallel alignment of permanent dipoles and their modification by induced polarization,  two different Kirkwood potentials of mean torques are deduced, for which Eq.\,\eqref{UsableGk} is solved to yield respective temperature dependent values of the static dielectric constant and the Kirkwood correlation factor. The models only contain physical quantities, like density, permanent dipole moment and refractive index or molecular polarizability, respectively, that are independently accessible by experiment. Only one single material specific and temperature independent parameter enters the calculation, which is connected to the molecular polarizability, that cannot be calculated from the latter in a straightforward manner and thus needs to be a fitting parameter. In that way, we are able to quantitatively compare the calculated values of $\epsilon(T)$ with experimental data, and it turns out that the derived model potentials seem to capture the underlying main physics of different system classes to a rather good accuracy, at least from the point of view of a static dielectric constant measurement.

A first important result from these calculations is the observation that a parallel alignment of the dipole pairs does not necessarily imply $g_{\rm K}>1$, and similarly an anti-parallel alignment of dipole pairs does not strictly imply $g_{\rm K}<1$ either. Rather, such alignment states are local minima of the effective pair interaction orientational potential of mean torques, for which not only permanent but also induced dipole moments play a decisive role. For example, applying Eq.\,\eqref{EqAngle} to TBP, we find that pairs of dipoles in this polar substance have a trend to make angles spreading between $0$ and $97^{\circ}$, explaining quantitatively the value of $g_{\rm K}\approx 2$ of TBP near its glass transition temperature and beyond.\cite{Pabst2020PRL}

We also discuss several examples of preferred antiparallel alignment, not only for acetonitrile, where $g_{\mathrm{K}}<1$ is found as expected, but also 
for acetone, nitrobenzene and dimethyl sulfoxide, where despite the antiparallel alignment clearly $g_{\mathrm{K}}>1$ due to the non-negligible influence of the molecular polarizability. This again underlines that the usual arguments relating parallel ($g_{\mathrm{K}}>1$) and antiparallel ($g_{\mathrm{K}}<1$) alignments based on Eq.\eqref{KirkwoodGk} is an oversimplification which hampers comparison of the results found from linear dielectric measurements concerning dipolar alignment with those obtained from other characterization techniques.

As examples for a preferred parallel alignment of dipoles we have investigated a series of linear monohydroxy alcohols, where our theory reproduces the experimental $\epsilon(T)$ for the full series from methanol to octanol with the importance of the polarizability component increasing with molecular volume, as expected. But also the static permittivity of liquid water from the melting temperature to the boiling point shows excellent agreement with the theory. This is quite remarkable, as the theory does not explicitly contain any particular H-bonding related mechanism. Thus, the idealization of a molecule which consists of its permanent and induced dipole moments only is enough to explain the temperature dependence of the static dielectric constant of these hydrogen bonding liquids 
as first Debye, Kirkwood and Fr\"{o}hlich assumed. \cite{Debye1929Book,Bottcher1973Book,Frohlich1958Book} Interestingly, the situation is different for the polyalcohol glycerol. Here, apparently our model for the pair potential is too simple to capture the actual electrostatic interaction. The reason for this is unlikely the specific role of hydrogen bonds, because our theory compares favorably with experimental data concerning water and monoalcohols. More likely, it may be suspected that since the dipole moment of glycerol is composed of the moments located in three different OH groups within the molecule, considerable intramolecular flexibility leads to a rather ill defined molecular dipole moment, resulting in turn in more complicated interactions. Work to develop appropriate interaction potentials for such associated molecular liquids is in progress.

In spite of the fact that our theory covers a large spectrum of values for $g_{\rm K}$, it still does not explain the experimental temperature variation of the dielectric constant of some carboxylic acids and also a couple of monohydroxy alcohols, where a minimum in the temperature dependence of the dielectric constant is observed. For example, the dielectric constant of acetic and caprylic acid\cite{Bottcher1973Book} or of certain octanol isomers\cite{Johari1968,dannhauser1968dielectric} first decreases with temperature, but then \textit{increases} again.  In fact, such unusual behavior of $\varepsilon(T)$ is usually explained by the simultaneous presence of hydrogen-bonded closed-ring structures, for which the net dipole moment is approximately zero,  together with linear multimer chains with various concentrations. \cite{Bottcher1973Book} A minimal modelling of such behavior may require to consider two different species in the sample and correspondingly different $\lambda$ factors appropriately weighted by the temperature dependent molar fractions of closed rings and linear chains, respectively. Testing of such ideas is currently in progress.

Certainly more demanding will it be to adapt the present theory to binary \textit{polar} mixtures. Here, a zero-order approximation for evaluating the static dielectric constant might be to consider the coupled Langevin equations for the overdamped nonlinear itinerant oscillator model \cite{Coffey2017Book} with a specific pair interaction potential and the corresponding equilibrium Smoluchowski equation \cite{Risken1989Book} to deduce equilibrium properties. Moreover, one may also try to extend the present model to dynamics, similar to previous work,\cite{Dejardin2019PRB} both in linear and nonlinear responses. Finally, one could also think of applying the present calculations to suspensions of magnetic nanoparticles similar to what was already pointed out previously.\cite{Dejardin2019PRB,Dejardin2011JAP} The development of the theory in all of these directions is currently in progress. 

\section*{Author Contributions}
\begin{itemize}
\item Conceptualization : P.M.D.
\item Data curation : P.M.D, F.P., A.H. and T.B. 
\item Funding acquisition : P.M.D. and T.B.
\item Methodology : All authors
\item Project administration : P.M.D. and T.B.
\item Resources : F.P., A.H. and T.B.
\item Software : P.M.D. Y.C. and C.C.
\item Supervision : P.M.D.
\item Validation : All authors
\item Visualization : P.M.D. and F.P.
\item Writing – original draft : All authors
\item Writing - review and editing : P.M.D.
\end{itemize}

\section*{Conflicts of interest}
There are no conflicts to declare.

\section*{Acknowledgements}
We thank Profs. Yu. P. Kalmykov, M.Sofonea (UPVD) and F. Affouard (University of Lille 1) for helpful conversations. P.M.D., Y.C., C.C., and R.B. gratefully acknowledge UPVD for financial support of LAMPS (EA 4217). T.\,B., F.\,P. and A.\,H. gratefully acknowledge financial support by the Deutsche Forschungsgemeinschaft under grant No. BL 1192/3.

\onecolumn

\section{Appendix A : Theoretical background and derivation of Eq. \eqref{eqDK}}
\subsection{Basic kinetic equations}
The starting point of our theory is the averaged rototranslational Dean equation \cite{Dean1996JPA,Cugliandolo2015PRE} for the one-body density $W(\mathbf{r},\mathbf{u},t)$ of having a polar molecule at $\mathbf{r}$ with orientation $\mathbf{u}$ at time $t$, viz.
\begin{eqnarray}
\nonumber
\frac{\partial W}{\partial t}(\mathbf{r},\mathbf{u},t)&=&D_{T}L_{\mathbf{r}}W(\mathbf{r},\mathbf{u},t)+D_{R}L_{\mathbf{u}}W(\mathbf{r},\mathbf{u},t)
+\beta D_{T}\nabla_{\mathbf{r}}\cdot\int\nabla_{\mathbf{r}}U_{int}(\mathbf{r},\mathbf{r}',\mathbf{u},\mathbf{u}')W_{2}(\mathbf{r},\mathbf{r}',\mathbf{u},\mathbf{u}',t)d\mathbf{r}'d\mathbf{u}'\\
&+&\beta D_{R}\nabla_{\mathbf{u}}\cdot\int\nabla_{\mathbf{u}}U_{int}(\mathbf{r},\mathbf{r}',\mathbf{u},\mathbf{u}')W_{2}(\mathbf{r},\mathbf{r}',\mathbf{u},\mathbf{u}',t)d\mathbf{r}'d\mathbf{u}'
\label{DeanEq1}
\end{eqnarray}
where $D_T$ and $D_R$ are the bare translation and rotational diffusion coefficients respectively, $U_{int}(\mathbf{r},\mathbf{r}',\mathbf{u},\mathbf{u}')$ is a generalized pair interaction potential, $W_{2}(\mathbf{r},\mathbf{r}',\mathbf{u},\mathbf{u}',t)$ is the rototranslational pair density, and
\begin{eqnarray}
L_{\mathbf{r}}W(\mathbf{r},\mathbf{u},t)&=&\nabla_{\mathbf{r}}\cdot(\nabla_{\mathbf{r}}W(\mathbf{r},\mathbf{u},t)+\beta W(\mathbf{r},\mathbf{u},t)\nabla_{\mathbf{r}}V_{1}(\mathbf{r},\mathbf{u},t))
\label{FPTrans}
\end{eqnarray}
\begin{eqnarray}
L_{\mathbf{u}}W(\mathbf{r},\mathbf{u},t)&=&\nabla_{\mathbf{u}}\cdot(\nabla_{\mathbf{u}}W(\mathbf{r},\mathbf{u},t)+\beta W(\mathbf{r},\mathbf{u},t)\nabla_{\mathbf{u}}V_{1}(\mathbf{r},\mathbf{u},t))
\label{FPRot}
\end{eqnarray}
define the actions of the one-particle Fokker-Planck operators $L_{\mathbf{r}}$ and $L_{\mathbf{u}}$ on  $W(\mathbf{r},\mathbf{u},t)$. In Eqs. \eqref{FPTrans} and \eqref{FPRot}, $V_{1}(\mathbf{r},\mathbf{u},t)$ is a single-particle generalized potential arising from external forces and torques. 
In order to solve Eq.\eqref{DeanEq1}, an equation governing the dynamics of $W_2$ is necessary. Using the techniques developed by Cugliandolo et al.,\cite{Cugliandolo2015PRE} we have after lengthy but trivial algebra
\begin{eqnarray}
\nonumber
\frac{\partial W_{2}}{\partial t}(\mathbf{r},\mathbf{r}',\mathbf{u},\mathbf{u}',t)&=&D_{T}(L_{\mathbf{r}}^{(2)}+L_{\mathbf{r}'}^{(2)}) W_{2}(\mathbf{r},\mathbf{r}',\mathbf{u},\mathbf{u}',t)+D_{R}(L_{\mathbf{u}}^{(2)}+L_{\mathbf{u}'}^{(2)}) W_{2}(\mathbf{r},\mathbf{r}',\mathbf{u},\mathbf{u}',t)\\
\nonumber
&+&\beta D_{T}\nabla_{\mathbf{r}}\cdot\int\nabla_{\mathbf{r}}U_{int}(\mathbf{r},\mathbf{r}'',\mathbf{u},\mathbf{u}'')W_{3}(\mathbf{r},\mathbf{r}',\mathbf{r}'',\mathbf{u},\mathbf{u}',\mathbf{u}'',t)d\mathbf{r}''d\mathbf{u}''\\
\nonumber
&+&\beta D_{T}\nabla_{\mathbf{r}'}\cdot\int\nabla_{\mathbf{r}'}U_{int}(\mathbf{r}',\mathbf{r}'',\mathbf{u}',\mathbf{u}'')W_{3}(\mathbf{r},\mathbf{r}',\mathbf{r}'',\mathbf{u},\mathbf{u}',\mathbf{u}'',t)d\mathbf{r}''d\mathbf{u}''\\
\nonumber
&+&\beta D_{R}\nabla_{\mathbf{u}}\cdot\int\nabla_{\mathbf{u}}U_{int}(\mathbf{r},\mathbf{r}'',\mathbf{u},\mathbf{u}'')W_{3}(\mathbf{r},\mathbf{r}',\mathbf{r}'',\mathbf{u},\mathbf{u}',\mathbf{u}'',t)d\mathbf{r}''d\mathbf{u}''\\
&+&\beta D_{R}\nabla_{\mathbf{u}'}\cdot\int\nabla_{\mathbf{u}'}U_{int}(\mathbf{r}',\mathbf{r}'',\mathbf{u}',\mathbf{u}'')W_{3}(\mathbf{r},\mathbf{r}',\mathbf{r}'',\mathbf{u},\mathbf{u}',\mathbf{u}'',t)d\mathbf{r}''d\mathbf{u}''
\label{DeanEq2}
\end{eqnarray}
where $W_{3}$ is the full rototranslational three-body density, the two-particle Fokker-Planck operators in Eq.\eqref{DeanEq2} are defined by their action on $W_2$, namely
\begin{eqnarray}
L_{\mathbf{r}}^{(2)}W_{2}(\mathbf{r},\mathbf{u},\mathbf{r}',\mathbf{u}',t)&=&\nabla_{\mathbf{r}}\cdot(\nabla_{\mathbf{r}}W_{2}(\mathbf{r},\mathbf{u},\mathbf{r}',\mathbf{u}',t)+\beta W_{2}(\mathbf{r},\mathbf{u},\mathbf{r}',\mathbf{u}',t)\nabla_{\mathbf{r}}V_{2}(\mathbf{r},\mathbf{u},\mathbf{r}',\mathbf{u}',t)),
\label{L2R}
\end{eqnarray}
\begin{eqnarray}
L_{\mathbf{r}'}^{(2)}W_{2}(\mathbf{r},\mathbf{u},\mathbf{r}',\mathbf{u}',t)&=&\nabla_{\mathbf{r}'}\cdot(\nabla_{\mathbf{r}'}W_{2}(\mathbf{r},\mathbf{u},\mathbf{r}',\mathbf{u}',t)+\beta W_{2}(\mathbf{r},\mathbf{u},\mathbf{r}',\mathbf{u}',t)\nabla_{\mathbf{r}'}V_{2}(\mathbf{r},\mathbf{u},\mathbf{r}',\mathbf{u}',t)),
\label{L2Rp}
\end{eqnarray}
\begin{eqnarray}
L_{\mathbf{u}}^{(2)}W_{2}(\mathbf{r},\mathbf{u},\mathbf{r}',\mathbf{u}',t)&=&\nabla_{\mathbf{u}}\cdot(\nabla_{\mathbf{u}}W_{2}(\mathbf{r},\mathbf{u},\mathbf{r}',\mathbf{u}',t)+\beta W_{2}(\mathbf{r},\mathbf{u},\mathbf{r}',\mathbf{u}',t)\nabla_{\mathbf{u}}V_{2}(\mathbf{r},\mathbf{u},\mathbf{r}',\mathbf{u}',t)),
\label{L2U}
\end{eqnarray}
\begin{eqnarray}
L_{\mathbf{u}'}^{(2)}W_{2}(\mathbf{r},\mathbf{u},\mathbf{r}',\mathbf{u}',t)&=&\nabla_{\mathbf{u}'}\cdot(\nabla_{\mathbf{u}'}W_{2}(\mathbf{r},\mathbf{u},\mathbf{r}',\mathbf{u}',t)+\beta W_{2}(\mathbf{r},\mathbf{u},\mathbf{r}',\mathbf{u}',t)\nabla_{\mathbf{u}'}V_{2}(\mathbf{r},\mathbf{u},\mathbf{r}',\mathbf{u}',t)).
\label{L2Up}
\end{eqnarray}
and the pair potential $V_2$ in Eqs.\eqref{L2R}-\eqref{L2Up} is given by
\begin{eqnarray}
V_{2}(\mathbf{r},\mathbf{u},\mathbf{r}',\mathbf{u}',t)=U_{int}(\mathbf{r},\mathbf{u},\mathbf{r}',\mathbf{u}')+V_{1}(\mathbf{r},\mathbf{u},t)+V_{1}(\mathbf{r}',\mathbf{u}',t)
\end{eqnarray}
\subsection{Formal transformation of Eqs.\eqref{DeanEq1} and \eqref{DeanEq2} into Fokker-Planck form}
This tranformation will be convenient later as Fokker-Planck equations can always be written as continuity equations in phase space. \cite{Risken1989Book} Therefore, we proceed by writing \cite{Hansen2006Book}
\begin{eqnarray}
W_{2}(\mathbf{r},\mathbf{u},\mathbf{r}',\mathbf{u}',t)=W(\mathbf{r},\mathbf{u},t)W(\mathbf{r}',\mathbf{u}',t)G_{2}(\mathbf{r},\mathbf{u},\mathbf{r}',\mathbf{u}',t)
\end{eqnarray}
and a similar equation for $W_3$, namely
\begin{eqnarray}
W_{3}(\mathbf{r},\mathbf{u},\mathbf{r}',\mathbf{u}',\mathbf{r}'',\mathbf{u}'',t)=W(\mathbf{r},\mathbf{u},t)W(\mathbf{r}',\mathbf{u}',t)W(\mathbf{r}'',\mathbf{u}'',t)
G_{3}(\mathbf{r},\mathbf{u},\mathbf{r}',\mathbf{u}',\mathbf{r}'',\mathbf{u}'',t)
\end{eqnarray}
where $G_{2}$ and $G_{3}$ are the pair and triplet distribution functions. \cite{Hansen2006Book} 
We further introduce the effective one and two-body effective potentials $\phi_1$ and $\phi_2 $ via the partial differential equations
\begin{eqnarray}
\nabla_{\mathbf{r}}\phi_{1}(\mathbf{r},\mathbf{u},t)=\int\nabla_{\mathbf{r}}U_{int}(\mathbf{r},\mathbf{r}',\mathbf{u},\mathbf{u}')W(\mathbf{r}',\mathbf{u}',t)G_{2}(\mathbf{r},\mathbf{u},\mathbf{r}',\mathbf{u}',t)d\mathbf{r}'d\mathbf{u}',
\label{1BPotDist}
\end{eqnarray}
\begin{eqnarray}
\nabla_{\mathbf{u}}\phi_{1}(\mathbf{r},\mathbf{u},t)=\int\nabla_{\mathbf{u}}U_{int}(\mathbf{r},\mathbf{r}',\mathbf{u},\mathbf{u}')W(\mathbf{r}',\mathbf{u}',t)G_{2}(\mathbf{r},\mathbf{u},\mathbf{r}',\mathbf{u}',t)d\mathbf{r}'d\mathbf{u}',
\label{1BPotOr}
\end{eqnarray}
\begin{eqnarray}
\nabla_{\mathbf{r}}\phi_{2}(\mathbf{r},\mathbf{u},\mathbf{r}',\mathbf{u}',t)=\nabla_{\mathbf{r}}U_{int}(\mathbf{r},\mathbf{r}',\mathbf{u},\mathbf{u}')+\int\nabla_{\mathbf{r}}U_{int}(\mathbf{r},\mathbf{r}'',\mathbf{u},\mathbf{u}')W(\mathbf{r}'',\mathbf{u}'',t)\frac{G_{3}(\mathbf{r},\mathbf{u},\mathbf{r}',\mathbf{u}',\mathbf{r}'',\mathbf{u}'',t)}{G_{2}(\mathbf{r},\mathbf{u},\mathbf{r}',\mathbf{u}',t)}d\mathbf{r}''d\mathbf{u}'',
\label{2BPotDist1}
\end{eqnarray}
\begin{eqnarray}
\nabla_{\mathbf{u}}\phi_{2}(\mathbf{r},\mathbf{u},\mathbf{r}',\mathbf{u}',t)=\nabla_{\mathbf{u}}U_{int}(\mathbf{r},\mathbf{r}',\mathbf{u},\mathbf{u}')+\int\nabla_{\mathbf{u}}U_{int}(\mathbf{r},\mathbf{r}'',\mathbf{u},\mathbf{u}'')W(\mathbf{r}'',\mathbf{u}'',t)\frac{G_{3}(\mathbf{r},\mathbf{u},\mathbf{r}',\mathbf{u}',\mathbf{r}'',\mathbf{u}'',t)}{G_{2}(\mathbf{r},\mathbf{u},\mathbf{r}',\mathbf{u}',t)}d\mathbf{r}''d\mathbf{u}'',
\label{2BPotRot1}
\end{eqnarray}
\begin{eqnarray}
\nabla_{\mathbf{r}'}\phi_{2}(\mathbf{r},\mathbf{u},\mathbf{r}',\mathbf{u}',t)=\nabla_{\mathbf{r}'}U_{int}(\mathbf{r},\mathbf{r}',\mathbf{u},\mathbf{u}')+\int\nabla_{\mathbf{r}'}U_{int}(\mathbf{r}',\mathbf{r}'',\mathbf{u}',\mathbf{u}'')W(\mathbf{r}'',\mathbf{u}'',t)\frac{G_{3}(\mathbf{r},\mathbf{u},\mathbf{r}',\mathbf{u}',\mathbf{r}'',\mathbf{u}'',t)}{G_{2}(\mathbf{r},\mathbf{u},\mathbf{r}',\mathbf{u}',t)}d\mathbf{r}''d\mathbf{u}'',
\label{2BPotDist2}
\end{eqnarray}
\begin{eqnarray}
\nabla_{\mathbf{u}'}\phi_{2}(\mathbf{r},\mathbf{u},\mathbf{r}',\mathbf{u}',t)=\nabla_{\mathbf{u}'}U_{int}(\mathbf{r},\mathbf{r}',\mathbf{u},\mathbf{u}')+\int\nabla_{\mathbf{u}'}U_{int}(\mathbf{r}',\mathbf{r}'',\mathbf{u}',\mathbf{u}'')W(\mathbf{r}'',\mathbf{u}'',t)\frac{G_{3}(\mathbf{r},\mathbf{u},\mathbf{r}',\mathbf{u}',\mathbf{r}'',\mathbf{u}'',t)}{G_{2}(\mathbf{r},\mathbf{u},\mathbf{r}',\mathbf{u}',t)}d\mathbf{r}''d\mathbf{u}''.
\label{2BPotRot2}
\end{eqnarray}
This allows us to formally rewrite Eqs.\eqref{DeanEq1} and \eqref{DeanEq2} using Eqs.\eqref{1BPotDist}-\eqref{2BPotRot2} in Fokker-Planck form, viz.
\begin{eqnarray}
\frac{\partial W}{\partial t}(\mathbf{r},\mathbf{u},t)=D_{T}\nabla_{\mathbf{r}}\cdot(\nabla_{\mathbf{r}}W(\mathbf{r},\mathbf{u},t)+\beta W(\mathbf{r},\mathbf{u},t)\nabla_{\mathbf{r}}\Psi_{1}(\mathbf{r},\mathbf{u},t))+D_{R}\nabla_{\mathbf{u}}\cdot(\nabla_{\mathbf{u}}W(\mathbf{r},\mathbf{u},t)+\beta W(\mathbf{r},\mathbf{u},t)\nabla_{\mathbf{u}}\Psi_{1}(\mathbf{r},\mathbf{u},t))
\label{FPEDean1}
\end{eqnarray}
and
\begin{eqnarray}
\nonumber
\frac{\partial W_{2}}{\partial t}(\mathbf{r},\mathbf{u},\mathbf{r}',\mathbf{u}',t)=D_{T}\nabla_{\mathbf{r}}\cdot(\nabla_{\mathbf{r}} W_{2}(\mathbf{r},\mathbf{u},\mathbf{r}',\mathbf{u}',t)+\beta W_{2}(\mathbf{r},\mathbf{u},\mathbf{r}',\mathbf{u}',t)\nabla_{\mathbf{r}}\Psi_{2}(\mathbf{r},\mathbf{u},\mathbf{r}',\mathbf{u}',t))\\
\nonumber
+D_{T}\nabla_{\mathbf{r}'}\cdot(\nabla_{\mathbf{r}'} W_{2}(\mathbf{r},\mathbf{u},\mathbf{r}',\mathbf{u}',t)+\beta W_{2}(\mathbf{r},\mathbf{u},\mathbf{r}',\mathbf{u}',t)\nabla_{\mathbf{r}'}\Psi_{2}(\mathbf{r},\mathbf{u},\mathbf{r}',\mathbf{u}',t))\\
\nonumber
+D_{R}\nabla_{\mathbf{u}}\cdot(\nabla_{\mathbf{r}} W_{2}(\mathbf{r},\mathbf{u},\mathbf{r}',\mathbf{u}',t)+\beta W_{2}(\mathbf{r},\mathbf{u},\mathbf{r}',\mathbf{u}',t)\nabla_{\mathbf{u}}\Psi_{2}(\mathbf{r},\mathbf{u},\mathbf{r}',\mathbf{u}',t))\\
+D_{R}\nabla_{\mathbf{u}'}\cdot(\nabla_{\mathbf{r}} W_{2}(\mathbf{r},\mathbf{u},\mathbf{r}',\mathbf{u}',t)+\beta W_{2}(\mathbf{r},\mathbf{u},\mathbf{r}',\mathbf{u}',t)\nabla_{\mathbf{u}'}\Psi_{2}(\mathbf{r},\mathbf{u},\mathbf{r}',\mathbf{u}',t))
\label{FPEDean2}
\end{eqnarray}
where
\begin{eqnarray}
\Psi_{1}(\mathbf{r},\mathbf{u},t)=V_{1}(\mathbf{r},\mathbf{u},t)+\phi_{1}(\mathbf{r},\mathbf{u},t),
\label{Psi1}
\end{eqnarray}
and
\begin{eqnarray}
\Psi_{2}(\mathbf{r},\mathbf{u},\mathbf{r}',\mathbf{u}',t)=\phi_{2}(\mathbf{r},\mathbf{u},\mathbf{r}',\mathbf{u}',t)+V_{1}(\mathbf{r},\mathbf{u},t)+V_{1}(\mathbf{r}',\mathbf{u}',t).
\label{Psi2}
\end{eqnarray}
\subsection{Transformation of variables in Eqs.\eqref{DeanEq1} and \eqref{DeanEq2} : an important and exact simplification}
The Fokker-Planck like forms Eqs.\eqref{FPEDean1} and \eqref{FPEDean2} will be useful later. 
Now, a pair interaction potential should not be a function of the center of mass coordinates of a pair, but only a function of the relative position coordinates of a pair. This is why we introduce the center of mass and relative position coordinates of a pair of identical molecules via the usual equations :
\begin{eqnarray}
\mathbf{R}=\frac{1}{2}(\mathbf{r}+\mathbf{r}'),\quad\vec{\rho}=\mathbf{r}-\mathbf{r}',
\end{eqnarray}
so that we have
\begin{eqnarray}
\mathbf{r}=\mathbf{R}+\frac{\vec{\rho}}{2},\quad\mathbf{r}'=\mathbf{R}-\frac{\vec{\rho}}{2},
\end{eqnarray}
and therefore we have
\begin{eqnarray}
\nabla_{\mathbf{r}}=\nabla_{\mathbf{R}}+\frac{1}{2}\nabla_{\vec{\rho}},\quad\nabla_{\mathbf{r}'}=\nabla_{\mathbf{R}}-\frac{1}{2}\nabla_{\vec{\rho}},\quad 2\nabla_{\mathbf{R}}&=&\nabla_{\mathbf{r}}+\nabla_{\mathbf{r}'},\quad\nabla_{\vec{\rho}}=\nabla_{\mathbf{r}}-\nabla_{\mathbf{r}'}
\label{DiffRef}
\end{eqnarray}
Next we explicitly state that $U_{int}$ is a function of the relative position coordinates of a pair, but not of its center of mass coordinate, namely we have
\begin{eqnarray}
U_{int}(\mathbf{r},\mathbf{u},\mathbf{r}',\mathbf{u}')=U_{int}(\vec{\rho},\mathbf{u},\mathbf{u}')=U_{int}(-\vec{\rho},\mathbf{u},\mathbf{u}')=U_{int}(\vec{\rho},\mathbf{u}',\mathbf{u})
\label{InteractionProperty}
\end{eqnarray}
It follows immediately that Eqs.\eqref{2BPotDist1} and \eqref{2BPotDist2} can be combined to obtain the equations
\begin{eqnarray}
\nonumber
\nabla_{\mathbf{R}}\phi_{2}(\mathbf{R},\vec{\rho},\mathbf{u},\mathbf{u}',t)=\frac{1}{4}\int\nabla_{\vec{\rho}}U_{int}(\mathbf{R}+\frac{\vec{\rho}}{2}-\mathbf{r}'',\mathbf{u},\mathbf{u}'')W(\mathbf{r}'',\mathbf{u}'',t)\frac{G_{3}(\mathbf{R}+\frac{\vec{\rho}}{2},\mathbf{R}-\frac{\vec{\rho}}{2},\mathbf{r}'',\mathbf{u},\mathbf{u}',\mathbf{u}'',t)}{G_{2}(\mathbf{R}+\frac{\vec{\rho}}{2},\mathbf{R}-\frac{\vec{\rho}}{2},\mathbf{u},\mathbf{u}',t)}d\mathbf{r}''d\mathbf{u}''\\
\nonumber
-\frac{1}{4}\int\nabla_{\vec{\rho}}U_{int}(\mathbf{R}-\frac{\vec{\rho}}{2}-\mathbf{r}'',\mathbf{u}',\mathbf{u}'')W(\mathbf{r}'',\mathbf{u}'',t)\frac{G_{3}(\mathbf{R}+\frac{\vec{\rho}}{2},\mathbf{R}-\frac{\vec{\rho}}{2},\mathbf{r}'',\mathbf{u},\mathbf{u}',\mathbf{u}'',t)}{G_{2}(\mathbf{R}+\frac{\vec{\rho}}{2},\mathbf{R}-\frac{\vec{\rho}}{2},\mathbf{u},\mathbf{u}',t)}d\mathbf{r}''d\mathbf{u}''\\
\nonumber
+\frac{1}{2}\int\nabla_{\mathbf{R}}U_{int}(\mathbf{R}+\frac{\vec{\rho}}{2}-\mathbf{r}'',\mathbf{u},\mathbf{u}'')W(\mathbf{r}'',\mathbf{u}'',t)\frac{G_{3}(\mathbf{R}+\frac{\vec{\rho}}{2},\mathbf{R}-\frac{\vec{\rho}}{2},\mathbf{r}'',\mathbf{u},\mathbf{u}',\mathbf{u}'',t)}{G_{2}(\mathbf{R}+\frac{\vec{\rho}}{2},\mathbf{R}-\frac{\vec{\rho}}{2},\mathbf{u},\mathbf{u}',t)}d\mathbf{r}''d\mathbf{u}''\\
+\frac{1}{2}\int\nabla_{\mathbf{R}}U_{int}(\mathbf{R}-\frac{\vec{\rho}}{2}-\mathbf{r}'',\mathbf{u}',\mathbf{u}'')W(\mathbf{r}'',\mathbf{u}'',t)\frac{G_{3}(\mathbf{R}+\frac{\vec{\rho}}{2},\mathbf{R}-\frac{\vec{\rho}}{2},\mathbf{r}'',\mathbf{u},\mathbf{u}',\mathbf{u}'',t)}{G_{2}(\mathbf{R}+\frac{\vec{\rho}}{2},\mathbf{R}-\frac{\vec{\rho}}{2},\mathbf{u},\mathbf{u}',t)}d\mathbf{r}''d\mathbf{u}''
\label{DelRPhi2}
\end{eqnarray}
and
\begin{eqnarray}
\nonumber
\nabla_{\vec{\rho}}\phi_{2}(\mathbf{R},\vec{\rho},\mathbf{u},\mathbf{u}',t)&=&\nabla_{\vec{\rho}}U_{int}(\vec{\rho},\mathbf{u},\mathbf{u}')\\
\nonumber
&+&\frac{1}{2}\int\nabla_{\vec{\rho}}U_{int}(\mathbf{R}+\frac{\vec{\rho}}{2}-\mathbf{r}'',\mathbf{u},\mathbf{u}'')W(\mathbf{r}'',\mathbf{u}'',t)\frac{G_{3}(\mathbf{R}+\frac{\vec{\rho}}{2},\mathbf{R}-\frac{\vec{\rho}}{2},\mathbf{r}'',\mathbf{u},\mathbf{u}',\mathbf{u}'',t)}{G_{2}(\mathbf{R}+\frac{\vec{\rho}}{2},\mathbf{R}-\frac{\vec{\rho}}{2},\mathbf{u},\mathbf{u}',t)}d\mathbf{r}''d\mathbf{u}''\\
\nonumber
&+&\frac{1}{2}\int\nabla_{\vec{\rho}}U_{int}(\mathbf{R}-\frac{\vec{\rho}}{2}-\mathbf{r}'',\mathbf{u}',\mathbf{u}'')W(\mathbf{r}'',\mathbf{u}'',t)\frac{G_{3}(\mathbf{R}+\frac{\vec{\rho}}{2},\mathbf{R}-\frac{\vec{\rho}}{2},\mathbf{r}'',\mathbf{u},\mathbf{u}',\mathbf{u}'',t)}{G_{2}(\mathbf{R}+\frac{\vec{\rho}}{2},\mathbf{R}-\frac{\vec{\rho}}{2},\mathbf{u},\mathbf{u}',t)}d\mathbf{r}''d\mathbf{u}''\\
\nonumber
&+&\int\nabla_{\mathbf{R}}U_{int}(\mathbf{R}+\frac{\vec{\rho}}{2}-\mathbf{r}'',\mathbf{u},\mathbf{u}'')W(\mathbf{r}'',\mathbf{u}'',t)\frac{G_{3}(\mathbf{R}+\frac{\vec{\rho}}{2},\mathbf{R}-\frac{\vec{\rho}}{2},\mathbf{r}'',\mathbf{u},\mathbf{u}',\mathbf{u}'',t)}{G_{2}(\mathbf{R}+\frac{\vec{\rho}}{2},\mathbf{R}-\frac{\vec{\rho}}{2},\mathbf{u},\mathbf{u}',t)}d\mathbf{r}''d\mathbf{u}''\\
&-&\int\nabla_{\mathbf{R}}U_{int}(\mathbf{R}-\frac{\vec{\rho}}{2}-\mathbf{r}'',\mathbf{u}',\mathbf{u}'')W(\mathbf{r}'',\mathbf{u}'',t)\frac{G_{3}(\mathbf{R}+\frac{\vec{\rho}}{2},\mathbf{R}-\frac{\vec{\rho}}{2},\mathbf{r}'',\mathbf{u},\mathbf{u}',\mathbf{u}'',t)}{G_{2}(\mathbf{R}+\frac{\vec{\rho}}{2},\mathbf{R}-\frac{\vec{\rho}}{2},\mathbf{u},\mathbf{u}',t)}d\mathbf{r}''d\mathbf{u}''
\label{DelrhoPhi2}
\end{eqnarray}
From now on, we assume a uniform spatial distribution of molecules. This means that $W$ does not depend on $\mathbf{r}$ and therefore that we have
\begin{eqnarray}
W(\mathbf{r},\mathbf{u},t)=W(\mathbf{u},t)
\label{UniformSpace}
\end{eqnarray}
Furthermore, using the change of variables
\begin{eqnarray}
\vec{\rho}''=\mathbf{r}''-\mathbf{R}-\frac{\vec{\rho}}{2}
\end{eqnarray}
\begin{eqnarray}
\vec{\rho}''=\mathbf{r}''-\mathbf{R}+\frac{\vec{\rho}}{2}
\end{eqnarray}
wherever appropriate, Eqs. \eqref{DelRPhi2} and \eqref{DelrhoPhi2} can be rewritten as follows
\begin{eqnarray}
\nabla_{\mathbf{R}}\phi_{2}(\mathbf{R},\vec{\rho},\mathbf{u},\mathbf{u}',t)=\frac{3}{8}\mathbf{I}_{1}(\mathbf{R},\vec{\rho},\mathbf{u},\mathbf{u}',t)+\frac{5}{8}\mathbf{I}_{2}(\mathbf{R},\vec{\rho},\mathbf{u},\mathbf{u}',t)
\label{DelR2}
\end{eqnarray}
and
\begin{eqnarray}
\nabla_{\vec{\rho}}\phi_{2}(\mathbf{R},\vec{\rho},\mathbf{u},\mathbf{u}',t)=\nabla_{\vec{\rho}}U_{int}(\vec{\rho},\mathbf{u},\mathbf{u}')+\frac{3}{4}\mathbf{I}_{1}(\mathbf{R},\vec{\rho},\mathbf{u},\mathbf{u}',t)-\frac{5}{4}\mathbf{I}_{2}(\mathbf{R},\vec{\rho},\mathbf{u},\mathbf{u}',t)
\label{DelRho2}
\end{eqnarray}
where in Eqs.\eqref{DelR2} and \eqref{DelRho2}, $\mathbf{I}_{1}$ and $\mathbf{I}_{2}$ are vectors defined by the integrals
\begin{eqnarray}
\mathbf{I}_{1}(\mathbf{R},\vec{\rho},\mathbf{u},\mathbf{u}',t)=\int\nabla_{\vec{\rho}''}U_{int}(\vec{\rho}'',\mathbf{u},\mathbf{u}'')W(\mathbf{u}'',t)\frac{G_{3}(\mathbf{R}+\frac{\vec{\rho}}{2},\mathbf{R}-\frac{\vec{\rho}}{2},\vec{\rho}''+\mathbf{R}+\frac{\vec{\rho}}{2},\mathbf{u},\mathbf{u}',\mathbf{u}'',t)}{G_{2}(\mathbf{R}+\frac{\vec{\rho}}{2},\mathbf{R}-\frac{\vec{\rho}}{2},\mathbf{u},\mathbf{u}',t)}d\vec{\rho}''d\mathbf{u}''
\label{I1}
\end{eqnarray}
and
\begin{eqnarray}
\mathbf{I}_{2}(\mathbf{R},\vec{\rho},\mathbf{u},\mathbf{u}',t)=\int\nabla_{\vec{\rho}''}U_{int}(\vec{\rho}'',\mathbf{u}',\mathbf{u}'')W(\mathbf{u}'',t)\frac{G_{3}(\mathbf{R}+\frac{\vec{\rho}}{2},\mathbf{R}-\frac{\vec{\rho}}{2},\vec{\rho}''+\mathbf{R}-\frac{\vec{\rho}}{2},\mathbf{u},\mathbf{u}',\mathbf{u}'',t)}{G_{2}(\mathbf{R}+\frac{\vec{\rho}}{2},\mathbf{R}-\frac{\vec{\rho}}{2},\mathbf{u},\mathbf{u}',t)}d\vec{\rho}''d\mathbf{u}''
\label{I2}
\end{eqnarray}
Now, because we have a single-component material made of identical molecules, we have $G_{3}(1,2,3)=G_{3}(2,1,3)$ where $(1,2,3)$  denote the degrees of freedom of any three identical molecules. This entails that we also have
\begin{eqnarray}
\nonumber
\mathbf{I}_{1}(\mathbf{R},-\vec{\rho},\mathbf{u},\mathbf{u}',t)&=&\mathbf{I}_{2}(\mathbf{R},\vec{\rho},\mathbf{u}',\mathbf{u},t)\\
\mathbf{I}_{2}(\mathbf{R},-\vec{\rho},\mathbf{u},\mathbf{u}',t)&=&\mathbf{I}_{1}(\mathbf{R},\vec{\rho},\mathbf{u}',\mathbf{u},t)
\label{BodyCommutationRelations}
\end{eqnarray}
Now, since  $\phi_{2}$ is an effective pair interaction potential, we suppose that ${\phi}_{2}$ has the same properties as $U_{int}$ given by Eq.\eqref{InteractionProperty}. Therefore, in particular, we have  $\nabla_{\mathbf{R}}\phi_{2}=\mathbf{0}$. This means first that $\mathbf{I}_{1}$ and $\mathbf{I}_{2}$ do not depend on $\mathbf{R}$ and that in all equations, $\mathbf{R}$ may be replaced by $\mathbf{0}$. It follows easily from $\nabla_{\mathbf{R}}\phi_{2}=\mathbf{0}$ that
\begin{eqnarray}
\mathbf{I}_{2}(\mathbf{R},\vec{\rho},\mathbf{u},\mathbf{u}',t)=-\frac{3}{5}\mathbf{I}_{1}(\mathbf{R},\vec{\rho},\mathbf{u},\mathbf{u}',t)
\end{eqnarray}
while from $\nabla_{\mathbf{R}}\phi_{2}(\mathbf{R},-\vec{\rho},\mathbf{u}',\mathbf{u},t)=\nabla_{\mathbf{R}}\phi_{2}(\mathbf{R},\vec{\rho},\mathbf{u},\mathbf{u}',t)$ and Eq.\eqref{BodyCommutationRelations}, we also have
\begin{eqnarray}
\mathbf{I}_{2}(\mathbf{R},\vec{\rho},\mathbf{u},\mathbf{u}',t)=\mathbf{I}_{1}(\mathbf{R},\vec{\rho},\mathbf{u},\mathbf{u}',t)
\label{I1I2}
\end{eqnarray}
Hence whichever their argument is, $\mathbf{I}_{1}=\mathbf{I}_{2}=\mathbf{0}$. It follows immediately that we have the surprisingly simple and remarkable equation
\begin{eqnarray}
\nabla_{\vec{\rho}}\phi_{2}(\vec{\rho},\mathbf{u},\mathbf{u}',t)=\nabla_{\vec{\rho}}U_{int}(\vec{\rho},\mathbf{u},\mathbf{u}')
\label{Simple}
\end{eqnarray}
so that $\nabla_{\vec{\rho}}\phi_{2}$ \textit{is time-independent} and, of course,
\begin{eqnarray}
\nabla_{\mathbf{R}}\phi_{2}(\vec{\rho},\mathbf{u},\mathbf{u}',t)=\mathbf{0}
\end{eqnarray}
where now $\mathbf{R}$ has been omitted in the argument of $\mathbf{\phi}_{2}$ as this dependence does no longer exist in this function. Finally, since $\mathbf{I}_{1}=\mathbf{I}_{2}$, we have
\begin{eqnarray}
G_{3}(\frac{\vec{\rho}}{2},-\frac{\vec{\rho}}{2},\vec{\rho}''+\frac{\vec{\rho}}{2},\mathbf{u},\mathbf{u}',\mathbf{u}'',t)=G_{3}(\frac{\vec{\rho}}{2},-\frac{\vec{\rho}}{2},\vec{\rho}''-\frac{\vec{\rho}}{2},\mathbf{u},\mathbf{u}',\mathbf{u}'',t)=G_{3}(\vec{\rho},\vec{\rho}'',\mathbf{u},\mathbf{u}',\mathbf{u}'',t).
\label{G3Prop2}
\end{eqnarray}
and, as usual,
\begin{eqnarray}
G_{2}(\mathbf{R}+\frac{\vec{\rho}}{2},\mathbf{R}-\frac{\vec{\rho}}{2},\mathbf{u},\mathbf{u}',t)&=&G_{2}(\frac{\vec{\rho}}{2},-\frac{\vec{\rho}}{2},\mathbf{u},\mathbf{u}',t)=G_{2}(\vec{\rho},\mathbf{u},\mathbf{u}',t)
\label{G2}
\end{eqnarray}
However, in spite of all our simplifications of the interaction forces Eqs.\eqref{2BPotDist1} and \eqref{2BPotDist2} , the interaction \textit{torques} Eqs.\eqref{2BPotRot1} and \eqref{2BPotRot2} remain complicated and become
\begin{eqnarray}
\nabla_{\mathbf{u}}\phi_{2}(\vec{\rho},\mathbf{u},\mathbf{u}',t)=\nabla_{\mathbf{u}}U_{int}(\vec{\rho},\mathbf{u},\mathbf{u}')+\int\nabla_{\mathbf{u}}U_{int}(\vec{\rho}'',\mathbf{u},\mathbf{u}'')W(\mathbf{u}'',t)\frac{G_{3}(\vec{\rho},\vec{\rho}'',\mathbf{u},\mathbf{u}',\mathbf{u}'',t)}{G_{2}(\vec{\rho},\mathbf{u},\mathbf{u}',t)}d\vec{\rho}''d\mathbf{u}'',
\label{2BPotRot1Bis}
\end{eqnarray}
and
\begin{eqnarray}
\nabla_{\mathbf{u}'}\phi_{2}(\vec{\rho},\mathbf{u},\mathbf{u}',t)=\nabla_{\mathbf{u}}U_{int}(\vec{\rho},\mathbf{u},\mathbf{u}')+\int\nabla_{\mathbf{u}'}U_{int}(\vec{\rho}'',\mathbf{u}',\mathbf{u}'')W(\mathbf{u}'',t)\frac{G_{3}(\vec{\rho},\vec{\rho}'',\mathbf{u},\mathbf{u}',\mathbf{u}'',t)}{G_{2}(\vec{\rho},\mathbf{u},\mathbf{u}',t)}d\vec{\rho}''d\mathbf{u}'',
\label{2BPotRot2Bis}
\end{eqnarray}
The fact that the one-body density has the property Eq. \eqref{UniformSpace} entails that the translational part in Eq. \eqref{FPEDean1} and the $\mathbf{r}$ dependence of $V_1$ can be ignored completely. Therefore, this equation reads
\begin{eqnarray}
\frac{\partial W}{\partial t}(\mathbf{u},t)=D_{R}\nabla_{\mathbf{u}}\cdot\left(\nabla_{\mathbf{u}}W(\mathbf{u},t)+\beta W(\mathbf{u},t)\nabla_{\mathbf{u}}V_{1}(\mathbf{u},t)
+\beta\int\nabla_{\mathbf{u}}U_{int}(\vec{\rho},\mathbf{u},\mathbf{u}')W_{2}(\vec{\rho},\mathbf{u},\mathbf{u}',t)d\vec{\rho}d\mathbf{u}'\right)
\label{DeanEq1Fin}
\end{eqnarray}
It follows from Eqs. \eqref{2BPotRot1Bis} and \eqref{2BPotRot2Bis} that Eq. \eqref{FPEDean2} becomes
\begin{eqnarray}
\nonumber
\frac{\partial W_{2}}{\partial t}(\vec{\rho},\mathbf{u},\mathbf{u}',t)=\frac{D_T}{2}\nabla_{\vec{\rho}}\cdot(\nabla_{\vec{\rho}} W_{2}(\vec{\rho},\mathbf{u},\mathbf{u}',t)+\beta W_{2}(\vec{\rho},\mathbf{u},\mathbf{u}',t)\nabla_{\vec{\rho}}V_{2}(\vec{\rho},\mathbf{u},\mathbf{u}',t))\qquad\qquad\qquad\qquad\\
\nonumber
+D_{R}\nabla_{\mathbf{u}}\cdot\left(\nabla_{\mathbf{u}} W_{2}(\vec{\rho},\mathbf{u},\mathbf{u}',t)+\beta W_{2}(\vec{\rho},\mathbf{u},\mathbf{u}',t)\nabla_{\mathbf{u}}V_{2}(\vec{\rho},\mathbf{u},\mathbf{u}',t)+\beta\int\nabla_{\mathbf{u}}U_{int}(\vec{\rho}'',\mathbf{u},\mathbf{u}'')W_{3}(\vec{\rho},\vec{\rho}'',\mathbf{u},\mathbf{u}',\mathbf{u}'',t)d\vec{\rho}''d\mathbf{u}''\right)\\
+D_{R}\nabla_{\mathbf{u}'}\cdot\left(\nabla_{\mathbf{u}'} W_{2}(\vec{\rho},\mathbf{u},\mathbf{u}',t)+\beta W_{2}(\vec{\rho},\mathbf{u},\mathbf{u}',t)\nabla_{\mathbf{u}'}V_{2}(\vec{\rho},\mathbf{u},\mathbf{u}',t)+\beta\int\nabla_{\mathbf{u}'}U_{int}(\vec{\rho}'',\mathbf{u}',\mathbf{u}'')W_{3}(\vec{\rho},\vec{\rho}'',\mathbf{u},\mathbf{u}',\mathbf{u}'',t)d\vec{\rho}''d\mathbf{u}''\right)
\label{DeanEq2Fin}
\end{eqnarray}
Therefore, the $\vec{\rho}$ drift coefficient of the above equation \textit{does no longer} contain the three-body density $W_3$ and is therefore a \textit{linear drift coefficient} of Fokker-Planck type.\cite{Risken1989Book} This result is general, exact and \textit{entirely new}.  
Indeed, we have now
\begin{eqnarray}
V_{2}(\vec{\rho},\mathbf{u},\mathbf{u}',t)&=&U_{int}(\vec{\rho},\mathbf{u},\mathbf{u}')+V_{1}(\mathbf{u},t)+V_{1}(\mathbf{u}',t)\\
W_{2}(\vec{\rho},\mathbf{u},\mathbf{u}',t)&=&W(\mathbf{u},t)W(\mathbf{u}',t)G_{2}(\vec{\rho},\mathbf{u},\mathbf{u}',t)\\
W_{3}(\vec{\rho},\vec{\rho}'',\mathbf{u},\mathbf{u}',\mathbf{u}'',t)&=&W(\mathbf{u},t)W(\mathbf{u}',t)W(\mathbf{u}'',t)G_{3}(\vec{\rho},\vec{\rho}'',\mathbf{u},\mathbf{u}',\mathbf{u}'',t)
\end{eqnarray}
\subsection{An important result concerning $G_2$}
Now, we define $w_2$ as follows : 
\begin{eqnarray}
\nonumber
w_{2}(\mathbf{u},\mathbf{u}',t)=\int W_{2}(\vec{\rho},\mathbf{u},\mathbf{u}',t)d\vec{\rho}
\end{eqnarray}
and introduce the function $g_{\rho}(\vec{\rho},\mathbf{u},\mathbf{u}',t)$ via the equation
\begin{eqnarray}
g_{\rho}(\vec{\rho},\mathbf{u},\mathbf{u}',t)=\frac{W_{2}(\vec{\rho},\mathbf{u},\mathbf{u}',t)}{w_{2}(\mathbf{u},\mathbf{u}',t)}
\label{GdefGen}
\end{eqnarray}
This function contains spatial correlations and therefore must have null $\vec{\rho}$ gradient outside the Fr\"{o}hlich inner sphere because outside this sphere, the medium is continuous and the spatial distribution outside the sphere is finite. Let $r_{C}$ be its radius. Furthermore, we introduce a hard sphere radius $R_H$ which encodes the fact that the molecules cannot approach one another \textit{beyond} this distance. We have clearly
\begin{eqnarray}
\nabla_{\vec{\rho}}g_{\rho}(\vec{\rho},\mathbf{u},\mathbf{u}',t)\vert_{\rho\geq r_{C}}&=&\nabla_{\vec{\rho}}g_{\rho}(\vec{\rho},\mathbf{u},\mathbf{u}',t)\vert_{\rho\leq R_{H}}=\mathbf{0}\\
g_{\rho}(\vec{\rho},\mathbf{u},\mathbf{u}',t)\vert_{\rho\leq R_{H}}&=&0
\label{BoundaryConditions}
\end{eqnarray} 
Because $g_{\rho}$ can always be expanded in products of spherical harmonics $Y_{LM}(\mathbf{u})Y_{JK}(\mathbf{u}')$ (or Wigner $D$ functions), we can split the above function as a sum of an orientation-independent term $g_{\rho}^{(0)}(\vec{\rho},t)$ and an other term $\delta g(\vec{\rho},\mathbf{u},\mathbf{u}',t)$ \textit{which is not small \textit{a priori}}. The normalization of $W_{2}$ to unity implies that
\begin{eqnarray}
\int g_{\rho}^{(0)}(\vec{\rho},t)d\vec{\rho}\int w_{2}(\mathbf{u},\mathbf{u}',t)d\mathbf{u}d\mathbf{u}'=
1-\int\delta g(\vec{\rho},\mathbf{u},\mathbf{u}',t)w_{2}(\mathbf{u},\mathbf{u}',t)d\vec{\rho}\mathbf{u}d\mathbf{u}'
\end{eqnarray}
Constraining the above problem does not restrict its generality, therefore we choose as normalization conditions
\begin{eqnarray}
\int g_{\rho}^{(0)}(\vec{\rho},t)d\vec{\rho}&=&1\\
\int w_{2}(\mathbf{u},\mathbf{u}',t)d\mathbf{u}d\mathbf{u}'&=&1
\label{NormCond2}
\end{eqnarray}
from which it follows that we must have, for arbitrary time, the constraint
\begin{eqnarray}
\int\delta g(\vec{\rho},\mathbf{u},\mathbf{u}',t)w_{2}(\mathbf{u},\mathbf{u}',t)d\vec{\rho}d\mathbf{u}d\mathbf{u}'=0
\label{Constraint}
\end{eqnarray}
Notice that the first of the normalizing integrals \eqref{NormCond2} is consistent with Eq. \eqref{GdefGen} and the definition of $w_2$ above, and this for arbitrary times. We may also take the conditions at the boundaries that $g_{\rho}$ and $g_{\rho}^{(0)}$ have null spatial gradient at the boundaries. 

Next, we can integrate Eq.\eqref{DeanEq2Fin} over the whole configuration space (intermolecular distances and orientations). This has the effect of nullifying the orientational terms in the resulting equation by Gauss's theorem, because the orientational probability density currents are tangent to the unit spheres of orientational representative points (hence the advantage of having cosmetically transformed the Dean equations into Fokker-Planck ones which in turn are continuity equations). After application of Gauss's theorem to the other integrals, we have
\begin{eqnarray}
\oint(\nabla_{\vec{\rho}}g_{\rho}^{(0)}(\vec{\rho},t)+\beta g_{\rho}^{(0)}(\vec{\rho},t)\nabla_{\vec{\rho}}\bar{V}_{2}(\vec{\rho},t))\cdot{\mathbf{e}}_{\vec{\rho}}dS+\int\nabla_{\vec{\rho}}\cdot(\nabla_{\vec{\rho}}\delta g(\vec{\rho},\mathbf{u},\mathbf{u}',t)+\beta\delta g(\vec{\rho},\mathbf{u},\mathbf{u}',t)\nabla_{\vec{\rho}}V_{2}(\vec{\rho},\mathbf{u},\mathbf{u}',t))w_{2}(\mathbf{u},\mathbf{u}',t)d\vec{\rho}d\mathbf{u}d\mathbf{u}'=0
\label{IntegralEqNew}
\end{eqnarray}
where $\bar{V}_{2}$ is given by
\begin{eqnarray}
\bar{V_{2}}(\vec{\rho},t)=\int V_{2}(\vec{\rho},\mathbf{u},\mathbf{u}',t)w_{2}(\mathbf{u},\mathbf{u}',t)d\mathbf{u}d\mathbf{u}'
\label{V2Radial}
\end{eqnarray}

The first integral in Eq.\eqref{IntegralEqNew} is a surface integral extending to the boundaries of the cavity comprising two concentric spheres : a large but finite sphere of radius $r_C$ and a small sphere included in the former large one of radius $R_H$. Since we have assumed that $g_{\rho}^{(0)}$ has null gradient at the boundaries, and furthermore that the cavity is large, this integral vanishes since the interaction force is negligible at large intermolecular distances. Therefore, we have
\begin{eqnarray}
\int\nabla_{\vec{\rho}}\cdot(\nabla_{\vec{\rho}}\delta g(\vec{\rho},\mathbf{u},\mathbf{u}',t)+\beta\delta g(\vec{\rho},\mathbf{u},\mathbf{u}',t)\nabla_{\vec{\rho}}V_{2}(\vec{\rho},\mathbf{u},\mathbf{u}',t))w_{2}(\mathbf{u},\mathbf{u}',t)d\vec{\rho}d\mathbf{u}d\mathbf{u}'=0
\label{KeyPoint}
\end{eqnarray}
and this equation \textit{must hold for arbitrary times}. Notice that in general, $V_2$ is explicitly time-dependent, at least because time-dependent external fields may be applied to the system (one may also involve time-dependent effective interactions for the purpose of modelling). The fact that time occurs in Eq. \eqref{KeyPoint} makes it difficult to be satisfied \textit{at arbitrary times} (it is quite important) for arbitrary time-dependent $V_2$, except if 
\begin{eqnarray}
\nabla_{\vec{\rho}}\cdot(\nabla_{\vec{\rho}}\delta g(\vec{\rho},\mathbf{u},\mathbf{u}',t)+\beta\delta g(\vec{\rho},\mathbf{u},\mathbf{u}',t)\nabla_{\vec{\rho}}V_{2}(\vec{\rho},\mathbf{u},\mathbf{u}',t))=0
\label{Lagrange}
\end{eqnarray}
The solution of the above equation is
\begin{eqnarray}
\delta g(\vec{\rho},\mathbf{u},\mathbf{u}',t)=K(\mathbf{u},\mathbf{u}',t)\exp(-\beta V_{2}(\vec{\rho},\mathbf{u},\mathbf{u}',t))+f(\vec{\rho},\mathbf{u},\mathbf{u}',t)
\label{LagrangeSolution}
\end{eqnarray}
where $f$ is an arbitrary solution of Eq.\eqref{Lagrange} with no singularities of Dirac delta type. Because $\delta g$ has zero gradient at the boundaries of the cavity, the function $K(\mathbf{u},\mathbf{u}',t)$ is overdetermined except if it is zero. Furthermore, this function cannot be overdetermined since Eq. \eqref{Lagrange} is a linear partial differential equation (given the boundary conditions, the solution of a linear partial differential equation is unique if it exists). Next, since $f$ is a solution of Eq. \eqref{Lagrange}, its  representation is also given by an equation of the form of Eq. \eqref{LagrangeSolution}, and has also null gradient at the boundaries. The $K$ function in this solution is also overdetermined, save if the gradient of the complementary function, $f_{1}$, say, has zero gradient at the boundaries. Iterating this procedure we are lead to the conclusion that $f=0$ everywhere and at arbitrary times. Therefore, $\delta g=0$ and we have
\begin{eqnarray}
W_{2}(\vec{\rho},\mathbf{u},\mathbf{u}',t)=g_{\rho}^{(0)}(\vec{\rho},t)w_{2}(\mathbf{u},\mathbf{u}',t)=g_{\rho}(\vec{\rho},t)w_{2}(\mathbf{u},\mathbf{u}',t)
\label{W2Ansatz}
\end{eqnarray}
This indeed entails
\begin{eqnarray}
G_{2}(\vec{\rho},\mathbf{u},\mathbf{u}',t)=g_{\rho}^{(0)}(\vec{\rho},t)g_{2}(\mathbf{u},\mathbf{u}',t)=g_{\rho}(\vec{\rho},t)g_{2}(\mathbf{u},\mathbf{u}',t)
\end{eqnarray}
where the orientational pair distribution function $g_{2}(\mathbf{u},\mathbf{u}',t)$ is defined via
\begin{eqnarray}
g_{2}(\mathbf{u},\mathbf{u}',t)=\frac{w_{2}(\mathbf{u},\mathbf{u}',t)}{W(\mathbf{u},t)W(\mathbf{u}',t)}
\label{OrPairDist}
\end{eqnarray}
This \textit{nontrivial} result concerning $G_{2}$ \textit{is extremely important, exact, new and general}. 
\subsection{Derivation of equations for $g_{\rho}$ and $w_{2}$ and final form of the kinetic equations}
First, we define the functions $g_{\rho}(\vec{\rho},t)$ and $w_{2}(\mathbf{u},\mathbf{u}',t)$ by the equations
\begin{eqnarray}
g_{\rho}(\vec{\rho},t)&=&\int W_{2}(\vec{\rho},\mathbf{u},\mathbf{u}',t)d\mathbf{u}d\mathbf{u}'\label{Grho}\\
w_{2}(\mathbf{u},\mathbf{u}',t)&=&\int W_{2}(\vec{\rho},\mathbf{u},\mathbf{u}',t)d\vec{\rho}\label{W2or}
\end{eqnarray} 
We now derive an equation for $g_{\rho}$ on one hand, and an equation for $w_{2}$ on an other hand. In fact, the two equations can readily be obtained by inserting Eq.\eqref{W2Ansatz} in Eq.\eqref{DeanEq2Fin} and integrating over the relevant variables. Combining Eqs.\eqref{DeanEq2Fin}, \eqref{W2Ansatz} and integrating the resulting equation over dipole orientations leads to the \textit{Fokker-Planck equation}
\begin{eqnarray}
\frac{\partial g_{\rho}}{\partial t}(\vec{\rho},t)=\frac{D_{T}}{2}\nabla_{\vec{\rho}}\cdot(\nabla_{\vec{\rho}}g_{\rho}(\vec{\rho},t)+\beta g_{\rho}(\vec{\rho},t)\nabla_{\vec{\rho}}\bar{V_{2}}(\vec{\rho},t))
\label{FPEGrho}
\end{eqnarray}
where $\bar{V}_{2}$ is given by Eq.\eqref{V2Radial}.
The obtaining of an equation for $w_2$ is more subtle. In order to derive it, it must be first mentioned that $g_{\rho}$ is the elementary probability density for two molecules to be separated by the vector $\vec{\rho}$ at time $t$. It was demonstrated on rather general grounds \cite{Dejardin2014JCP} that such theories as the one developed here have a meaning only if the minimal distance between molecules is larger than a molecular diameter (this is the hard sphere radius $R_H$) while it is also clear that density correlations cannot have an infinite length range $r_{C}$. The $r_{C}$ we are dealing with here is nevertheless much larger than that of Madden and Kivelson\cite{Madden1984Book} as the sphere of radius $r_{C}$ we consider is the \textit{sole cavity} in an infinite dielectric and contains a sufficiently \textit{large} number of molecules to allow statistical mechanics to be applied \textit{in this sphere}, while in the theory of Madden and Kivelson, $r_{C}$ is of the order of $2-4$ molecular diameters, so that $2$ molecules are contained in it at best (therefore many cavities separated by continuous dielectric regions exist in previous treatments). 
Since in our theory molecules cannot approach each other below a minimal distance $R_{H}$, it follows that (hard sphere assumption)
\begin{eqnarray}
g_{\rho}(\vec{\rho},t)=0\quad \mathrm{when}\quad \vert\vec{\rho}\vert\leq R_{H}.
\label{HardSphereCondition}
\end{eqnarray} 
Furthermore, \textit{we assume} that the pair density correlation length $r_{C}$ is \textit{large, but finite}, and therefore treat $r_{C}$ as a parameter of the model (both $R_H$ and $r_{C}$ are therefore related to the temperature-dependent mass density of the specimen but no longer to molecular parameters). This means that we must, for Eq.\eqref{FPEGrho}, take as a necessary boundary condition 
\begin{eqnarray}
g_{\rho}(\vec{\rho},t)=A\quad \mathrm{when}\quad \vert\vec{\rho}\vert\geq r_{C},
\label{GrhoBC}
\end{eqnarray}
where $A$ is a constant having the dimensions of inverse volume, so that as long as $R_{H}<\vert\vec{\rho}\vert\leq r_{C}$ (meaning $g_{\rho}(\mathbf{R}_{H}^{+},t)\neq 0$) $g_{\rho}$ is continuous, has continuous first derivatives, is twice differentiable and is bounded in this distance interval. Those two finite nonzero bounds allow us to normalize $g_{\rho}(\vec{\rho},t)$ inside the domain $R_H<\vert\vec{\rho}\vert\leq r_{C}$ for all $\hat{\rho}$ orientations, so that the integration commutation problem alluded to by Madden and Kivelson \cite{Madden1984Book} is therefore \textit{not relevant in our work}. Like Onsager and Fr\"{o}hlich \cite{Onsager1936JACS,Frohlich1958Book}  assumed implicitly, for $\vert\vec{\rho}\vert>r_{C}$, the distribution of bodies in a polar fluid is uniform in space and we can treat the surroundings  of the cavity (which are uniformly polarized) by  classical macroscopic electromagnetism. Therefore, with this proviso, we can write
\begin{eqnarray}
\int g_{\rho}(\vec{\rho},t)d\vec{\rho}=1
\label{NormGrho}
\end{eqnarray}
entailing that the time-independent solution of Eq.\eqref{FPEGrho} \textit{is not proportional to the trivial} $\exp(-\beta\bar{V}_{2}(\rho))$, but is more complicated as non-trivial boundary conditions must be chosen (i.e., on the probability current, see Reference \citenum{Risken1989Book}, Chapter $11$ for an example regarding a quite different subject) in order to solve this equation. However, since we are not interested in the precise calculation of the temperature-dependent mass density of the fluid here, the explicit knowledge of the time-independent (and also time-dependent) $g_{\rho}$ as a function of temperature is presently of little importance (see Appendix D). 

By combining Eqs.\eqref{DeanEq2Fin}, \eqref{W2Ansatz}, and integrating the resulting equation over $\vec{\rho}$ we obtain the equation
\begin{eqnarray}
\nonumber
\frac{\partial w_2}{\partial t}=\frac{\beta D_{T}}{2}w_{2}(\mathbf{u},\mathbf{u}',t)\int_{S_{r_C}} g_{\rho}(\vec{\rho},t)\nabla_{\vec{\rho}}(U_{int}(\vec{\rho},\mathbf{u},\mathbf{u}')-\bar{V_{2}}(\vec{\rho},t))\cdot\mathbf{e_{\rho}}dS\qquad\qquad\qquad\qquad\qquad\\
\nonumber
+D_{R}\nabla_{\mathbf{u}}\cdot(\nabla_{\mathbf{u}} w_{2}(\mathbf{u},\mathbf{u}',t)+\beta w_{2}(\mathbf{u},\mathbf{u}',t)\nabla_{\mathbf{u}}U_{m}(\mathbf{u},\mathbf{u}',t)+\beta w_{2}(\mathbf{u},\mathbf{u}',t)\int\nabla_{\mathbf{u}}U_{int}(\vec{\rho}'',\mathbf{u},\mathbf{u}'')W(\mathbf{u}'',t)\frac{G_{3}(\vec{\rho},\vec{\rho}'',\mathbf{u},\mathbf{u}',\mathbf{u}'',t)}{g_{2}(\mathbf{u},\mathbf{u}',t)}d\vec{\rho}d\vec{\rho}''d\mathbf{u}'')\\
+D_{R}\nabla_{\mathbf{u}'}\cdot(\nabla_{\mathbf{u}'} w_{2}(\mathbf{u},\mathbf{u}',t)+\beta w_{2}(\mathbf{u},\mathbf{u}',t)\nabla_{\mathbf{u}'}U_{m}(\mathbf{u},\mathbf{u}',t)+\beta w_{2}(\mathbf{u},\mathbf{u}',t)\int\nabla_{\mathbf{u}'}U_{int}(\vec{\rho}'',\mathbf{u}',\mathbf{u}'')W(\mathbf{u}'',t)\frac{G_{3}(\vec{\rho},\vec{\rho}'',\mathbf{u},\mathbf{u}',\mathbf{u}'',t)}{g_{2}(\mathbf{u},\mathbf{u}',t)}d\vec{\rho}d\vec{\rho}''d\mathbf{u}'')
\label{w2Orexact}
\end{eqnarray}
where we have used Eq.\eqref{FPEGrho}, $\mathbf{e}_{\rho}$ is the unit normal to the sphere of radius $r_{C}$, the integral $\int_{S_{r_C}}$ which arises from the application of Gauss's theorem ($\int_{S_{R_{H}}}=0$, where $S_{R_{H}}$ is the hard sphere of radius $R_H$ for which $g_{\rho}=0$)  and 
\begin{eqnarray}
U_{m}(\mathbf{u},\mathbf{u}',t)=\int U_{int}(\vec{\rho},\mathbf{u},\mathbf{u}',t)g_{\rho}(\vec{\rho},t)d\vec{\rho}
\label{UmDef1}
\end{eqnarray}
contains the memory of structural relaxation, and is a function of dipole orientations only. Note that since we have Eqs.\eqref{HardSphereCondition} and the integral is extended up to a large, but finite distance, the integral in Eq.\eqref{UmDef1} \textit{converges}. Furthermore, since $g_{\rho}(\vec{\rho},t)=A$ on this surface, we may let $r_C\rightarrow\infty$ in the integral  $\int_{S_{r_C}}$ so that it vanishes. As already alluded to above, this is the main difference between our treatment and previous ones, which implicitly consider an ensemble of tiny cavities containing a few molecules only, while we consider a \textit{unique} large spherical cavity which contains a sufficient number of molecules to be able to define a statistical mechanical ensemble onto which statistical mechanics can be applied. Note also that because of this argument, \textit{we need only a unique spherical cavity of large but finite finite size immersed in an infinite dielectric} as first Kirkwood and Fr\"{o}hlich implicitly assumed. 
\noindent
At last, little is known on three-body correlations, therefore little is known on $G_3$. Yet, we may make an ansatz concerning this function. In a similar spirit with what we have obtained for $G_2$, we write
\begin{eqnarray}
G_{3}(\vec{\rho},\vec{\rho}'',\mathbf{u},\mathbf{u}',\mathbf{u}'',t)=g_{\rho,\rho''}(\vec{\rho},\vec{\rho}'',t)g_{3}(\mathbf{u},\mathbf{u}',\mathbf{u}'',t)
\label{G3Ansatz}
\end{eqnarray}
and $g_{\rho,\rho''}$ is the probability density to find three molecules with vector separations $\vec{\rho}$ and $\vec{\rho}''$. Since furthermore intermolecular distances are in between the hard sphere and the Kirkwood radii (therefore larger than the hard sphere radius $R_H$ in general), it is reasonable to make the mean-field estimate 
\begin{eqnarray}
g_{\rho,\rho''}(\vec{\rho},\vec{\rho}'',t)\approx g_{\rho}(\vec{\rho},t)g_{\rho}(\vec{\rho}'',t)
\label{3BodyAnsatzMeanField}
\end{eqnarray}
which is tantamount to saying that we neglect three-body \textit{spatial} correlations (more precisely, we neglect the \textit{spatial} correlation of the third body in regard to that of the two others). Hence Eq.\eqref{w2Orexact} becomes finally, with $\tau_{D}=(2D_{R})^{-1}$
\begin{eqnarray}
\nonumber
2\tau_{D}\frac{\partial w_2}{\partial t}(\mathbf{u},\mathbf{u}',t)&=&\nabla_{\mathbf{u}}\cdot(\nabla_{\mathbf{u}} w_{2}(\mathbf{u},\mathbf{u}',t)+\beta w_{2}(\mathbf{u},\mathbf{u}',t)\nabla_{\mathbf{u}}U_{m}(\mathbf{u},\mathbf{u}',t)\quad\\
\nonumber
&+&\beta\int\nabla_{\mathbf{u}}U_{m}(\mathbf{u},\mathbf{u}'',t)w_{3}(\mathbf{u},\mathbf{u}',\mathbf{u}'',t)d\mathbf{u}'')\\
\nonumber
&+&\nabla_{\mathbf{u}'}\cdot(\nabla_{\mathbf{u}'} w_{2}(\mathbf{u},\mathbf{u}',t)+\beta w_{2}(\mathbf{u},\mathbf{u}',t)\nabla_{\mathbf{u}'}U_{m}(\mathbf{u},\mathbf{u}',t)\\
&+&\beta\int\nabla_{\mathbf{u}'}U_{m}(\mathbf{u}',\mathbf{u}'',t)w_{3}(\mathbf{u},\mathbf{u}',\mathbf{u}'',t)d\mathbf{u}'')
\label{RotationalDK2}
\end{eqnarray}
where the three-body orientational density $w_3$ is given by
\begin{eqnarray}
w_{3}(\mathbf{u},\mathbf{u}',\mathbf{u}'',t)=W(\mathbf{u},t)W(\mathbf{u}',t)W(\mathbf{u}'',t)g_{3}(\mathbf{u},\mathbf{u}',\mathbf{u}'',t)
\label{3Bodydensity}
\end{eqnarray}
we also notice, using Eqs.\eqref{W2Ansatz} in conjunction with Eq.\eqref{UmDef1} that \textit{without any approximation}, Eq.\eqref{DeanEq1Fin} reads
\begin{eqnarray}
2\tau_{D}\dfrac{\partial W}{\partial t}(\mathbf{u},t)=
\nabla_{\mathbf{u}}\cdot\left(\nabla_{\mathbf{u}} W (\mathbf{u},t)+\beta W(\mathbf{u},t)\nabla_{\mathbf{u}}V_{1}(\mathbf{u},t)
+\beta\int\nabla_{\mathbf{u}}U_{m}(\mathbf{u},\mathbf{u}',t)w_{2}(\mathbf{u},\mathbf{u}',t)d\mathbf{u}'\right)
\label{RotationalDK}
\end{eqnarray}
In the static regime, all quantities are time-independent and Eq. \eqref{eqDK} results. Eqs.\eqref{RotationalDK2} and \eqref{RotationalDK} were used by some of us recently \cite{Dejardin2018JCP,Dejardin2019PRB}, however with a static $U_{m}$. Having given the general theoretical background, we are ready to apply our formalism to the calculation of the dielectric constant of simple polar fluids. However, before accomplishing this, we discuss in the next Appendix relation of our work with theoretical results obtained previously by several milestone authors. 
  
\section{Appendix B : comparison with several benchmark results for purely polar fluids}
Here, all quantities considered in the previous section are time-independent. The static equations are all equivalent to a Yvon-Born-Green hierarchy member.\cite{Hansen2006Book}  
Below we compare the outcomes of our theory with benchmark results concerning the dielectric constant of purely polar fluids and also some others in contiguous areas. Whenever the comparison can be made quantitative, this is justified by equations.

\subsection{Onsager's theory}
Here, we derive Onsager's dielectric equation of state \cite{Onsager1936JACS} for a purely polar liquid with our method. Inspired by Lorentz's work \cite{Bottcher1973Book}, he tried to relate the microscopic behavior of the dielectric to the macroscopic one by defining a spherical cavity inside an infinite dielectric into which \textit{it may appear that only one molecule is present inside it}, and apparently applied statistical mechanics to this molecule (meaning there is an ensemble of cavities, i.e., "holes" separated by continuous regions). In fact, the statistical treatment is best understood if the cavity is large \cite{Bottcher1973Book} (as ours is) and made of non-interacting molecules at the microscopic level so that only \textit{one cavity} inside an infinite dielectric is necessary, and the molecules do not interact at the microscopic level even if they are many. The surroundings of the cavity are continuous and treated on a macroscopic basis \cite{Evans1982Book}.  
Then, if the cavity is empty and an external field is applied, a field appears in the cavity which is not the applied field (this is Onsager's cavity field). Then, if a dipole is inserted in the cavity, it orients through the effect of the cavity field (the directing field in Onsager's theory) and by orienting itself in the field, necessarily acts on its surroundings to polarize them. The surroundings, by reaction, exert a field on the dipole in the cavity, hence the name "reaction field". The total field seen by a dipole in the cavity is the sum of the cavity field and the reaction field (the reaction field being proportional to the molecular dipole vector, it does not orient the dipole \cite{Onsager1936JACS,Bottcher1973Book}).
Thus, Onsager's theory ignores intermolecular interactions  \textit{in the statistical treatment}. Therefore, Eq.\eqref{RotationalDK} in its time-independent version is used with $U_{m}=0$. The external potential $V_{1}$ is given by
\begin{eqnarray}
V_{1}(\mathbf{u})=-\mu E(\mathbf{u}\cdot\mathbf{e})
\label{V1potential}
\end{eqnarray} 
where $E$ denotes the amplitude of Onsager's cavity field \cite{Onsager1936JACS}
\begin{eqnarray}
E=\frac{9\varepsilon E_{0}}{(2\varepsilon+1)(\varepsilon+2)},
\label{cavityfield}
\end{eqnarray}
$E_{0}$ is the amplitude of the electrostatic field created by charges external to the dielectric and $\mathbf{e}$ is a unit vector along the externally applied field. We have, in the linear response regime \cite{Debye1929Book,Coffey2017Book} 
\begin{eqnarray}
W(\mathbf{u})\approx\frac{1}{4\pi}(1+\xi(\mathbf{u}\cdot\mathbf{e}))
\label{WDebyeOnsager}
\end{eqnarray}
where $\xi=\beta\mu E$ and where we have used that in linear response to the external field, $\xi<<1$. The statistical calculation of the polarization in the direction of the applied field proceeds as follows. We have
\begin{eqnarray}
\mathrm{P}&=&\rho_{0}\mu \int (\mathbf{u}\cdot\mathbf{e})W(\mathbf{u})d\mathbf{u}\\
&=&\frac{\rho_{0}\mu\xi}{3}
\label{StatpolDebye}
\end{eqnarray}
The macroscopic linear polarization of a homogeneously polarized sphere in the direction of the applied field is
\begin{eqnarray}
P=3\varepsilon_{0}\frac{\varepsilon-1}{\varepsilon+2}E_{0}
\label{MacropolSphere}
\end{eqnarray}
where $\varepsilon_{0}$ is the permittivity of vacuum.  On using $\mathrm{P}=P$ and combining Eqs.\eqref{StatpolDebye} and \eqref{MacropolSphere} we have
\begin{eqnarray}
\frac{(\varepsilon-1)(2\varepsilon+1)}{3\varepsilon}=\lambda
\label{Onsager}
\end{eqnarray}
where $\lambda$ is the susceptibility of an ideal gas of dipoles, viz.
\begin{eqnarray}
\lambda=\frac{\rho_{0}\mu^{2}}{3\varepsilon_{0}kT}
\end{eqnarray}
\subsection{Maier-Saupe theory of the isotropic-nematic phase transition in nematic liquid crystals}
A nematic liquid crystal is made of rigid cylindrically (or axially) symmetric molecules. In the nematic phase, a typical molecule sees the interaction of its nearest neighbours mainly through their orientations. The only interactions are due to the induced dipolar effects (van der Waals forces, attractive) and steric effects (repulsive). These effects are therefore at least quadrupolar. On the contrary in the isotropic phase, thermal agitation destroys all interactions. The orientational order is tracked by the variation of $\langle P_{2}\rangle$ at equilibrium, which is zero in the isotropic phase and jumps to a non-zero value by crossing the isotropic to nematic phase transition \cite{MaierSaupe1960ZNaturF,Chandrasekhar1973Book}.
In order to recover this theory, one may remark that since $U_m$ given by Eq.\eqref{UmDef1} is a function of the orientations of the dipoles only, one may always expand the time-independent $U_m$ in spherical harmonics as follows
\begin{eqnarray}
\nonumber
U_{m}(\mathbf{u},\mathbf{u}')=\sum_{l=0}^{\infty}\sum_{m=-l}^{l}\sum_{j=0}^{\infty}\sum_{k=-j}^{j}a_{lmjk}Y_{lm}(\mathbf{u})Y_{jk}(\mathbf{u}')\\
\label{UmExp}
\end{eqnarray}
In the Maier-Saupe theory, there is only one non-zero term in the infinite series Eq.\eqref{UmExp} which corresponds to $a_{2020}$, and is a pure quadrupolar effect (dispersion). We write 
\begin{eqnarray}
\nonumber
U_{m}(\mathbf{u},\mathbf{u}')=-KP_{2}(\mathbf{u})P_{2}(\mathbf{u}')\\
\label{UmExpMaierSaupe}
\end{eqnarray}
where $P_{n}(\mathbf{u})$ is a Legendre polynomial. Furthermore, the mean-field approximation is used for $w_2$. We have
\begin{eqnarray}
w_2(\mathbf{u},\mathbf{u}')=W(\mathbf{u})W(\mathbf{u}')
\label{w2MeanField}
\end{eqnarray}
so that Eq.\eqref{RotationalDK} in the static regime becomes
\begin{eqnarray}
\nabla_{\mathbf{u}}\cdot(\nabla_{\mathbf{u}} W (\mathbf{u})+\beta W(\mathbf{u})\nabla_{u}V_{1}^{eff}(\mathbf{u}))=0
\label{RotationalDKStatMFNematic}
\end{eqnarray}
where
\begin{eqnarray}
V_{1}^{eff}(\mathbf{u})&=&-KP_{2}(\mathbf{u})\int P_{2}(\mathbf{u})W(\mathbf{u}')d\mathbf{u}'\\
&=&-KP_{2}(\mathbf{u})\langle P_{2}\rangle_{0}
\end{eqnarray}
where $K$ is a constant (and therefore a parameter of the Maier-Saupe model) and the angular brackets $\langle\rangle_{0}$ denote an equilibrium (and external field free) statistical average over $W$, which is given by
\begin{eqnarray}
W(\mathbf{u})&=&\frac{1}{Z_{MS}}e^{\beta KP_{2}(\mathbf{u})\langle P_{2}\rangle_{0}}\\
Z_{MS}&=&\int{W(\mathbf{u})d\mathbf{u}}
\end{eqnarray} 
The nematic order $\langle P_{2}\rangle_{0}$ is determined by self-consistently solving the equation
\begin{eqnarray}
\langle P_{2}\rangle_{0}=\int P_{2}(\mathbf{u})W(\mathbf{u})d\mathbf{u}
\end{eqnarray}
while the right hand side of this equation can be expressed in terms of a ratio of confluent hypergeometric (Kummer) functions. \cite{Coffey2017Book} Solving this equation leads indeed to the isotropic to nematic phase transition results of Maier and Saupe where there a pure quadrupolar interaction contribution is accounted for. \cite{Chandrasekhar1973Book}
\subsection{A preliminary test of our decorrelation procedure : the Debye-Fr\"{o}hlich model of dielectric relaxation from the averaged Dean equation}
This model is somewhat equivalent to the Maier-Saupe theory described above, using $\sigma=\beta K\langle P_{2}\rangle_{0}$ so that $\sigma$ is no longer \textit{explicitly} temperature-dependent \cite{Frohlich1958Book,Coffey2005PRE,Wei2016PRE,Coffey2017Book}. In order to derive a form of it, we assume in Eq. \eqref{RotationalDK} that $\beta U_{m}$ is time-independent and given by
\begin{eqnarray}
\beta U_{m}(\mathbf{u},\mathbf{u}')=-\gamma\cos\vartheta\cos\vartheta',
\label{UmDebyeFrohlich}
\end{eqnarray}
$\gamma$ being some parameter. The split of a statistical sample made of pairs does not pose the questions we address to in a later Appendix (the split of a triplet into a doublet and a singlet and recombination of a pair of singlets to make a doublet), because a pair is automatically split into two singlets, so that the probability density of the ensemble is automatically an effective one-body density which obeys a \textit{true} one-body Fokker-Planck equation. Such an equation was exhaustively used by Coffey and co-workers (see Reference \citenum{Coffey2017Book} for a review) for both evaluating the linear and nonlinear dipolar responses to a uniform external field. We have, neglecting all time dependences (the structure of a solid is rigid, hence time-independent in a first approximation)
\begin{eqnarray}
\nabla_{\mathbf{u}}V_{1}^{eff}(\mathbf{u})=\nabla_{\mathbf{u}}V_{1}(\mathbf{u})+\int\nabla_{\mathbf{u}}U_{m}(\mathbf{u},\mathbf{u}')W(\mathbf{u}')g_{2}(\mathbf{u},\mathbf{u}')d\mathbf{u}'
\label{V1eff}
\end{eqnarray}
The approximation which follows, which is made in this context is clearly $g_{2}=1$ and $W(\mathbf{u}')=\delta(\mathbf{u}-\mathbf{u}')$, yielding, ($z=\cos\vartheta$),
\begin{eqnarray}
\frac{\partial\phi_{1}}{\partial z}=\left(\frac{\partial U_{m}}{\partial z}(z,z')\right)_{z'=z}
\end{eqnarray}
where $\phi_{1}=V_{1}^{eff}-V_1$, leading to
\begin{eqnarray}
\beta\phi_{1}(\mathbf{u})=-\frac{\gamma}{2}\cos^{2}\vartheta,
\end{eqnarray}
hence to
\begin{eqnarray}
\beta V_{1}^{eff}(\mathbf{u})=\beta V_{1}(\mathbf{u})-\frac{\gamma}{2}\cos^{2}\vartheta
\label{DebyeFrohlichPotential}
\end{eqnarray}
which is a plausible version of the Debye-Fr\"{o}hlich model with $\sigma=\gamma/2$. Notice that Eq.\eqref{DebyeFrohlichPotential} may also be formally transposed to yield the Stoner-Wohlfarth potential (see Reference \citenum{Coffey2017Book} for a discussion) by giving $\sigma$ another meaning which includes magnetoelastic and magnetocrystalline anisotropy together with local demagnetizing and surface anisotropy effects in one and the same model. In order to accomplish this, one only needs to define an effective $\gamma$ which, given our theoretical state of knowledge on magnetic particle assemblies, is largely sufficient at the present stage of development. 

\section{Appendix C : Derivation of Eq.\eqref{KirkwoodGkGeneral} and comparison with previous results}
In the first part of this Appendix, we derive Eq.\eqref{KirkwoodGkGeneral} while in the second part, we derive from Eq. \eqref{KirkwoodGkGeneral} previously obtained results and demonstrate that they are valid for weak densities only, independently of which weak density $G_2$ is used to compute $g_{\rm K}$.

\subsection{Derivation of Eq. \eqref{KirkwoodGkGeneral} from Eq.\eqref{eqDK}}

In order to achieve this, we need Eq.\eqref{RotationalDK} at equilibrium, where all quantities are time-independent. Thus we have \cite{Cugliandolo2015PRE}
\begin{eqnarray}
{{\nabla }_{\mathbf{u}}}\cdot \left[ {{\nabla }_{\mathbf{u}}}W\left( \mathbf{u} \right)+\beta W\left( \mathbf{u} \right){{\nabla }_{\mathbf{u}}}{{V}_{1}}\left( \mathbf{u} \right) \right]+\beta {{\nabla }_{\mathbf{u}}}\cdot \int{{{\nabla }_{\mathbf{u}}}{{U}_{m}}\left( \mathbf{u},\mathbf{{u}'} \right){{w}_{2}}\left( \mathbf{u},\mathbf{{u}'} \right)d\mathbf{{u}'}}=0
\end{eqnarray}
where $\mathbf{u}$ is a unit vector along a molecular dipole moment of constant magnitude $\mu$, $W(\mathbf{u})$ is the one-body orientational probability density, $V_{1}(\mathbf{u})=-\mu\mathbf{u}\cdot\mathbf{E}$ is a one-body potential containing the effect of the directing electric field $\mathbf{E}$ which is assumed uniform,  The mean polarization in the direction of the directing field is given by
\begin{eqnarray}
\nonumber
\left\langle \mathbf{P}\cdot \mathbf{e} \right\rangle ={{\rho }_{0}}\mu \int{\left( \mathbf{u}\cdot \mathbf{e} \right)W\left( \mathbf{u} \right)d\mathbf{u}}={{\rho }_{0}}\mu \left\langle {{P}_{1}}\left( \mathbf{u}\cdot \mathbf{e} \right) \right\rangle
\label{PolMicro}
\end{eqnarray}
where $\rho_0$ is the number of molecules per unit volume, $\mathbf{P}$ is the polarization of the sample, $\mathbf{e}$ is a unit vector along the directing field $\mathbf{E}$, $P_{1}(z)$ is the first Legendre polynomial and the angular brackets denote a statistical average over $W(\mathbf{u})$. On multiplying Eq.\eqref{eqDK} by the Legendre polynomial $P_{n}(\mathbf{u}\cdot\mathbf{e})$ of order $n$ and integrating the resulting equation on the unit sphere of representative points of a dipole with orientation $\mathbf{u}$, we arrive after some algebra at the set of moment equations
\begin{eqnarray}
n\left( n+1 \right)\left\langle {{P}_{n}} \right\rangle =\frac{n\left( n+1 \right)\xi }{2n+1}\left( \left\langle {{P}_{n-1}} \right\rangle -\left\langle {{P}_{n+1}} \right\rangle  \right)
-\beta \iint{{{\nabla }_{\mathbf{u}}}{{U}_{m}}\left( \mathbf{u},\mathbf{{u}'} \right)\cdot {{\nabla }_{\mathbf{u}}}{{P}_{n}}\left( \mathbf{u} \right){{w}_{2}}\left( \mathbf{u},\mathbf{{u}'} \right)d\mathbf{u}d\mathbf{{u}'}}
\label{GeneralMomentEqs}
\end{eqnarray}
where $\xi=\beta\mu E$ and where we have implicitly assumed that $\mathbf{E}$ is directed along the \textit{Z} axis of the laboratory frame, so that we may write $P_{n}(\mathbf{u}\cdot\mathbf{e})=P_{n}(\mathbf{u})$. In particular, for $n=1$ and $n=2$, Eqs.\eqref{GeneralMomentEqs} read respectively
\begin{eqnarray}
\left\langle {{P}_{1}} \right\rangle =\frac{\xi }{3}\left( 1-\langle {{P}_{2}}\rangle  \right)
-\frac{\beta }{2}\iint{{{\nabla }_{\mathbf{u}}}{{U}_{m}}\left( \mathbf{u},\mathbf{{u}'} \right)\cdot {{\nabla }_{\mathbf{u}}}{{P}_{1}}\left( \mathbf{u} \right){{w}_{2}}\left( \mathbf{u},\mathbf{{u}'} \right)d\mathbf{u}d\mathbf{{u}'}}
\label{P1Gen}
\end{eqnarray}
\begin{eqnarray}
\left\langle {{P}_{2}} \right\rangle &=&\frac{\xi }{5}\left( \left\langle {{P}_{1}} \right\rangle -\left\langle {{P}_{3}} \right\rangle  \right)
-\frac{\beta }{6}\iint{{{\nabla }_{\mathbf{u}}}{{U}_{m}}\left( \mathbf{u},\mathbf{{u}'} \right)\cdot {{\nabla }_{\mathbf{u}}}{{P}_{2}}\left( \mathbf{u} \right){{w}_{2}}\left( \mathbf{u},\mathbf{{u}'} \right)d\mathbf{u}d\mathbf{{u}'}}
\label{P2Gen}
\end{eqnarray}
Now, Eqs.\eqref{P1Gen} and \eqref{P2Gen} pertain to response to arbitrary order in the field strength. However, we are solely interested in linear response. Hence we expand the various quantities involved in these equations in powers of the field strength and retain linear terms only. Explicitly, on using the following expansions
\begin{eqnarray}
\left\langle {{P}_{n}} \right\rangle ={{\left\langle {{P}_{n}} \right\rangle }^{\left( 0 \right)}}+\xi {{\left\langle {{P}_{n}} \right\rangle }^{\left( 1 \right)}}+\ldots
\label{perturbPn}
\end{eqnarray}
\begin{eqnarray}
{{w}_{2}}\left( \mathbf{u},\mathbf{{u}'} \right)=w_{2}^{\left( 0 \right)}\left( \mathbf{u},\mathbf{{u}'} \right)+\xi w_{2}^{\left( 1 \right)}\left( \mathbf{u},\mathbf{{u}'} \right)+\ldots
\end{eqnarray}
Eqs.\eqref{P1Gen} and \eqref{P2Gen} become the perturbation equations
\begin{eqnarray}
\langle P_{1}\rangle^{(0)}=-\frac{\beta }{2}\iint{{{\nabla }_{\mathbf{u}}}{{U}_{m}}\left( \mathbf{u},\mathbf{{u}'} \right)\cdot {{\nabla }_{\mathbf{u}}}{{P}_{1}}\left( \mathbf{u} \right)w_{2}^{\left( 0 \right)}\left( \mathbf{u},\mathbf{{u}'} \right)d\mathbf{u}d\mathbf{{u}'}}=0
\label{P1zero}
\end{eqnarray}
which is the interaction torque balance equation in the absence of directing field at equilibrium,
\begin{eqnarray}
{{\left\langle {{P}_{2}} \right\rangle }^{\left( 0 \right)}}=-\frac{\beta }{6}\iint{{{\nabla }_{\mathbf{u}}}{{U}_{m}}\left( \mathbf{u},\mathbf{{u}'} \right)\cdot {{\nabla }_{\mathbf{u}}}{{P}_{2}}\left( \mathbf{u} \right)w_{2}^{\left( 0 \right)}\left( \mathbf{u},\mathbf{{u}'} \right)d\mathbf{u}d\mathbf{{u}'}}
\label{P2zero}
\end{eqnarray}
and
\begin{eqnarray}
\langle P_{1}\rangle^{(1)}=\frac{1}{3}\left( 1-{{\left\langle {{P}_{2}} \right\rangle }^{\left( 0 \right)}} \right)-\frac{\beta }{2}\iint{{{\nabla }_{\mathbf{u}}}{{U}_{m}}\left( \mathbf{u},\mathbf{{u}'} \right)\cdot {{\nabla }_{\mathbf{u}}}{{P}_{1}}\left( \mathbf{u} \right)w_{2}^{\left( 1 \right)}\left( \mathbf{u},\mathbf{{u}'} \right)d\mathbf{u}d\mathbf{{u}'}}.
\label{P11}
\end{eqnarray}
We can now use Eqs.\eqref{P2zero} and \eqref{P11} to obtain the equation
\begin{eqnarray}
\langle P_{1}(\mathbf{u})\rangle^{(1)}&=&\frac{1}{3}+\frac{\beta}{18}\iint\nabla_{\mathbf{u}}U_{m}(\mathbf{u},\mathbf{u}')\cdot\Phi(\mathbf{u},\mathbf{u}')d\mathbf{u}d\mathbf{u}',
\label{LinearP1}
\end{eqnarray}
where
\begin{eqnarray}
\Phi(\mathbf{u},\mathbf{u}')=w_{2}^{(0)}(\mathbf{u},\mathbf{u}')\nabla_{\mathbf{u}}P_{2}(\mathbf{u})-9w_{2}^{(1)}(\mathbf{u},\mathbf{u}')\nabla_{\mathbf{u}}P_{1}(\mathbf{u}).
\label{Phi01bis}
\end{eqnarray}
Now, on using Eqs. \eqref{PolMicro}, \eqref{perturbPn} with $n=1$,  \eqref{P1zero}, and equating the resulting linear microscopic polarization with $\varepsilon_{0}(\varepsilon-\varepsilon_{\infty})E_{M}$ ($E_{M}$ being the Maxwell field amplitude), we have
\begin{eqnarray}
\nonumber
\rho_{0}\mu\xi\langle P_{1}\rangle^{(1)}=\varepsilon_{0}(\varepsilon-\varepsilon_{\infty})E_{M}.
\end{eqnarray}
Finally, on combining the above equation with Eqs.\eqref{LinearP1} and \eqref{Phi01bis} we have
\begin{eqnarray}
(\varepsilon-\varepsilon_{\infty})\frac{\partial E_M}{\partial E}=\lambda\left(1+\frac{\beta}{6}\iint\nabla_{\mathbf{u}}U_{m}(\mathbf{u},\mathbf{u}')\cdot\Phi(\mathbf{u},\mathbf{u}')d\mathbf{u}d\mathbf{u}'\right)
\label{EqnStateDielectrics}
\end{eqnarray}
The derivative in the left-hand side of Eq.\eqref{EqnStateDielectrics} can be written using Fr\"{o}hlich's field \cite{Bottcher1973Book}, viz.
\begin{eqnarray}
\frac{\partial {{E}_{M}}}{\partial E}=\frac{2\varepsilon +{{\varepsilon }_{\infty }}}{3\varepsilon}
\label{FrohlichField}
\end{eqnarray}
By identification with Eq.\eqref{KFEquation} we obtain the following integral representation for $g_{\rm K}$, viz.
\begin{eqnarray}
\nonumber
{{g}_{\rm K}}=1+\frac{\beta }{6}\iint{{{\nabla }_{\mathbf{u}}}{{U}_{m}}(\mathbf{u},\mathbf{{u}'})\cdot\Phi(\mathbf{u},\mathbf{{u}'})d\mathbf{u}d\mathbf{{u}'}}
\end{eqnarray}
which is Eq.\eqref{KirkwoodGkGeneral}, where in the body text we have used the notation $W_{2}^{(0)}$ en lieu and in place of $w_{2}^{(0)}$.
\subsection{Comparison with previous results}
In this second part, we obtain a formula for $g_{\rm K}$ which was derived earlier by many authors (see for example References \citenum{Madden1984Book} and \citenum{Tani1983MolPhys}) which we demonstrate to be valid for weak densities only. This formula is \cite{Madden1984Book,Tani1983MolPhys}
\begin{eqnarray}
g_{\rm K}=1+\frac{\rho_{0}}{(4\pi)^2}\int(\mathbf{u}\cdot\mathbf{u}')G_{2}(\vec{\rho},\mathbf{u},\mathbf{u}')d\vec{\rho}d\mathbf{u}d\mathbf{u}'
\label{CurrentgK}
\end{eqnarray}
In the presence of the external field, we recall  that the static pair density $W_{2}$ is given by
\begin{eqnarray}
W_{2}(\vec{\rho},\mathbf{u},\mathbf{u}')=w(\mathbf{u})w(\mathbf{u}')G_{2}(\vec{\rho},\mathbf{u},\mathbf{u}')
\label{WW2Gen}
\end{eqnarray}
and since $G_{2}$ is related to intermolecular interactions, it is not affected by the external field. We write, in linear response
\begin{eqnarray}
w(\mathbf{u})=w^{(0)}(\mathbf{u})+\xi w^{(1)}(\mathbf{u}')
\label{WWpert}
\end{eqnarray}
By combining Eqs.\eqref{WW2Gen} and \eqref{WWpert} we may write
\begin{eqnarray}
W_{2}(\vec{\rho},\mathbf{u},\mathbf{u}')=W_{2}^{(0)}(\vec{\rho},\mathbf{u},\mathbf{u}')+\xi W_{2}^{(1)}(\vec{\rho},\mathbf{u},\mathbf{u}')
\label{WW2Pert}
\end{eqnarray}
where
\begin{eqnarray}
W_{2}^{(0)}(\vec{\rho},\mathbf{u},\mathbf{u}')&=&w^{(0)}(\mathbf{u})w^{(0)}(\mathbf{u}')G_{2}(\vec{\rho},\mathbf{u},\mathbf{u}')\\
W_{2}^{(1)}(\vec{\rho},\mathbf{u},\mathbf{u}')&=&(w^{(0)}(\mathbf{u})w^{(1)}(\mathbf{u}')+w^{(1)}(\mathbf{u})w^{(0)}(\mathbf{u}'))G_{2}(\vec{\rho},\mathbf{u},\mathbf{u}')
\label{WW20WW21}
\end{eqnarray}
so that, because $G_{2}=g_{\rho}g_{2}$, we have
\begin{eqnarray}
W_{2}^{(0)}(\vec{\rho},\mathbf{u},\mathbf{u}')&=&w_{2}^{(0)}(\mathbf{u},\mathbf{u}')g_{\rho}(\vec{\rho})\\
W_{2}^{(1)}(\vec{\rho},\mathbf{u},\mathbf{u}')&=&w_{2}^{(1)}(\mathbf{u},\mathbf{u}')g_{\rho}(\vec{\rho})
\label{WW20WW21Bis}
\end{eqnarray}
with the notations
\begin{eqnarray}
w_{2}^{(0)}(\mathbf{u},\mathbf{u}')&=&w^{(0)}(\mathbf{u})w^{(0)}(\mathbf{u}')g_{2}(\mathbf{u},\mathbf{u}')=\int W_{2}^{(0)}(\vec{\rho},\mathbf{u},\mathbf{u}')d\vec{\rho}\\
w_{2}^{(1)}(\mathbf{u},\mathbf{u}')&=&(w^{(0)}(\mathbf{u})w^{(1)}(\mathbf{u}')+w^{(1)}(\mathbf{u})w^{(0)}(\mathbf{u}'))g_{2}(\mathbf{u},\mathbf{u}')=\int W_{2}^{(1)}(\vec{\rho},\mathbf{u},\mathbf{u}')d\vec{\rho}
\end{eqnarray}
Now, we use (see Appendix A for a definition of $g_{\rho}$)
\begin{eqnarray}
\nabla_{\mathbf{u}}U_{m}(\mathbf{u},\mathbf{u}')=\int \nabla_{\mathbf{u}}U_{int}(\vec{\rho},\mathbf{u},\mathbf{u}')g_{\rho}(\vec{\rho})d\vec{\rho}
\label{DelUm22}
\end{eqnarray}
and use all the above equations in conjunction with Eq. \eqref{KirkwoodGkGeneral} and \eqref{Phi01bis} to obtain
\begin{eqnarray}
g_{\rm K}=1+\frac{\rho_{0}}{6}A_{1}-\frac{3\rho_{0}}{2}A_{2}
\label{gKInterm}
\end{eqnarray} 
with (here, we explicitly use $P_{n}(\mathbf{u})=P_{n}(\mathbf{u}\cdot\mathbf{e})$, where $\mathbf{e}$ is a unit vector along the $Z$ axis of the Laboratory frame)
\begin{eqnarray}
A_{1}&=&\beta\int\nabla_{\mathbf{u}}U_{int}(\vec{\rho},\mathbf{u},\mathbf{u}')\cdot\nabla_{\mathbf{u}}P_{2}(\mathbf{u}\cdot\mathbf{e})W_{2}^{(0)}(\vec{\rho},\mathbf{u},\mathbf{u}')d\vec{\rho}d\mathbf{u}d\mathbf{u}'\label{A1}\\
A_{2}&=&\beta\int\nabla_{\mathbf{u}}U_{int}(\vec{\rho},\mathbf{u},\mathbf{u}')\cdot\nabla_{\mathbf{u}}P_{1}(\mathbf{u}\cdot\mathbf{e})W_{2}^{(1)}(\vec{\rho},\mathbf{u},\mathbf{u}')d\vec{\rho}d\mathbf{u}d\mathbf{u}'\label{A2}
\end{eqnarray}
Now, we make \textit{the weak density approximation} which consists in first replacing $w^{(0)}$ and $w^{(1)}$ by their\textit{ Debye approximations} (see Appendix B) so that we have
\begin{eqnarray}
w^{(0)}(\mathbf{u})=\frac{1}{4\pi},\qquad
w^{(1)}(\mathbf{u})=\frac{\mathbf{u}\cdot\mathbf{e}}{4\pi},
\end{eqnarray}
so that Eqs.\eqref{WW20WW21} become
\begin{eqnarray}
W_{2}^{(0)}(\vec{\rho},\mathbf{u},\mathbf{u}')&\approx & \frac{G_{2}(\vec{\rho},\mathbf{u},\mathbf{u}')}{(4\pi)^{2}}\label{WW20WeakD}\\
W_{2}^{(1)}(\vec{\rho},\mathbf{u},\mathbf{u}')&\approx &\frac{(\mathbf{u}+\mathbf{u}')\cdot\mathbf{e}}{(4\pi)^{2}}G_{2}(\vec{\rho},\mathbf{u},\mathbf{u}')
\label{WW21WeakD}
\end{eqnarray}
Furthermore, we also have
\begin{eqnarray}
G_{2}(\vec{\rho},\mathbf{u},\mathbf{u}')=\exp(-\beta U_{int}(\vec{\rho},\mathbf{u},\mathbf{u}'))
\label{G2WeakDens}
\end{eqnarray}
Hence $A_1$ given by Eq. \eqref{A1} becomes
\begin{eqnarray}
\nonumber
A_{1}\approx -\frac{1}{(4\pi)^{2}}\int\nabla_{\mathbf{u}}G_{2}(\vec{\rho},\mathbf{u},\mathbf{u}')\cdot\nabla_{\mathbf{u}}P_{2}(\mathbf{u}\cdot\mathbf{e})d\vec{\rho}d\mathbf{u}d\mathbf{u}'
\end{eqnarray}
which, by use of Gauss's theorem and $\nabla_{\mathbf{u}}^{2}P_{2}(\mathbf{u}\cdot\mathbf{e})=-6P_{2}(\mathbf{u}\cdot\mathbf{e})$, becomes
\begin{eqnarray}
A_{1}\approx -\frac{6}{(4\pi)^{2}}\int P_{2}(\mathbf{u}\cdot\mathbf{e})G_{2}(\vec{\rho},\mathbf{u},\mathbf{u}')d\vec{\rho}d\mathbf{u}d\mathbf{u}'
\label{A1Approx}
\end{eqnarray}
Using exactly the same procedures, $A_{2}$ in the weak density approximation becomes
\begin{eqnarray}
A_{2}\approx -\frac{2}{(4\pi)^{2}}\int (P_{2}(\mathbf{u}\cdot\mathbf{e})+(\mathbf{u}\cdot\mathbf{e})(\mathbf{u}'\cdot\mathbf{e}))G_{2}(\vec{\rho},\mathbf{u},\mathbf{u}')d\vec{\rho}d\mathbf{u}d\mathbf{u}'
\label{A2Approx}
\end{eqnarray}
By combining Eq. \eqref{gKInterm} with Eqs.\eqref{A1Approx} and \eqref{A2Approx}, we have
\begin{eqnarray}
g_{\rm K}\approx 1+\frac{3}{(4\pi)^{2}}\int\left((\mathbf{u}\cdot\mathbf{e})(\mathbf{u}'\cdot\mathbf{e})+\frac{2}{3}P_{2}(\mathbf{u}\cdot\mathbf{e})\right)G_{2}(\vec{\rho},\mathbf{u},\mathbf{u}')d\vec{\rho}d\mathbf{u}d\mathbf{u}'
\end{eqnarray}
which, on using as is usual
\begin{eqnarray}
(\mathbf{u}\cdot\mathbf{e})(\mathbf{u}'\cdot\mathbf{e})=\frac{1}{3}\mathbf{u}\cdot\mathbf{u}'
\end{eqnarray}
(where in this equation the "$=$" sign means "has the same effect as")  becomes
\begin{eqnarray}
g_{\rm K}\approx 1+\frac{\rho_{0}}{(4\pi)^{2}}\int\left(\mathbf{u}\cdot\mathbf{u}'+2 P_{2}(\mathbf{u}\cdot\mathbf{e})\right)G_{2}(\vec{\rho},\mathbf{u},\mathbf{u}')d\vec{\rho}d\mathbf{u}d\mathbf{u}'
\end{eqnarray}
A further expansion of $G_{2}$ in rotational invariants\cite{Stell1981Book} leads to the conclusion that the $P_{2}$ term cancels since the pair correlation function has no $h_{200}$ component, hence finally leading to
\begin{eqnarray}
g_{\rm K}\approx 1+\frac{\rho_{0}}{(4\pi)^{2}}\int\left(\mathbf{u}\cdot\mathbf{u}'\right)G_{2}(\vec{\rho},\mathbf{u},\mathbf{u}')d\vec{\rho}d\mathbf{u}d\mathbf{u}'
\end{eqnarray}
which is \textit{a weak density estimate} of $g_{\rm K}$ for \textit{purely polar assemblies} because we have assumed that Eqs. \eqref{WW20WeakD} and \eqref{WW21WeakD} hold.
\section{Appendix D : Dipole-dipole interactions for purely polar molecules}
Here, we derive the form of $U_m$ arising from pure dipole-dipole interactions, where we assume that all densities are normalized to unity according to the theory we described in Appendix A and where no commutation problem in the order of integration arise in the moment integrals. Thus, the dipole-dipole interaction for a pair of identical molecules is
\begin{eqnarray}
U_{int}(\vec{\rho},\mathbf{u},\mathbf{u}')=\frac{\mu^{2}}{4\pi\varepsilon_{0}\rho^{3}}\mathbf{u}\cdot\mathbf{T}(\hat{\rho})\cdot\mathbf{u}'
\label{DipoleDipoleInt}
\end{eqnarray}
where $\mathbf{T}(\hat{\rho})$ is a tensor that can be written in dyadic form as follows
\begin{eqnarray}
\mathbf{T}(\hat{\rho})=\mathbf{I}-3\hat{\rho}\hat{\rho},
\label{T}
\end{eqnarray}
$\mathbf{I}$ is the unit tensor and $\hat{\rho}$ is a unit vector along $\vec{\rho}$. We recall the definition of $U_{m}$ in its time-independent version, viz.
\begin{eqnarray}
U_{m}(\mathbf{u},\mathbf{u}')=\int_{R_H}^{r_C}\int_{0}^{\pi}\int_{0}^{2\pi} U_{int}(\vec{\rho},\mathbf{u},\mathbf{u}')g_{\rho}(\vec{\rho})\rho^{2}\sin\vartheta_{\rho}d\rho d\vartheta_{\rho}d\varphi_{\rho}
\label{UmDDIStatic}
\end{eqnarray}
where $(\vartheta_{\rho},\varphi_{\rho})$ are the spherical polar angles specifying the orientation of $\vec{\rho}$. We expand $g_{\rho}$ in spherical harmonics so that
\begin{eqnarray}
g_{\rho}(\vec{\rho})=\sum_{n=0}^{\infty}\sum_{m=-n}^{n}g_{\rho}^{n,m}(\rho)Y_{n,m}(\hat{\rho})
\end{eqnarray}
so that Eq.\eqref{UmDDIStatic} becomes
\begin{eqnarray}
U_{m}(\mathbf{u},\mathbf{u}')&=&\frac{\mu^{2}}{4\pi\varepsilon_{0}}\mathbf{u}\cdot\sum_{n=0}^{\infty}\sum_{m=-n}^{n}\int_{R_H}^{r_C}d\rho\frac{g_{\rho}^{n,m}(\rho)}{\rho}\int_{0}^{\pi}\int_{0}^{2\pi} \mathbf{T}(\hat{\rho})Y_{n,m}(\hat{\rho})\sin\vartheta_{\rho}d\vartheta_{\rho}d\varphi_{\rho}\cdot\mathbf{u}'
\label{UmDDIStatic2}
\end{eqnarray}
It is obvious that only $n=2$ terms will contribute to the double sum in Eq.\eqref{UmDDIStatic2}. Therefore, we have
\begin{eqnarray}
U_{m}(\mathbf{u},\mathbf{u}')&=&\frac{\mu^{2}}{4\pi\varepsilon_{0}}\mathbf{u}\cdot\sum_{m=-2}^{2}\int_{R_H}^{r_C}d\rho\frac{g_{\rho}^{2,m}(\rho)}{\rho}\int_{0}^{\pi}\int_{0}^{2\pi} \mathbf{T}(\hat{\rho})Y_{2,m}(\hat{\rho})\sin\vartheta_{\rho}d\vartheta_{\rho}d\varphi_{\rho}\cdot\mathbf{u}'
\label{UmDDIStatic3}
\end{eqnarray}
Because $g_{\rho}$ is real, we must have
\begin{eqnarray}
g_{\rho}^{n,-m}(\rho)=(-1)^{m}\bar{g}_{\rho}^{n,m}(\rho)
\end{eqnarray}
where in this last equation only the overbar denotes the complex conjugate. Separating real and imaginary parts in $g_{\rho}^{n,m}(\rho)$, we write
\begin{eqnarray}
g_{\rho}^{n,m}(\rho)=g_{\rho}^{'(n,m)}(\rho)+ig_{\rho}^{''(n,m)}(\rho)
\end{eqnarray} 
so that Eq. \eqref{UmDDIStatic3} reads
\begin{eqnarray}
\nonumber
U_{m}(\mathbf{u},\mathbf{u}')=\frac{\mu^{2}}{4\pi\varepsilon_{0}}(-4\sqrt{\frac{\pi}{5}}G_{20}(u_{Z}u'_{Z}-\frac{1}{2}(u_{X}u'_{X}+u_{Y}u'_{Y}))+2\sqrt{\frac{6\pi}{5}}G'_{21}(u_{X}u'_{Z}+u_{Z}u'_{X})\\
\nonumber
-2\sqrt{\frac{6\pi}{5}}G''_{21}(u_{Y}u'_{Z}+u_{Z}u'_{Y})-2\sqrt{\frac{6\pi}{5}}G'_{22}(u_{X}u'_{X}+u_{Y}u'_{Y})+2\sqrt{\frac{6\pi}{5}}G''_{22}(u_{X}u'_{X}+u_{Y}u'_{Y}))
\label{UmIntegrated}
\end{eqnarray}
where 
\begin{eqnarray}
G'_{nm}&=&\int_{R_H}^{r_c}g_{\rho}^{'(n,m)}(\rho)d(\ln\rho)\\
G''_{nm}&=&\int_{R_H}^{r_c}g_{\rho}^{''(n,m)}(\rho)d(\ln\rho)
\label{GprimeGsecond}
\end{eqnarray}
are both real \textit{converging} integrals for $R_H$ and $r_C$ nonvanishing, positive and finite. 

Now, we come to modelling dipole-dipole interactions, because, we insist, solving the general problem generated by Eqs. \eqref{FPEGrho}, \eqref{RotationalDK} and \eqref{RotationalDK2} is a very difficult one. In order to model dipole-dipole interactions, let us consider the terms in Eq. \eqref{UmIntegrated}. To this purpose, we assume that the directing electric field is applied along the $Z$ axis of the laboratory frame so that $\mathbf{e}=\hat{Z}$. We largely anticipate below that $U_m$ as given by Eq. \eqref{UmIntegrated} will occur under the integral sign in Eq. \eqref{KirkwoodGkGeneral} so that the "$=$" in the following equations mean "has the same effect as", and does not represent an equality in the strict mathematical sense. The term proportional to $G_{20}$ may be written ($\mathbf{u}\cdot\mathbf{e}=u_{Z}$) 
\begin{eqnarray}
u_{Z}u'_{Z}-\frac{1}{2}(u_{X}u'_{X}+u_{Y}u'_{Y})&=&(\mathbf{u}\cdot\mathbf{e})(\mathbf{u}'\cdot\mathbf{e})-\frac{1}{2}(\mathbf{u}\cdot\mathbf{u}'-(\mathbf{u}\cdot\mathbf{e})(\mathbf{u}'\cdot\mathbf{e}))\\
&=&\frac{3}{2}(\mathbf{u}\cdot\mathbf{e})(\mathbf{u}'\cdot\mathbf{e})-\frac{1}{2}(\mathbf{u}\cdot\mathbf{u}')\\
&=&\frac{3}{2}(\mathbf{u}\cdot\mathbf{e})(\mathbf{u}'\cdot\mathbf{e})-\frac{3}{2}(\mathbf{u}\cdot\mathbf{e})(\mathbf{u}'\cdot\mathbf{e})\\
&=& 0.
\label{LongCont}
\end{eqnarray}
Hence, the term proportional to $G_{20}$ does not contribute to the dielectric constant. The terms proportional to $G'_{21}$ and $G''_{21}$ involve products between different dipole components. Since the field is directed along the $Z$ axis, these terms have no effect. Therefore these terms drop out as they are useless. Most interesting are the terms proportional to  $G'_{22}$ and $G''_{22}$. Both invoke dipolar components that are perpendicular to the externally applied electric field. With the same meaning for the "$=$" sign as in Eq. \eqref{LongCont}, we have
\begin{eqnarray}
\nonumber
u_{X}u'_{X}+u_{Y}u'_{Y}&=&\mathbf{u}\cdot\mathbf{u}'-(\mathbf{u}\cdot\mathbf{e})(\mathbf{u}'\cdot\mathbf{e})\\
&=&3(\mathbf{u}\cdot\mathbf{e})(\mathbf{u}'\cdot\mathbf{e})-(\mathbf{u}\cdot\mathbf{e})(\mathbf{u}'\cdot\mathbf{e})\\
&=&2(\mathbf{u}\cdot\mathbf{e})(\mathbf{u}'\cdot\mathbf{e})
\end{eqnarray}
so that $U_{m}$ becomes, using all the above results (and restoring the usual meaning for the "$=$" sign)
\begin{eqnarray}
U_{m}(\mathbf{u},\mathbf{u}')=\frac{\mu^{2}}{\varepsilon_{0}}\sqrt{\frac{6}{5\pi}}(\mathbf{u}\cdot\mathbf{e})(\mathbf{u}'\cdot\mathbf{e})\int_{R_H}^{r_C}(g_{\rho}^{''(2,2)}(\rho)-g_{\rho}^{'(2,2)}(\rho))d(\ln\rho)
\label{UmCrucial}
\end{eqnarray}
We may compare this expression with Berne's equation for $U_{m}$ which is \cite{Berne1975JCP,Dejardin2011JAP,Dejardin2014JCP}
\begin{eqnarray}
U_{m}(\mathbf{u},\mathbf{u}')=\pm\frac{\rho_{0}\mu^{2}}{3\varepsilon_{0}}(\mathbf{u}\cdot\mathbf{e})(\mathbf{u}'\cdot\mathbf{e}),
\label{BerneUm}
\end{eqnarray}
the $\pm$ sign being present because the integral in Eq.\eqref{UmCrucial} may be positive or negative, depending on the microstructure of the liquid \cite{Dejardin2018JCP}. This leads to a definition of $\rho_{0}$, namely
\begin{eqnarray}
\rho_{0}=\sqrt{\frac{54}{5\pi}}\;\left\vert\int_{R_{H}}^{r_{C}}(g_{\rho}^{''(2,2)}(\rho)-g_{\rho}^{'(2,2)}(\rho))d(\ln\rho)\right\vert
\label{Rho0Def}
\end{eqnarray} 
Whether Eq.\eqref{Rho0Def} can be checked against experiment depends upon solving Eq. \eqref{FPEGrho}, which, as we have already alluded to, is an outstandingly difficult task. Furthermore, we can easily argue with Eq.\eqref{Rho0Def} why actually $\rho_{0}$ is temperature-dependent (so that the volume of the cavity is also temperature-dependent). This actually happens for two reasons : 
\begin{itemize}
\item $r_{C}$ being the (finite) correlation length of density correlations, it depends on $g_{\rho}$, which a priori makes it \textit{a priori} a temperature-dependent parameter since $g_{\rho}$ is temperature-dependent as provided by Eq.\eqref{FPEGrho}, 
\item No reliable calculation of the temperature dependence of the mass density $M_{v}(T)$ of any body, had it been solid, liquid or gaseous has ever been achieved in the static regime from first principles, and the precise technique for achieving this task is yet unknown. Since $\rho_{0}$ can furthermore be easily related to the experimental temperature-dependent $M_{v}(T)$, the molar mass of a constituent molecule and Avogadro's number, it is much easier to use empirical formulas for representing this quantity. This justifies the employment of $a_{1}=\pm\lambda$ for the leading term in the expansion of $U_{m}$ in terms of direction cosines of the dipole pairs as a model potential for dipole-dipole interactions. 
\end{itemize}
The "quadrupolar term" we consider (i.e. the term proportional to $a_2$) loosely corresponds to induction, (i.e. dipole-induced dipole interactions), and even more loosely to dispersion (i.e. induced dipole-induced dipole interactions, see the Maier-Saupe theory of nematics in Appendix B or in Reference \citenum{Chandrasekhar1973Book}) which are "quadrupolar" effects (this is definitely so for the Maier-Saupe model, where only dispersion is present\cite{Chandrasekhar1973Book}). The problem with these interactions is that although being pairwise, they are not pairwise additive \cite{Stone2013Book}. This is the reason why we set $a_{2}=\kappa a_{1}$ : in effect, this is the easiest way to experimentally understand how intermolecular interactions deviate from pure dipole-dipole interactions in the context of our theory.  
\section{Appendix E : Decorrelation of the third body}

In this Appendix we detail the decorrelation procedure followed by D\'{e}jardin et al. \cite{Dejardin2019PRB} in order to compute $V_{2}^\text{eff}$ from $U_m$ as a lot of mathematical-physical tricks are only vaguely alluded to in their paper. 

\begin{figure}[h!]
\centering
\includegraphics[height=7cm]{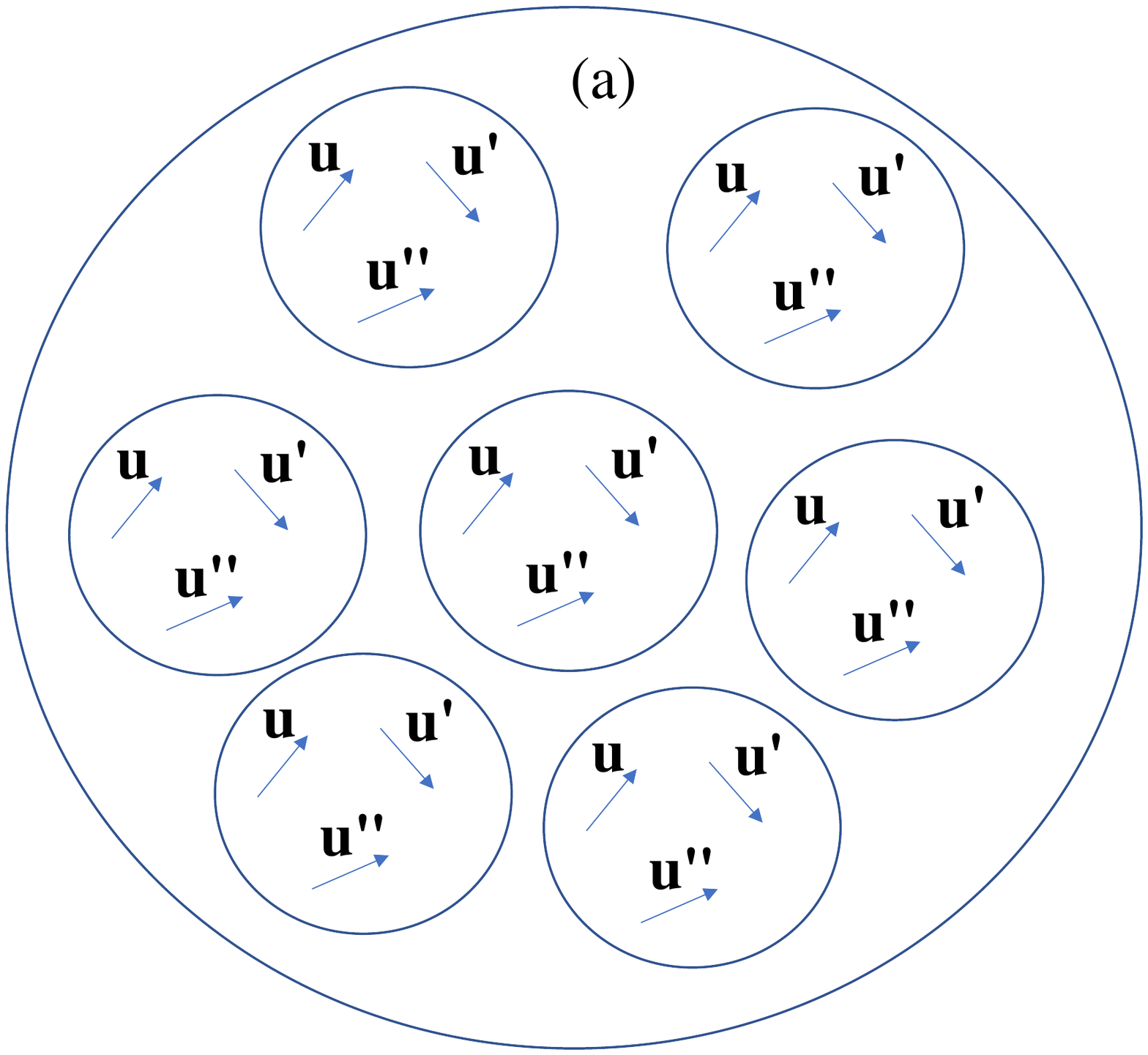}
\caption{Decorrelation step (a) -step $0$ -: A schematic representation of a statistical ensemble made of samples of identical interacting three dipoles.} 
\label{DecStepA}
\end{figure}

\begin{figure}[h!]
\centering
\includegraphics[height=7cm]{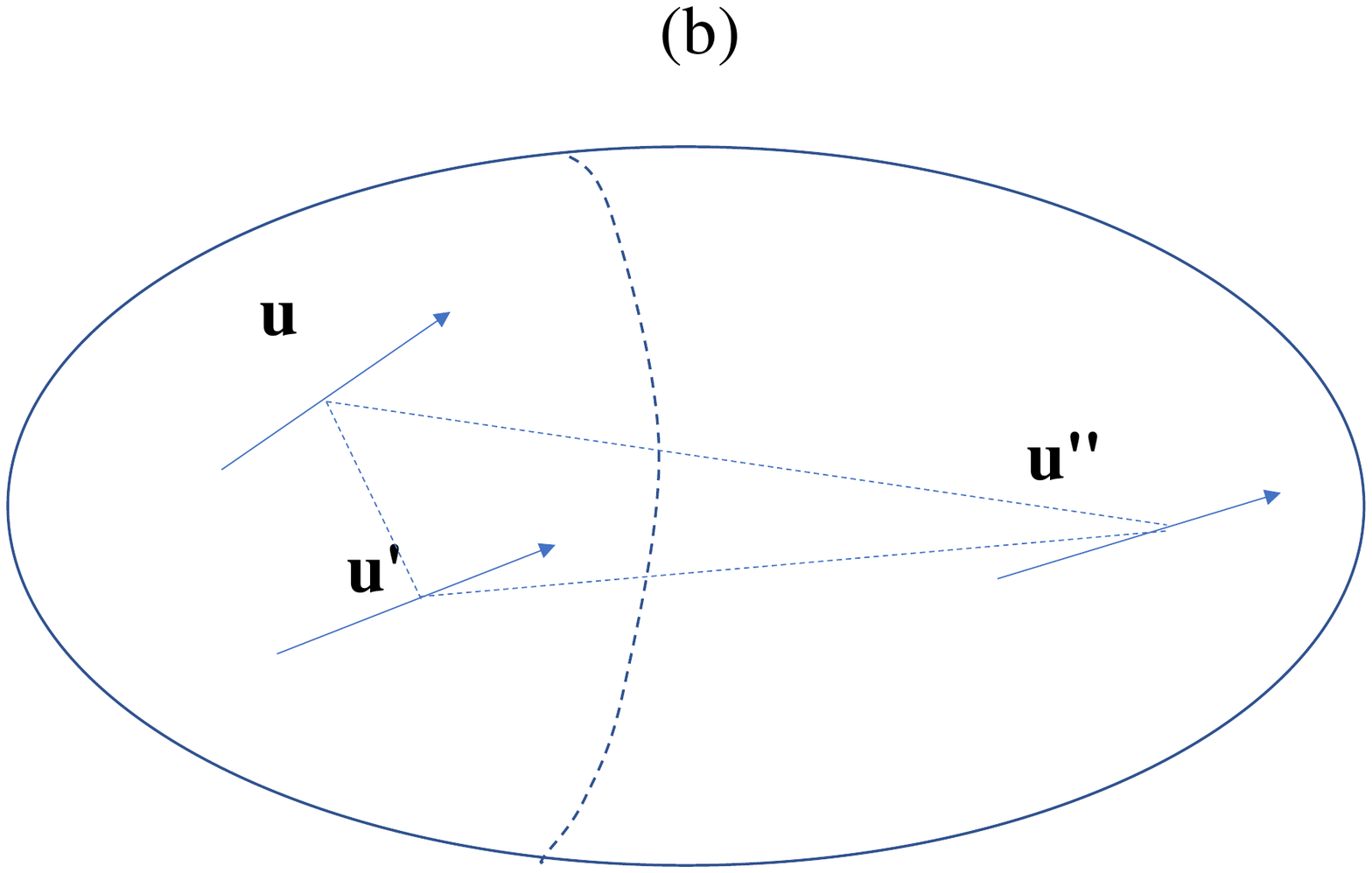}
\caption{Decorrelation step (b) : in each statistical representative of the ensemble, the dipole with orientation $\mathbf{u}''$ is taken away from the influence of the interacting pair with orientations $(\mathbf{u},\mathbf{u}')$, splitting the statistical representative into a doublet and a singlet.} 
\label{DecStepB}
\end{figure}

It was implicitly suggested by D\'{e}jardin et al. \cite{Dejardin2019PRB} that even if it is \textit{a priori} impossible to truncate the BBGKY (for Bogolyubov-Born-Green-Kirkwood-Yvon) hierarchical process, \cite{Hansen2006Book} yet it is possible to \textit{decide} to \textit{stop} the aforementioned process at rank $p$ by making appropriate hypotheses concerning the $(p+1)$-body orientational density, in particular by analyzing all partial densities as \textit{probability densities}. They further showed that at rank $p=2$, the corresponding equilibrium BBGKY member could be mapped upon an equilibrium two-body Smoluchowski equation after the \textit{decorrelation of the third body} from the two others has been achieved. To this aim, they introduced the Kirkwood potential of mean torques $V_{2}^\text{eff}$ having the essential (but vague) constraint of grossly (if not exactly) describing the same physics as that contained in $U_m$. This means in particular that the location and nature of the stationary points of $V_{2}^\text{eff}$ should approximately, if not exactly, be the same as those of $U_{m}$. Of course, this implies that $V_{2}^\text{eff}$ must have the essential property of global rotational invariance, viz.
\begin{eqnarray}
V_{2}^\text{eff}(-\mathbf{u},-\mathbf{u}')=V_{2}^\text{eff}(\mathbf{u},\mathbf{u}')
\end{eqnarray}
Furthermore, it has been shown by D\'{e}jardin et al. \cite{Dejardin2019PRB} that
\begin{eqnarray}
V_{2}^\text{eff}(\mathbf{u},\mathbf{u}')=U_{m}(\mathbf{u},\mathbf{u}')+V_{c}(\mathbf{u},\mathbf{u}')
\end{eqnarray}
where the complementary term $V_c$ obeys the \textit{exact} partial differential equations
\begin{eqnarray}
\nabla_{\mathbf{u}}V_{c}(\mathbf{u},\mathbf{u}')=
\int \nabla_{\mathbf{u}}U_{m}(\mathbf{u},\mathbf{u}'')W^{(0)}(\mathbf{u}'')\frac{g_{3}(\mathbf{u},\mathbf{u}',\mathbf{u}'')}{g(\mathbf{u},\mathbf{u}')}d\mathbf{u}''
\label{Vc1Eq}
\end{eqnarray}
and
\begin{eqnarray}
\nabla_{\mathbf{u}'}V_{c}(\mathbf{u},\mathbf{u}')=\int \nabla_{\mathbf{u}'}U_{m}(\mathbf{u}',\mathbf{u}'')W^{(0)}(\mathbf{u}'')\frac{g_{3}(\mathbf{u},\mathbf{u}',\mathbf{u}'')}{g(\mathbf{u},\mathbf{u}')}d\mathbf{u}''
\label{Vc2Eq}
\end{eqnarray}
where $g_3$ denotes the orientational three-body distribution function, $g$ is the two-body distribution function and $W^{(0)}$ obeys Eq. \eqref{eqDK} with $V_{1}=0$. From the global rotational invariance of $V_c$, we must have
\begin{eqnarray}
g_{3}(-\mathbf{u},-\mathbf{u}',\mathbf{u}'')=g_{3}(\mathbf{u},\mathbf{u}',\mathbf{u}'')
\label{RotInvg3}
\end{eqnarray}
and indeed, 
\begin{eqnarray}
g(-\mathbf{u},-\mathbf{u}')=g(\mathbf{u},\mathbf{u}')
\label{RotInvg2}
\end{eqnarray}
\textit{provided} $\mathbf{u}$ and $\mathbf{u}'$ \textit{are describing the orientation of a pair of molecules belonging to the same representative sample of a statistical ensemble}. 
Next, since generally $U_m$ consists of a superposition of $n$ terms, we can write with obvious notations
\begin{eqnarray}
U_{m}(\mathbf{u},\mathbf{u}')=\sum_{i=1}^{n}U_{m}^{(i)}(\mathbf{u},\mathbf{u}')
\end{eqnarray}
so that we can also postulate that $V_c$ also consists of a superposition of $n$ terms, viz.
\begin{eqnarray}
V_{c}(\mathbf{u},\mathbf{u}')=\sum_{i=1}^{n}V_{c}^{(i)}(\mathbf{u},\mathbf{u}')
\end{eqnarray}
where
\begin{eqnarray}
\nabla_{\mathbf{u}}V_{c}^{(i)}(\mathbf{u},\mathbf{u}')=
\int \nabla_{\mathbf{u}}U_{m}^{(i)}(\mathbf{u},\mathbf{u}'')W^{(0)}(\mathbf{u}'')\frac{g_{3}(\mathbf{u},\mathbf{u}',\mathbf{u}'')}{g(\mathbf{u},\mathbf{u}')}d\mathbf{u}''
\label{VcI1Eq}
\end{eqnarray}
and
\begin{eqnarray}
\nabla_{\mathbf{u}'}V_{c}^{(i)}(\mathbf{u},\mathbf{u}')=\int \nabla_{\mathbf{u}'}U_{m}^{(i)}(\mathbf{u}',\mathbf{u}'')W^{(0)}(\mathbf{u}'')\frac{g_{3}(\mathbf{u},\mathbf{u}',\mathbf{u}'')}{g(\mathbf{u},\mathbf{u}')}d\mathbf{u}''
\label{VcI2Eq}
\end{eqnarray}
It indeed follows that both $V_c$ and $V_{2}^\text{eff}$ can be determined term by term. The statistical ensemble we are interested in is therefore one made of identical samples of three interacting bodies, i.e., the one depicted in Figure \ref{DecStepA}. 
Then the decorrelation procedure consists in writing the Kirkwood superposition approximation (KSA) for $g_3$ and examining further some of the consequences of this approximation. Thus, the KSA is \cite{Hansen2006Book}
\begin{eqnarray}
g_{3}(\mathbf{u},\mathbf{u}',\mathbf{u}'')\approx g(\mathbf{u},\mathbf{u}')g(\mathbf{u}',\mathbf{u}'')g(\mathbf{u},\mathbf{u}'')
\label{KirkSupApprox}
\end{eqnarray}
This superposition approximation is the result of approximating the three-body interaction potential as a superposition of pair interactions, due to the impossibility to even write an exact analytical expression for this three-body interaction.  It has several immediate consequences, the first one being
\begin{eqnarray}
g_{3}(-\mathbf{u},-\mathbf{u}',\mathbf{u}'')\approx g(\mathbf{u},\mathbf{u}')g(-\mathbf{u}',\mathbf{u}'')g(-\mathbf{u},\mathbf{u}'')
\label{g3Prop1}
\end{eqnarray}
This last equation is difficult to interpret in statistical and physical terms because $g(-\mathbf{u},\mathbf{u}'')$ and $g(-\mathbf{u}',\mathbf{u}'')$ are a priori unknown \textit{if no extra hypothesis} is made on the behavior of the third body. One reasonable hypothesis is to \textit{assume} that a dipole with orientation $\mathbf{u}''$ is substracted from the influence of a pair of dipoles with orientations $(\mathbf{u},\mathbf{u}')$ so that a representative triplet of the initial ensemble depicted in Figure \ref{DecStepA} $(\mathbf{u},\mathbf{u}',\mathbf{u}'')$ is \emph{split} into a doublet $(\mathbf{u},\mathbf{u}')$ and a singlet $(\mathbf{u}'')$. This is depicted in Figure \ref{DecStepB}.

\begin{figure}[h!]
\centering
\includegraphics[height=7cm]{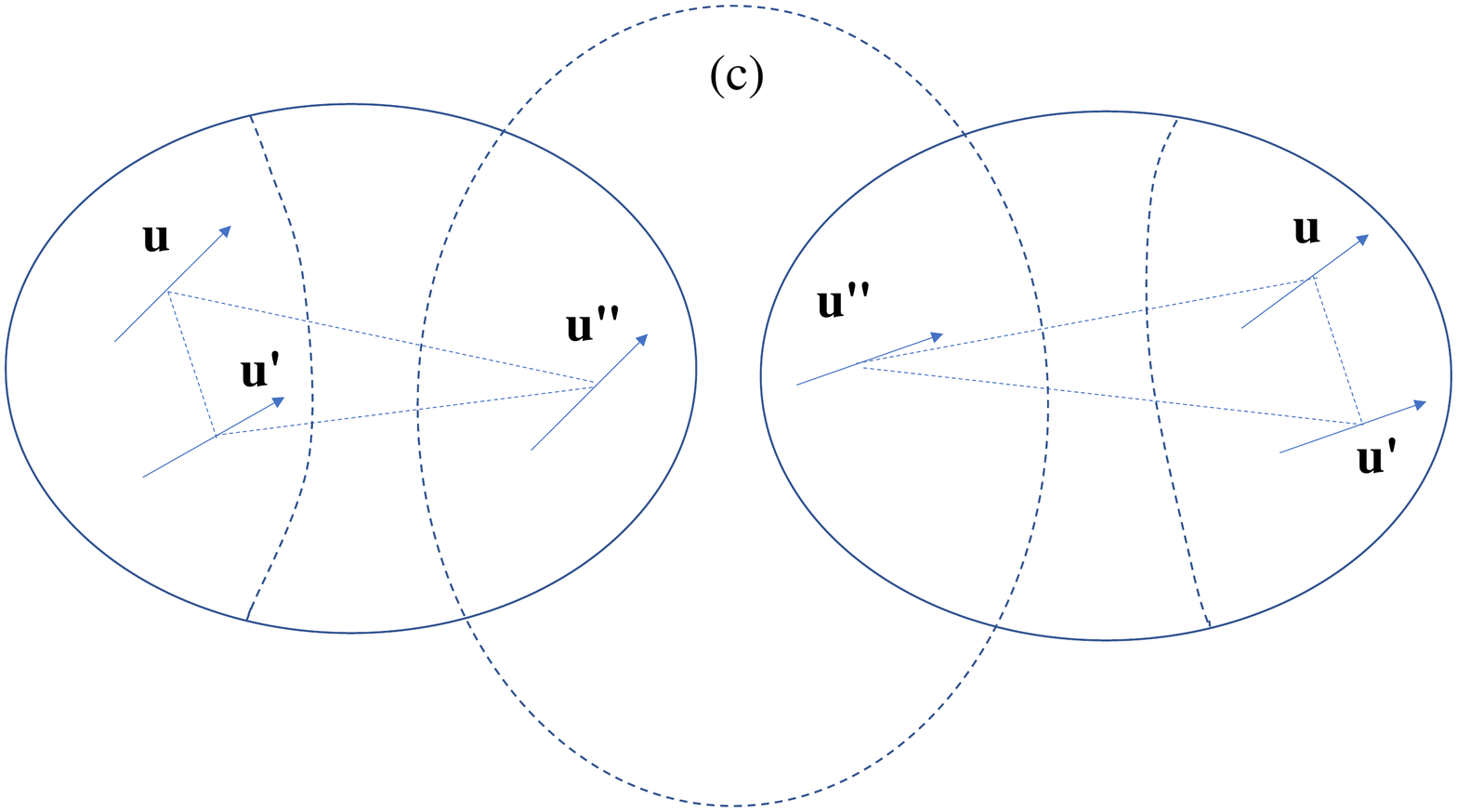}
\caption{Decorrelation step (c) : the dipole with orientation $\mathbf{u}''$ in one sample is under the influence of the dipole with orientation $\mathbf{u}''$ of an other statistical sample. These two dipoles form a pair with orientations $(\mathbf{u},\mathbf{u}')$ so that two triplets of interacting bodies form three pairs of interacting bodies.} 
\label{DecStepC}
\end{figure}

At this stage, the newly generated ensemble is no longer an ensemble onto which a \textit{unique} probability density can be defined so easily, because this new ensemble is not made of \emph{identical} non-interacting representatives, so that statistics are not easy on such an ensemble, or impossible. In order to recover a \textit{usual} statistical ensemble from this newly generated one, the sole chance we have is to consider \textit{pairs of split triplets} a typical pair of which is such that for one sample $\mathbf{u}''$ will coincide with $\mathbf{u}$ and in the other $\mathbf{u}''$ will coincide with $\mathbf{u}'$. This decomposition is partly depicted in Figure \ref{DecStepC}. 

\begin{figure}[h!]
\centering
\includegraphics[height=7cm]{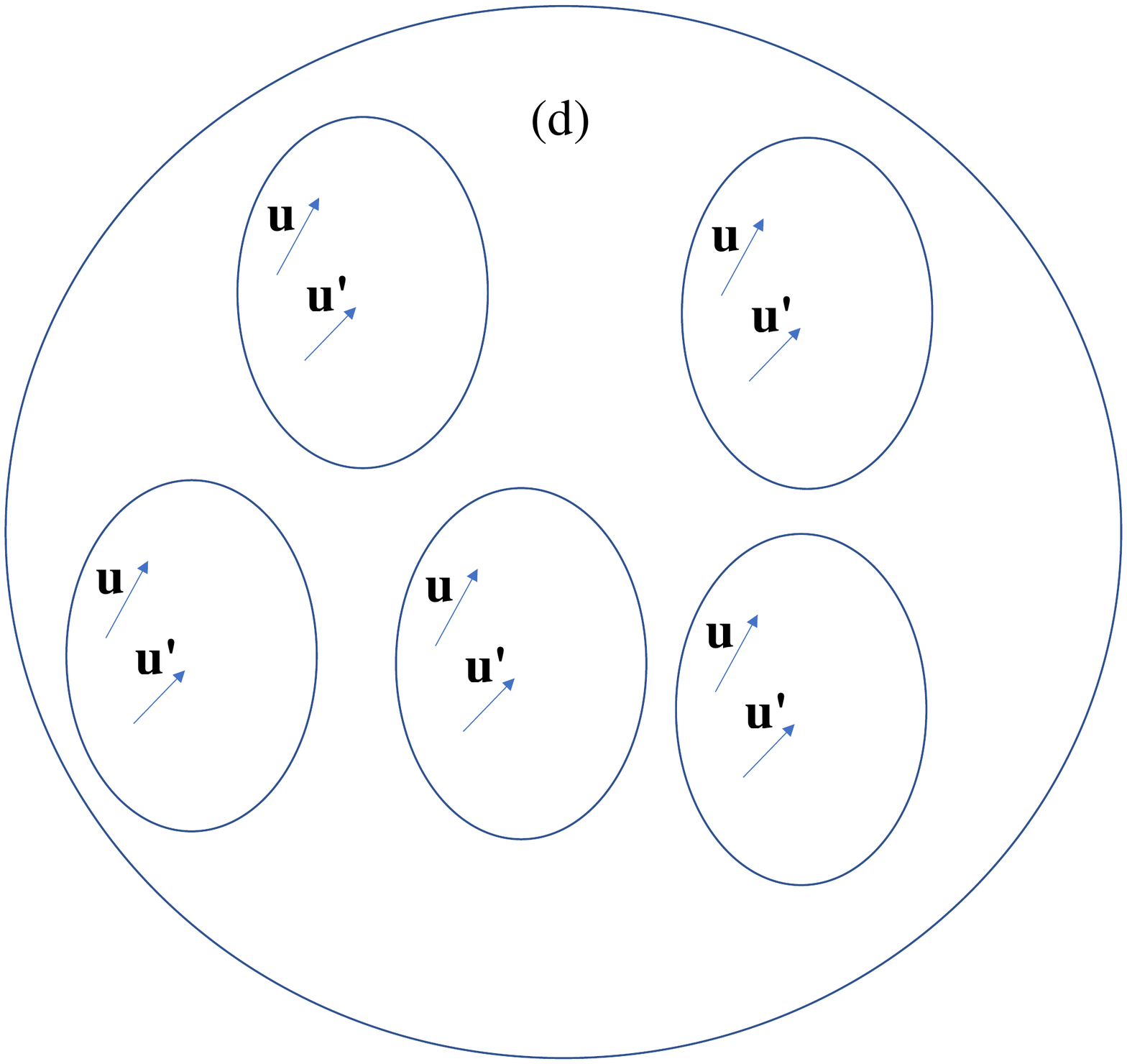}
\caption{Decorrelation step (d): the statistical ensemble made of interacting triplets has become an effective statistical ensemble made of pairs interacting via the effective pair potential $V_{2}^\text{eff}$.} 
\label{DecStepD}
\end{figure}

Hence a pair of triplets interacting through some three-body interaction effectively form three identical doublets interacting through an effetive two-body interaction $V_{2}^\text{eff}$. The structure of this final newly formed ensemble is depicted in Figure \ref{DecStepD}. 

This implies three important consequences. First, now the ensemble is made of identical representative samples of interacting pairs through $V_{2}^\text{eff}$ and each sample does not interact with one another, so that statistics \textit{in the sense of usual statistical mechanics can be made}. This consequence is extremely important for further purposes. The second one may be expressed via the two equations
\begin{eqnarray}
g(\mathbf{u},\mathbf{u}'')=1
\end{eqnarray}
if $(\mathbf{u})$ and $(\mathbf{u}'')$ are orientations of dipoles which do not belong to the same representative of the newly formed statistical ensemble, and 
\begin{eqnarray}
g(\mathbf{u},\mathbf{u}'')=e^{-\beta U_{m}(\mathbf{u},\mathbf{u}'')}\neq 1
\end{eqnarray}
if $(\mathbf{u})$ and $(\mathbf{u}'')$ are orientations belonging to the same representative of the newly formed statistical ensemble. Using the KSA and the two above equations, then we have
\begin{eqnarray}
\nonumber
\nabla_{\mathbf{u}}V_{c}^{(i)}(\mathbf{u},\mathbf{u}')\approx\int\nabla_{\mathbf{u}}U_{m}^{(i)}(\mathbf{u},\mathbf{u}'')W^{(0)}(\mathbf{u}'')d\mathbf{u}''\\
\label{Vc1EqBis}
\end{eqnarray}
and
\begin{eqnarray}
\nonumber
\nabla_{\mathbf{u}'}V_{c}^{(i)}(\mathbf{u},\mathbf{u}')\approx\int\nabla_{\mathbf{u}'}U_{m}^{(i)}(\mathbf{u}',\mathbf{u}'')W^{(0)}(\mathbf{u}'')d\mathbf{u}''\\
\label{Vc2EqBis}
\end{eqnarray}
so that \textit{in the KSA}, $V_{c}^{(i)}$ can be written down as a sum of single-body potentials $U_{an}^{(i)}$, viz.
\begin{eqnarray}
V_{c}^{(i)}(\mathbf{u},\mathbf{u}')=U_{an}^{(i)}(\mathbf{u})+U_{an}^{(i)}(\mathbf{u}')
\end{eqnarray}
since Eqs. \eqref{Vc1EqBis} and \eqref{Vc2EqBis} have identical mathematical form. Because $V_c$ has the global rotational invariance property, it immediately follows that $U_{an}^{(i)}$ is an even function of $\mathbf{u}$, viz.
\begin{eqnarray}
U_{an}^{(i)}(-\mathbf{u})=U_{an}^{(i)}(\mathbf{u})
\end{eqnarray} 
The third immediate consequence in the decorrelation scheme proposed by D\'{e}jardin et al. \cite{Dejardin2019PRB} is that in Eq.\eqref{Vc1EqBis}, one must set (by analogy with the Debye-Fr\"{o}hlich model derived in Appendix B)
\begin{eqnarray}
W^{(0)}(\mathbf{u}'')=\delta (\mathbf{u}''-\mathbf{u})=W^{(0)}(\mathbf{u}''|\mathbf{u})
\label{ConditionalW1}
\end{eqnarray}
where $W^{(0)}(\mathbf{u}''|\mathbf{u})$ is a conditional probability density (therefore it \textit{conditionates} the third body), while in Eq. \eqref{Vc2EqBis}, we must use
\begin{eqnarray}
W^{(0)}(\mathbf{u}'')=\delta (\mathbf{u}''-\mathbf{u}')=W^{(0)}(\mathbf{u}''|\mathbf{u}')
\label{ConditionalW2}
\end{eqnarray}
These last two equations finally yield Eq.\eqref{Uan}, viz.
\begin{eqnarray}
\nabla_{\mathbf{u}}U_{an}^{(i)}(\mathbf{u})=\nabla_{\mathbf{u}}U_{m}^{(i)}(\mathbf{u},\mathbf{u}'')|_{\mathbf{u}''=\mathbf{u}}
\label{Eq11}
\end{eqnarray}
Yet, this is an incomplete solution of the problem as if one takes Eq. \eqref{Eq11} literally, then the reverse interaction torque is not explicitly included as a possible solution of the potential theory problem generated by Eqs. \eqref{Vc1Eq} and \eqref{Vc2Eq}. This in turn may cause stationary points of $V_{2}^\text{eff}$ not to coincide with those of $U_m$.
In order to show that the reverse torque is also included in Eq.\eqref{Eq11} for a solution, we rewrite Eq. \eqref{Vc1EqBis} as
\begin{eqnarray}
\nonumber
\int(\nabla_{\mathbf{u}}U_{an}^{(i)}(\mathbf{u})-\nabla_{\mathbf{u}}U_{m}^{(i)}(\mathbf{u},\mathbf{u}''))\delta(\mathbf{u}''-\mathbf{u})d\mathbf{u}''=\mathbf{0},\\
\nonumber
\end{eqnarray}
where Eq.\eqref{ConditionalW1} has been used. This in turn can be rewritten
\begin{eqnarray}
\nonumber
\int(\nabla_{\mathbf{u}''}U_{an}^{(i)}(\mathbf{u}'')-\nabla_{\mathbf{u}''}U_{m}^{(i)}(\mathbf{u}'',\mathbf{u}))\delta(\mathbf{u}''-\mathbf{u})d\mathbf{u}''=\mathbf{0},\\
\nonumber
\end{eqnarray}
Now, by Newton's third law, we have
\begin{eqnarray}
\nabla_{\mathbf{u}''}U_{m}^{(i)}(\mathbf{u}'',\mathbf{u})=-\nabla_{\mathbf{u}''}U_{m}^{(i)}(\mathbf{u},\mathbf{u}'')
\end{eqnarray}
so that we have
\begin{eqnarray}
\nonumber
\int(\nabla_{\mathbf{u}''}U_{an}^{(i)}(\mathbf{u}'')+\nabla_{\mathbf{u}''}U_{m}^{(i)}(\mathbf{u},\mathbf{u}''))\delta(\mathbf{u}''-\mathbf{u})d\mathbf{u}''=\mathbf{0},\\
\nonumber
\end{eqnarray}
hence leading to a second possibility for $U_{an}$, viz.
\begin{eqnarray}
\nabla_{\mathbf{u}}U_{an}^{(i)}(\mathbf{u})=-\nabla_{\mathbf{u}}U_{m}^{(i)}(\mathbf{u},\mathbf{u}'')|_{\mathbf{u}''=\mathbf{u}}
\label{Eq11Bis}
\end{eqnarray}
Hence, we interpret Eq. \eqref{Eq11} as really meaning
\begin{eqnarray}
\nabla_{\mathbf{u}}U_{an}^{(i)}(\mathbf{u})=\mp\nabla_{\mathbf{u}}U_{m}^{(i)}(\mathbf{u},\mathbf{u}'')\vert_{\mathbf{u}''=\mathbf{u}}
\label{Eq11Meaning}
\end{eqnarray}
where the $\mp$ sign in this last equation has \emph{nothing to do} with the one we choose in a given $U_{m}^{(i)}$ term, and has to be interpreted as the lack of knowledge we have regarding the effect of the third body on the two others. As a consequence, this generates plethora of possibilities for $V_{2}^\text{eff}$ as all combinations of signs are possible. However, all $V_{2}^\text{eff}$ expressions save one can be eliminated by using the two criteria a) $V_{2}^\text{eff}$ and $U_m$ must approximately describe the same physics (by investigating the nature and location of the stationary points for both potentials) and b) the resulting values of $g_{\rm K}$ must be \textit{positive} across the \textit{widest possible range} of parameters involved in the expression for $U_m$. Again, all these considerations have been vaguely alluded to in the paper of D\'{e}jardin et al. \cite{Dejardin2019PRB}. For simplicity, these authors stated Eq. \eqref{Eq11} only without the details given here.

\section{Appendix F : Comparison with Wertheim's theories}\label{app:AppWertheim}
\begin{figure}[h!]
\centering
\includegraphics[height=7cm]{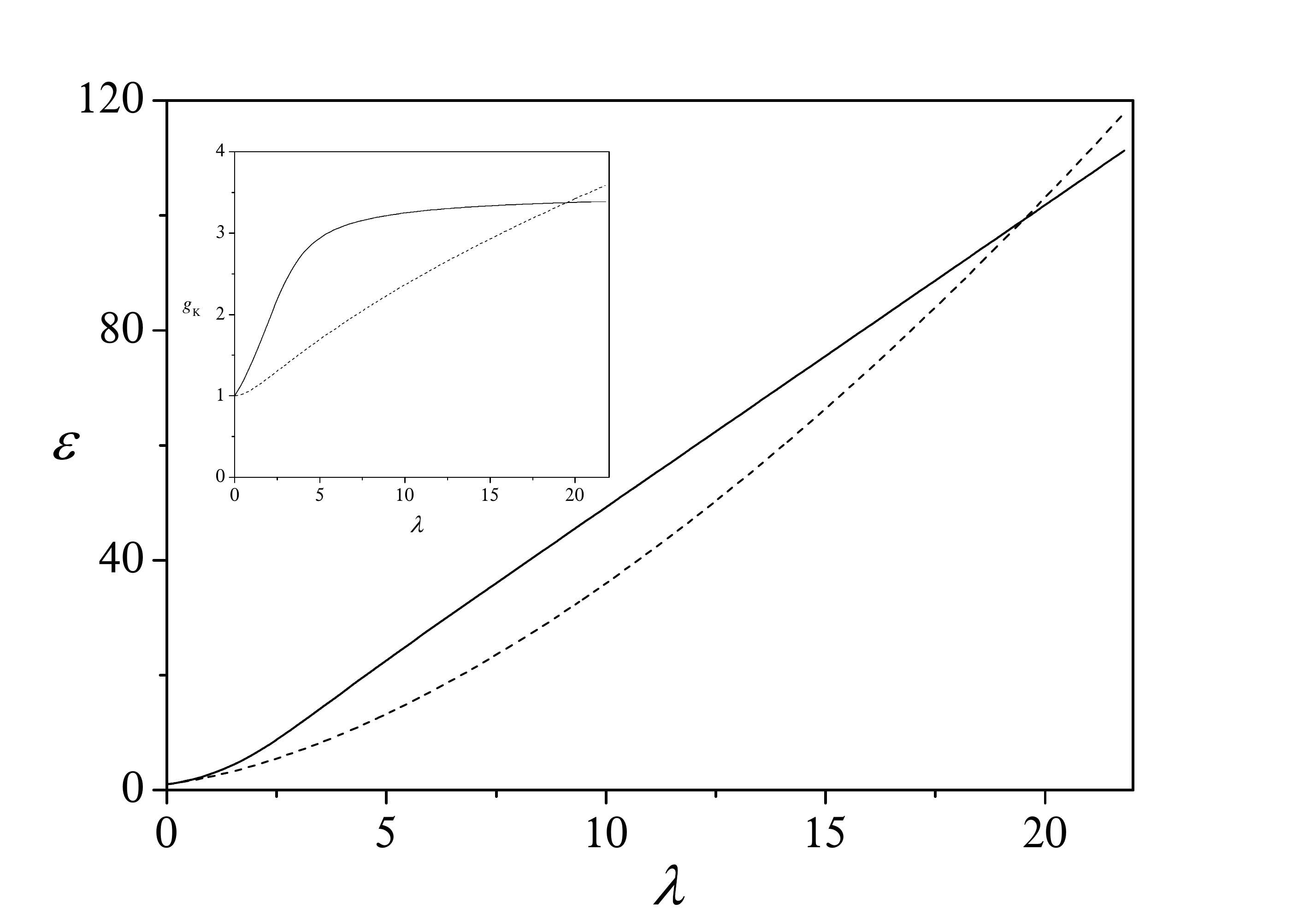}
\caption{$\varepsilon$ vs $\lambda$. Solid line : $\varepsilon$ computed with Eq.\eqref{UsableGk} and $\kappa = 0$. Dashed line : Wertheim's Eq.\eqref{WertheimEps}. Inset : $g_{\rm K}$ vs $\lambda$. Solid line : Eq.\eqref{UsableGk} with $\kappa = 0$. Dashed line : Eq. \eqref{WertheimGk}.} 
\label{plotsWertTheor}
\end{figure}

In Wertheim's 1971 statistical theory of the dielectric constant for purely polar fluids \cite{Wertheim1971JCP}, $\varepsilon$ is given by the simple expression
\begin{eqnarray}
\varepsilon=\frac{q(2x)}{q(-x)}
\label{WertheimEps}
\end{eqnarray}
where $q(x)$ is the inverse isothermal compressibility of the fluid given in the Percus-Yevick approximation by \cite{Wertheim1971JCP,Hansen2006Book}
\begin{eqnarray}
q(x)=\frac{(1+2x)^{2}}{(1-x)^{4}}
\label{PYmodulus}
\end{eqnarray}
and $x$ is determined by $\lambda$ via the self-consistency equation
\begin{eqnarray}
\lambda=q(2x)-q(-x).
\label{SelfCEq}
\end{eqnarray}
If one inserts Eqs.\eqref{WertheimEps}-\eqref{SelfCEq} in Eq.\eqref{KFEquation} with $\varepsilon_{\infty}=1$, one obtains $g_{\rm K}$ in terms of $q(x)$, viz.
\begin{eqnarray}
g_{\rm K}=\frac{1}{3}\left(\frac{1}{q(2x)}+\frac{2}{q(-x)}\right)
\label{WertheimGk}
\end{eqnarray}
Comparison between our results for the dielectric constant and $g_{\rm K}$ with $\varepsilon_{\infty} = 1,\:\kappa=0$ and Eqs.\eqref{WertheimEps} and \eqref{WertheimGk} is given in Figure \ref{plotsWertTheor}. One may see from the inset in this Figure that Eqs.\eqref{UsableGk} and \eqref{WertheimGk} predict fairly different values at all $\lambda$. In particular, Eq.\eqref{WertheimGk} predicts a value of $g_{\rm K}$ that grows almost linearly with $\lambda$ with no limiting value. In contrast, Eq.\eqref{UsableGk} grows linearly with $\lambda$ at low lambda values as predicted by Eq.\eqref{GkMF} (although with a slope different than that predicted by Eq.\eqref{WertheimGk}), but tends to a \textit{finite} value at large $\lambda$ (here, $3.5$ for $\kappa=0$ \cite{Dejardin2018JCP}). We note in passing that the fact that $g_{\rm K}$ tends to a finite value at low temperatures is in qualitative agreement with a number of experimental data \cite{Bottcher1973Book} on various polar fluids. In spite of this discrepancy, and as can be seen on Figure \ref{plots5}, the two theories predict nearly the same dielectric constant for $0\leq\lambda\leq2$. The relative deviation between the two theories is at most $40\%$ for $\lambda=10$ provided $\lambda\leq20$. This may be explained firstly by the fact that the two methods of calculation indeed differ. In effect, in Wertheim's MSA theory, the dependence of the pair density on orientational degrees of freedom of the molecules is trivial, and special caution is brought to handling the dependence of the pair density correlation function on the intermolecular distance in the best possible fashion, while in the present theory, the intermolecular positions are averaged over $g_{\rho}$ \textit{by nature}, and rotational degrees of freedom of the molecules are treated beyond the mean-field approximation, i.e., within the Kirkwood superposition approximation for the three-body orientational distribution function (which describes thermal equilibrium correctly, as shown by Singer \cite{Singer2004JCP}). Secondly, in the present treatment, the field-free one-body density depends on the rotational degrees of freedom of the molecules in a non-trivial way. Altogether, this leads to a linear variation of the dielectric constant with $\lambda$ as is often the case. Thus, Wertheim's MSA theory is essentially a mean-field one in handling long-range forces in polar fluids. Hence, it can only be applied in the dilute case, where $\lambda$ is of order unity, giving no special emphasis on orientational cross-correlations.

In order to test Wertheim's fluctuation law Eq.\eqref{WertheimFluctuationLaw} \cite{Wertheim1978MolPhys} , we assume that $g_{\rm K}$ can be calculated in the same way as in the Kirkwood-Fr\"{o}hlich theory, and therefore, that $g_{\rm K}$ is the same as in the present work. We write
\begin{eqnarray}
\frac{(\varepsilon-\varepsilon_{\infty})(2\varepsilon+\varepsilon_{\infty}^{-1})}{3\varepsilon}=\lambda_{W}g_{\rm K},\qquad\lambda_{W}=\frac{\beta\rho_{0}\mu_{W}^{2}}{3\varepsilon_{0}}
\label{WertheimKirkwood}
\end{eqnarray}
where $\mu_{W}$ is Wertheim's dipole. The positive root of the above equation is 
\begin{eqnarray}
\nonumber
\varepsilon=\frac{3\lambda_{W} g_{\rm K}+2\varepsilon_{\infty}-\varepsilon_{\infty}^{-1}}{4}+\\
\frac{1}{4}\sqrt{8+\left(3\lambda_{W} g_{\rm K}+2\varepsilon_{\infty}-\varepsilon_{\infty}^{-1}\right)^{2}}
\label{WertheimFlEps}
\end{eqnarray}
Since both theories are expected to differ substantially when $\varepsilon/\varepsilon_{\infty}$ approaches unity \cite{Wertheim1978MolPhys}, we choose three substances exhibiting a wide range of values of the dielectric constant as a function of temperature, namely, water, methanol and TBP.
\begin{figure}[h!]
\centering
\includegraphics[height=7cm]{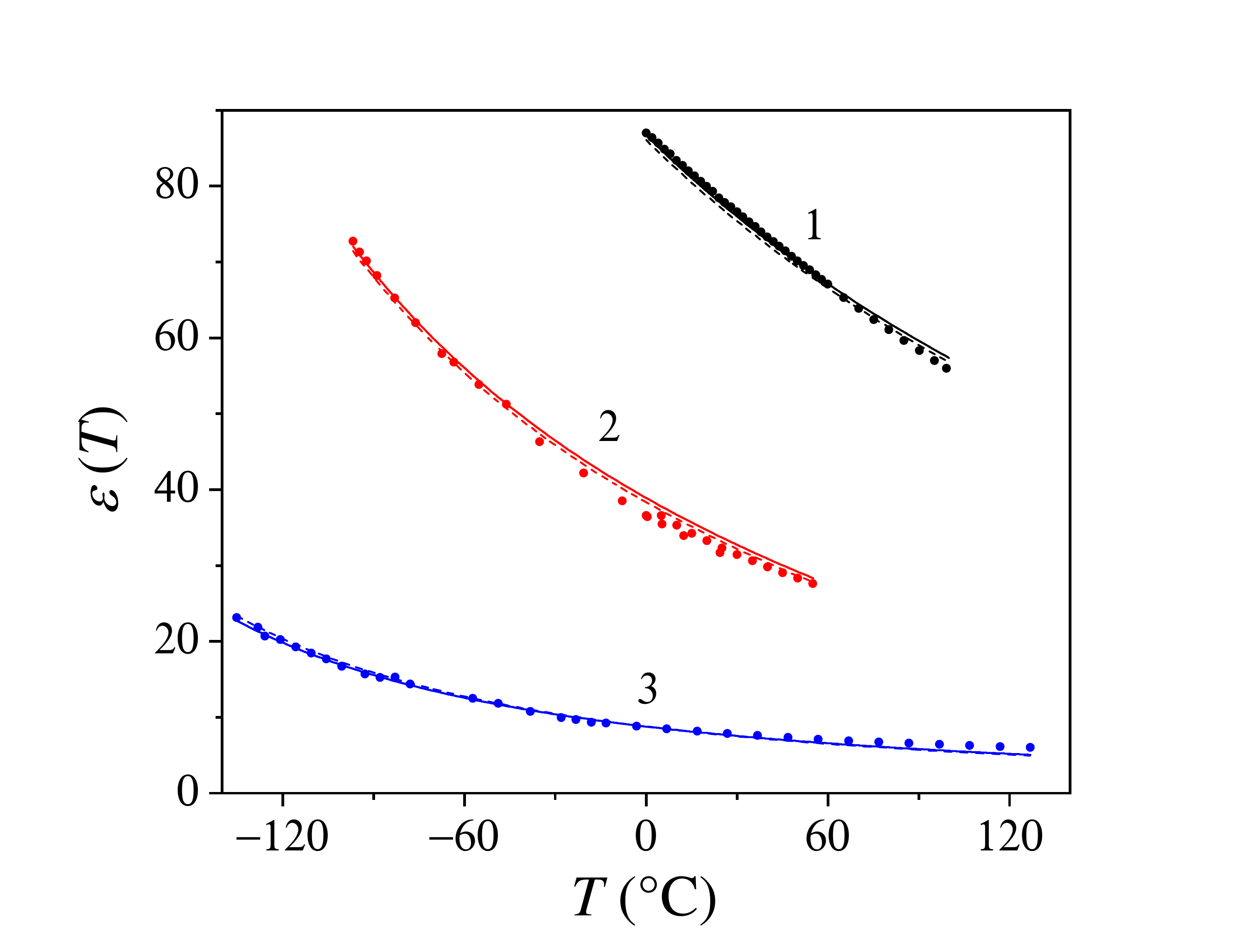}
\caption{Experimental temperature dependence of the linear static permittivity of Water (1), Methanol (2) and TBP (3). Solid lines : Eq.\eqref{WertheimFlEps}. Dots : Experimental points. Dashed lines : Eq.\eqref{FrohlichFEps}.} 
\label{plotsWertheimFrohlich}
\end{figure}
The results of the comparison between the two theories and experimental data are shown in Figure \ref{plotsWertheimFrohlich}. As can be remarked from this Figure, there is no substantial difference between the predictions of Wertheim's and Fr\"{o}hlich's equations. However, a controversy started with the publication of Wertheim's work \cite{Wertheim1978MolPhys}, since his result Eq.\eqref{WertheimFluctuationLaw} is different from Fr\"{o}hlich's Eq. \eqref{FEquation} which, at the time of writing, was believed to be a rigorous one. Attempting to resolve this controversy, Felderhof used a formalism developed by him \cite{Felderhof1977JCP} and explained on a macroscopic basis \cite{Felderhof1979JPhysC} that \textit{both fluctuation theories are correct}, but do not refer to the same macro-moments. A very much detailed discussion about this was later given by Madden and Kivelson \cite{Madden1984Book}. In effect, Madden and Kivelson derived an equation for the linear static permittivity and arrived at the conclusions that Wertheim's dipoles feel not only the effect of their neighbors, but also those of all other dipoles \textit{outside} the region of volume $\upsilon$. This conclusion is similar with Felderhof's macroscopic reasoning \cite{Felderhof1979JPhysC}. Nevertheless, Figure \ref{plotsWertheimFrohlich} demonstrates that in practice, the quantitative difference between both theories is negligible. Therefore, either theory can be used for the sake of comparing theoretical findings with experimental data, with the \textit{same} definition for the Kirkwood correlation factor and with the \textit{same} effective dipole, in agreement with Zhang et al. \cite{Zhang2016JPCL}. 
At last, it is straightforward to check that both Wertheim's and Fr\"{o}hlich's fluctuation laws agree if in Wertheim's fluctuation equation one sets the quantity $\mu_{W}$ 
\begin{eqnarray}
\mu_W=\sqrt{\frac{2\varepsilon\varepsilon_{\infty}+1}{2\varepsilon\varepsilon_{\infty}+\varepsilon_{\infty}^2}}\mu_{eff},\qquad\mu_{eff}\approx\frac{\varepsilon_{\infty}+2}{3}\mu_g,
\label{muWertheim}
\end{eqnarray}
because \textit{both} Wertheim's and Fr\"{o}hlich's fluctuation laws are correct. This illustrates Felderhof's and Madden and Kivelson's theoretical conclusions. Again, the quantitative difference between Wertheim's and Fr\"{o}hlich's laws can be ignored in practice as Figure \ref{plotsWertheimFrohlich} unambiguously demonstrates.
\section{Appendix G : The OZ and YBG approaches to the many-body problem}
In this last Appendix we describe the relation between the OZ and YBG treatments of the many-body statistical-mechanical problem.
In the OZ approach, one introduces a hierarchy of direct many-body density correlation functions via the equation \cite{Hansen2006Book}
\begin{eqnarray}
c^{(n+1)}(\mathbf{r}_{1},\mathbf{u}_{1},\cdots,\mathbf{r}_{n+1},\mathbf{u}_{n+1})=\frac{\delta c^{(n)}(\mathbf{r}_{1},\mathbf{u}_{1},\cdots,\mathbf{r}_{n},\mathbf{u}_{n})}{\delta W(\mathbf{r}_{n+1},\mathbf{u}_{n+1})},\quad n=1,2,...
\label{directCFsgen}
\end{eqnarray}
where $W(\mathbf{r},\mathbf{u})\propto\exp{c^{(1)}(\mathbf{r},\mathbf{u})}$ is the solution of the first member of the YBG hierarchy (in the absence of externally applied fields) and $c^{(1)}(\mathbf{r},\mathbf{u})$ is the excess chemical potential function. The direct pair correlation function obeys the well-known OZ relation, viz.
\begin{eqnarray}
h(\mathbf{r},\mathbf{r}',\mathbf{u},\mathbf{u}')=c^{(2)}(\mathbf{r},\mathbf{r}',\mathbf{u},\mathbf{u}')+\int c^{(2)}(\mathbf{r},\mathbf{r}',\mathbf{u},\mathbf{u}'')W(\mathbf{r}'',\mathbf{u}'')h(\mathbf{r}'',\mathbf{r}',\mathbf{u}'',\mathbf{u}')d\mathbf{r}''d\mathbf{u}''
\label{OZ1}
\end{eqnarray}
where $h=G_{2}-1$ is the pair density correlation function. Higher OZ$n$ equations ($n>1$) can be obtained by functional differentiation of Eq.\eqref{OZ1} with respect to $W$ and generate very complicated integral equations (OZ2 equations were derived by Lee \cite{Lee1973JCP} and has four equivalent forms). Let us just remark that for the moment, Eq. \eqref{OZ1} may also be taken as a definition for $c^{(2)}$, the computation of which is possible if $h$ and $W$ are known. In the context of the present work, Eq.\eqref{OZ1} can readily be simplified because $W$ does not depend on $\mathbf{r}$ and pair interactions are functions of the relative position of a pair of identical molecules, but not of the center of gravity of a pair. It follows immediately that the simpler OZ equation holds, viz.
\begin{eqnarray}
h(\vec{\rho},\mathbf{u},\mathbf{u}')=c^{(2)}(\vec{\rho},\mathbf{u},\mathbf{u}')+\rho_{0}\int c^{(2)}(\vec{\rho}-\vec{\rho}'',\mathbf{u},\mathbf{u}'')W(\mathbf{u}'')h(\vec{\rho}'',\mathbf{u}'',\mathbf{u}')d\vec{\rho}''d\mathbf{u}''
\label{OZ1SPolFl}
\end{eqnarray}
The first YBG member may be written (see the Introduction and Appendix A)
\begin{eqnarray}
\nabla_{\mathbf{u}}(\ln W(\mathbf{u})+\beta V_{1}(\mathbf{u}))=-\beta\int\nabla_{\mathbf{u}}U_{int}(\vec{\rho},\mathbf{u},\mathbf{u}')W(\mathbf{u}')h(\vec{\rho},\mathbf{u},\mathbf{u}')d\vec{\rho}d\mathbf{u}'
\label{YBG2bisbi}
\end{eqnarray}
This equation may be used to determine $W$ and constitutes one closure of the OZ equation \eqref{OZ1SPolFl} since $\nabla_{\mathbf{u}}(\ln W(\mathbf{u}))=\nabla_{\mathbf{u}}(c^{(1)}(\mathbf{u}))$ in the absence of the external potential $V_1$. However, this is insufficient since $c^{(2)}$ depends on $\vec{\rho}$ so that Eq.\eqref{directCFsgen} cannot be used for $n=1$ directly. Actually, a closure of the OZ equation \eqref{OZ1SPolFl} can be derived from the second YBG equation when the latter is transformed into an equation for $G_2$, and an approximation is inserted in order to replace $G_{3}$, the three-body distribution function. For example, it was shown by Hu et al. \cite{Hu1992Physica} using the KSA in the context of simple liquids, that all standard closures of the OZ equation (MSA, PY, HNC) could be derived from it. We think that this is also possible here for non-constant $W$ but this would bring us too far. Let us notice that if the polar fluid is considered as a simple liquid (i.e., a constant $W$), then Eq.\eqref{YBG2bisbi} becomes
\begin{eqnarray}
\int\nabla_{\mathbf{u}}U_{int}(\vec{\rho},\mathbf{u},\mathbf{u}')h(\vec{\rho},\mathbf{u},\mathbf{u}')d\vec{\rho}d\mathbf{u}'=\mathbf{0}
\label{YBG2bisbi2}
\end{eqnarray}
leading to a constant $c^{(1)}$. Indeed, if the simple liquid hypothesis is abandoned, the OZ calculation route becomes a very complicated one.

If, instead of the OZ approach, one prefers to attack the many-body problem by the YBG, then Eq. \eqref{OZ1SPolFl} becomes \textit{a definition} for $c^{(2)}$. Thus, since in principle we can calculate $W_{2}(\vec{\rho},\mathbf{u},\mathbf{u}')$ and $W(\mathbf{u})$, we can also compute $G_{2}(\vec{\rho},\mathbf{u},\mathbf{u}')$ via
\begin{eqnarray}
G_{2}(\vec{\rho},\mathbf{u},\mathbf{u}')=\frac{W_{2}(\vec{\rho},\mathbf{u},\mathbf{u}')}{W(\mathbf{u})W(\mathbf{u})}=1+h(\vec{\rho},\mathbf{u},\mathbf{u}')=g_{\rho}(\vec{\rho})w_{2}(\mathbf{u},\mathbf{u}')
\end{eqnarray}
It follows that, since $W(\mathbf{u})$ is known, we can insert the aforementioned quantities in order to compute $c^{(2)}(\vec{\rho},\mathbf{u},\mathbf{u}')$. In this context, the OZ equation \eqref{OZ1SPolFl} has yet to be solved since this function remains under the integral sign. In practice, this is difficult however, because $g_{\rho}$ has to be calculated, and this is a) far from a triviality and b) out of scope of our work. 

\balance
\bibliographystyle{unsrt}
\bibliography{biblio_pm-fp-tb}

\end{document}